\documentclass[prd,aps,twocolumn,showpacs,amsmath,amssymb,floatfix,
preprintnumbers,letterpaper,
superscriptaddress,nofootinbib]{revtex4}

\usepackage{graphicx}
\usepackage{url}
\usepackage{xspace}
\usepackage{color}
\usepackage{rotating}
\usepackage[utf8]{inputenc}
\usepackage[T1]{fontenc} 
\usepackage{bm} 
\usepackage[nolist,nohyperlinks]{acronym}
\usepackage{float}
\usepackage[caption=false]{subfig} 
\usepackage{hyperref}
\usepackage{overpic}

\definecolor{orange}{RGB}{255,127,0}

\renewcommand{\Re}{\operatorname{Re}}

\begin{document}

\preprint{LIGO-P1500142-v2}

\newcommand{\Cardiff}{School of Physics and Astronomy, Cardiff
University, Queens Building, CF24 3AA, Cardiff, United Kingdom}
\newcommand{\UIB}{Departament de F\'isica Universiat de les Illes Balears and Institut d'Estudis Espacials de Catalunya,
Crta. Valldemossa km 7.5, E-07122 Palma, Spain}
\newcommand{\AEI}{Max Planck Institute for Gravitational Physics (Albert Einstein Institute), 
Am M\"uhlenberg 1, Potsdam-Golm 14476, Germany}

\newcommand{\ICTS}{International Centre for Theoretical Sciences, Tata Institute of Fundamental Research, IISc Campus, Bangalore 560012, India}

\title{Frequency-domain gravitational waves from non-precessing black-hole binaries. \\ 
II. A phenomenological model for the advanced detector era}

\author{Sebastian Khan}
\affiliation{\Cardiff}
\author{Sascha Husa}
\affiliation{\UIB}
\affiliation{\ICTS}
\author{Mark Hannam}
\affiliation{\Cardiff}
\affiliation{\ICTS}
\author{Frank Ohme}
\affiliation{\Cardiff}
\author{Michael P\"urrer}
\affiliation{\Cardiff}
\author{Xisco Jim\'enez Forteza}
\affiliation{\UIB}
\author{Alejandro Boh\'e}
\affiliation{\UIB}
\affiliation{\AEI}

\begin{abstract}
We present a new frequency-domain phenomenological model of the 
gravitational-wave signal from the inspiral, merger and ringdown of non-precessing (aligned-spin) 
black-hole binaries. The model is calibrated to 19 hybrid effective-one-body--numerical-relativity
 waveforms 
up to mass ratios of 1:18 and black-hole spins of $|a/m| \sim 0.85$ ($0.98$ for
equal-mass systems). 
The inspiral part of the model consists of an extension of frequency-domain post-Newtonian expressions,
using higher-order terms fit to the hybrids. The merger-ringdown is based on a phenomenological 
ansatz that has been significantly improved over previous models. The model exhibits mismatches of
typically less than 1\% against all 19 calibration hybrids, and an additional 29 verification hybrids, 
which provide strong evidence that, over the calibration region, the model is sufficiently accurate for all relevant 
gravitational-wave astronomy applications with the Advanced LIGO and Virgo detectors. Beyond the 
calibration region the model produces physically reasonable results, although we recommend caution in 
assuming that \emph{any} merger-ringdown waveform model is accurate outside its calibration region. As an example, 
we note that an alternative non-precessing model, SEOBNRv2 (calibrated up to spins of only 0.5 for unequal-mass
systems), exhibits mismatch errors of up to 10\% for high spins outside its calibration region. We conclude that
 waveform models would benefit most from a larger number of numerical-relativity simulations of high-aligned-spin
unequal-mass binaries. 
\end{abstract}

\maketitle

\acrodef{PN}{post-Newtonian}
\acrodef{EOB}{effective-one-body}
\acrodef{NR}{numerical relativity}
\acrodef{GW}{gravitational wave}
\acrodef{BBH}{binary black hole}
\acrodef{BH}{black hole}
\acrodef{BNS}{binary neutron star}
\acrodef{NSBH}{neutron star-black hole}
\acrodef{SNR}{signal-to-noise ratio}
\acrodef{aLIGO}{Advanced Laser Interferometer Gravitational-wave Observatory}
\acrodef{AdV}{Advanced Virgo}
\acrodef{IMR}{inspiral-merger-ringdown}

\newcommand{\PN}[0]{\ac{PN}\xspace}
\newcommand{\EOB}[0]{\ac{EOB}\xspace}
\newcommand{\NR}[0]{\ac{NR}\xspace}
\newcommand{\BBH}[0]{\ac{BBH}\xspace}
\newcommand{\BH}[0]{\ac{BH}\xspace}
\newcommand{\BNS}[0]{\ac{BNS}\xspace}
\newcommand{\NSBH}[0]{\ac{NSBH}\xspace}
\newcommand{\GW}[0]{\ac{GW}\xspace}
\newcommand{\SNR}[0]{\ac{SNR}\xspace}
\newcommand{\aLIGO}[0]{\ac{aLIGO}\xspace}
\newcommand{\AdV}[0]{\ac{AdV}\xspace}
\newcommand{\IMR}[0]{\ac{IMR}\xspace}

\section{Introduction} 
\label{sec:intro}

The first direct \GW detection is anticipated sometime during the operation of the 
\aLIGO~\cite{Abbott:2007kv, 2010CQGra..27h4006H, advLIGO} and 
\AdV~\cite{Accadia:2011zzc}
detectors~\cite{Aasi2013}, beginning with \aLIGO in 2015. One of the most likely sources for the first detections,
and a rich source of scientific information about both fundamental physics and astrophysics~\cite{Sathyaprakash:2009xs} 
are the inspiral and merger of \BBH systems. Observations and measurements of \BBH will rely on accurate theoretical 
models of their \GW signal, and the construction of such models is currently an active research 
topic~\cite{Hannam:2013pra}.

To date most effort has focussed on binaries where the spin of each \BH
is either zero, or aligned with
the binary's orbital angular momentum. In these configurations the orbital plane and spin directions remain
fixed, and the resulting \GW signal is far simpler than in generic (precessing) configurations. Recent work has suggested
that aligned-spin models may allow detection of most (even precessing) 
binaries~\cite{Ajith2011,Harry:2013tca,Canton:2014uja}, and also that accurate
approximate generic models can be constructed based on an underlying aligned-spin 
model~\cite{Schmidt:2012rh}. 

Aligned-spin models that include the two \acp{BH}' inspiral, their merger and
the ringdown of the final \BH, are based on a combination of analytic \PN and
\EOB methods to describe the inspiral, and the calibration
of phenomenological merger-ringdown models to \NR simulations. The two classes of models are the
\emph{phenomenological} (``Phenom'') models~\cite{Ajith2007,Ajith:2007kx,Ajith2011,Santamaria2010}, which began as 
phenomenological treatments of both the inspiral 
and merger-ringdown, and \EOB
models~\cite{Buonanno:2007pf,Buonanno:2009qa,Damour:2007vq,
Damour:2008te,Damour:2009kr,Pan:2009wj,Yunes:2009ef,Pan:2011gk,
Taracchini:2012ig,Damour:2012ky,Pan2014,Taracchini2014,Damour2014,
Nagar:2015xqa}, which have used successively more sophisticated
versions of the \EOB approach to describe the inspiral all the way
to  merger, followed by the smooth connection of
a ringdown portion; \NR waveforms are used to calibrate unknown \EOB coefficients and 
free parameters in the merger-ringdown.

The original motivation of the Phenom approach was to produce an approximate and efficient waveform 
family suitable for \GW searches (the models are written as closed-form analytic expressions in the frequency 
domain), and indeed this practical approach allowed the construction of the
first aligned-spin model, often referred to as 
``PhenomB''~\cite{Ajith2011}. Although some aspects of the model were made more accurate in the succeeding 
``PhenomC'' model~\cite{Santamaria2010}, the Phenom approach is still regarded by many as approximate, 
and in particular not suitable for 
parameter estimation. This perception has been reinforced by the limited region of parameter space over which 
the aligned-spin PhenomC model was calibrated --- up to binary mass ratios of
only 1:4 (spinning up to 1:3), and \BH spins
of only $a/m \sim 0.75$ ($0.85$ for equal-mass systems). In this work (and its
companion article, which we will refer to as Paper 1), we show that
the phenomenological approach is capable of describing \BBH waveforms with a high degree of physical 
fidelity, well within the requirements of \aLIGO and \AdV, and we construct a model that is calibrated to the
largest region of parameter space to date ---  up to mass ratios of 1:18, and
spins up to $a/m \sim 0.85$ ($0.98$ for equal-mass systems). 
This constitutes the main purpose of this paper, to present our new ``PhenomD'' model, and demonstrate its
accuracy. 

In contrast, the most recent \EOB-\NR
(SEOBNRv2)~\cite{Taracchini2014} model is calibrated to \NR waveforms up to
mass ratio 1:8, and spins up to $a/m \sim 0.5$. It has been shown to be
extremely accurate within its calibration region, and it also appears to produce
physically reasonable waveforms over the full range of \BH spins, and up to much
higher mass ratios~\cite{Kumar:2015tha}..
In this work, however, we find that the SEOBNRv2
model may not accurately 
describe the 
merger-ringdown regime for high spins $a/m \gtrsim 0.7$. This finding motivates a second purpose of this 
paper: to make clear that the accuracy of \emph{any} merger-ringdown model, Phenom, EOB-NR, or otherwise,
is only as good as its \NR calibration region. The model may give physically plausible results, but its 
accuracy cannot be guaranteed until it has been checked against fully general relativistic \NR calculations,
and its accuracy may well be poor until it has been calibrated to those simulations. This seemingly obvious
observation bears emphasising. It also motivates efforts to quantify the accuracy of \PN and \EOB calculations
increasingly far back into the inspiral~\cite{Szilagyi:2015rwa,Kumar:2015tha}. 

Another important contribution of the Phenom programme has been to isolate which combinations of physical
binary parameters will be measurable in \GW observations. For example, the previous aligned-spin 
Phenom models~\cite{Ajith2011,Santamaria2010} exploited the observation that the dominant spin effect
on the \GW phase is due to a weighted combination of the individual \BH spins, and the models depend on
only two physical parameters, the symmetric mass ratio and this single effective spin parameter. The 
identification of a simple combination of the in-plane spin components in generic binaries~\cite{Schmidt:2014iyl} 
in turn led to a simple extension of PhenomC to produce a generic-binary model, 
PhenomP~\cite{Hannam:2013oca}.

A corollary
of this parameter-space reduction is that individual spins are expected to be difficult to measure from \GW
observations, even if we have a two-spin model to hand. Based on previous 
studies~\cite{Ajith:2011ec,Puerrer2013}, and a forthcoming study that illustrates in detail the difficulty of 
measuring individual spins with an aligned-spin model~\cite{Puerrer2015}, we also use an effective 
reduced spin parameter in certain parts of the PhenomD model. We will
nonetheless pursue the extension of the Phenom approach
to two spins in future work. 

An additional feature of the PhenomD model is its modularity. The separate inspiral and merger-ringdown 
parts of the model are connected by the requirement of continuity in the phase and amplitude. This 
simple construction makes it straightforward to improve and change either part of the model independently.
We make use of this feature to compare versions with alternative choices for the inspiral part of the model.

This paper is organized as follows. In Paper 1 we discussed in detail the numerical simulations we have used,
and in particular presented studies of the accuracy of the new \NR waveforms that we have produced. 
In this paper we re-visit these waveforms, but from the point of view of \GW applications, and assess their
accuracy in terms of their noise-weighted inner product (match). The match is defined in Sec.~\ref{sec:match},
along with techniques that we use to estimate the match between \NR waveforms over frequency ranges that
extend beyond those where we have \NR data. In Sec.~\ref{sec:NR} we summarize the waveforms that we 
use, and present our match-based accuracy analysis. In Secs.~\ref{sec:MRmodel} and
\ref{sec:Inspiral} we give details of procedure we use to
construct our models of the signal phase and amplitude, over three frequency regions. More details are 
provided in Paper 1, but here we summarize the approach, and its use across all of the waveforms used to 
calibrate our model, and the accuracy of the final models for each of its six constituent parts (three phase
parts, and three amplitude parts). In Sec.~\ref{sec:modelval} we assess the final complete model's accuracy 
by calculating matches against both the waveforms used for calibration, and an additional set of waveforms that
were \emph{not} used for calibration. We discuss the accuracy of our single-reduced-spin approximation, 
and our choice of the minimal set of waveforms necessary for the model. In Sec.~\ref{sec:modelcomparisons}
we compare against the SEOBNRv2 model, illustrating the high-spin, unequal-mass
region
where we find disagreement between the two models; this is outside the calibration region of 
SEOBNRv2.
In Appendix~\ref{app:td} we revisit the agreement between our new model and the original
NR data by transforming PhenomD to the time domain, and in Appendix~\ref{sec:app_pncoeffs} we list the
\PN inspiral coefficients used in our model.

\section{Preliminaries}
\label{sec:prelim}

\subsection{Outline of the model}

We describe a \BBH system by the following parameters. The masses are $m_1$ and $m_2$, where we choose
$m_1 > m_2$, and the total mass is $M = m_1 + m_2$. The mass ratio of the binary is denoted $q = m_1/m_2 \geq 1$, 
and the symmetric mass ratio is $\eta = m_1 m_2 / M^2$. The \BH spin angular
momenta are $\vec S_1$ and
$\vec S_2$ which we assume to be parallel to the direction of the orbital
angular momentum, $\hat L$. In this work we 
restrict ourselves to aligned-spin (non-precessing) systems, and so are only
concerned with the dimensionless spin parameters defined as
\begin{equation}
 \chi_i = \frac{\vec S_i \cdot \hat L}{m_i^2},
\end{equation}
with $\chi_i \in [-1,1]$. 

In previous aligned-spin Phenom models, we have parameterized the spin depedence of the model by a single
effective spin parameter \cite{Ajith2011,Santamaria2010},
\begin{equation}
\chi_{\rm eff} = \frac{m_1 \chi_1 + m_2
\chi_2}{M}. 
\end{equation}
This is based on the 
observation that it is a weighted sum of the spins that constitutes the dominant effect of the spin on the inspiral
of the binary. In \PN theory, the leading-order spin effect on the binary's phasing is in 
fact~\cite{Cutler:1994ys,Poisson:1995ef,Ajith:2011ec}, \begin{equation}
\chi_{\rm PN} = \chi_{\rm eff} - \frac{38 \eta}{113} (\chi_1 + \chi_2),
\end{equation} and we have seen evidence in previous work that this is in general a better parameter to use also
in \IMR models~\cite{Puerrer2013}.
In this work we use a combination of spin parametrizations. The phenomenological calibrations to \NR 
waveforms are parameterized by $\chi_{\rm PN}$,  normalized
such that its range is from -1 to 1 for all mass ratios, \begin{equation}
\hat\chi = \frac{\chi_{\rm PN}}{1 - 76 \eta/113}.
\label{eqn:chihat}
\end{equation} 

The final \BH is correctly parameterized by the final mass $M_f$
and 
spin $a_f$, and for this reason the final mass 
and spin estimates that we use (see Paper 1), are parameterized by a different spin 
combination, $S_1 + S_2$. Finally, our inspiral model is based on the standard frequency-domain \PN approximant,
``TaylorF2''~\cite{Damour:2000zb, Damour:2002kr, Arun:2004hn}, and this is
parameterized by \emph{both} spins, $\chi_1$ and $\chi_2$. The final result is a
model that depends on both spins $\chi_1$ and $\chi_2$, but the calibration to hybrid EOB+NR waveforms is parameterized
by different combinations of $\chi_1$ and $\chi_2$ for the inspiral, merger and ringdown parts of the model. 
Most of the hybrid waveforms are for equal-spin
$\hat\chi = \chi_1 = \chi_2$ systems, so we can guarantee our model's accuracy only for these configurations.
However, as we discuss in Sec.~\ref{sec:chiapprox}, the $\hat\chi$ approximation is extremely accurate for
most regions of parameter space, and in those where it is not (higher mass ratios and high parallel spins), the 
innacurracy is unlikely to have any influence on \GW astronomy applications with \aLIGO or \AdV. 

The PhenomD model provides expressions for the $\ell =2, |m| = 2$ spin-weighted spherical-harmonic modes
of the \GW signal, since these are the dominant modes in aligned-spin systems. The full signal as a function of the
physical parameters $\Xi \in (M, \eta, \chi_1, \chi_2)$ and the observer's orientation $(\theta, \phi)$ with respect to 
the orbital angular momentum of the binary, is given by, 
\begin{align}
\tilde{h}(f;\Xi, \theta, \phi) &= \tilde h_+(f;\Xi, \theta, \phi) - i \tilde
h_\times(f;\Xi, \theta, \phi) \\ &=\sum_{m=-2,2} \tilde{h}_{2m} (f; \Xi) \
^{-2}Y_{2m} (\theta,\phi),
\end{align} 
where $\tilde{h}_{2,-2}(f) = \tilde{h}^*_{2,2}(-f)$. We express
$\tilde{h}_{22}(f)$ in terms of the
signal amplitude and phase by \begin{equation}
\tilde{h}_{22}(f; \Xi) = A(f; \Xi) e^{-i \phi(f; \Xi)}, 
\end{equation} and it is models of $A(f; \Xi)$ and $\phi(f; \Xi)$ that we provide.
Note also that the 
total mass $M$ provides an overall scale for our waveforms, so the physical parameters over which the model
has been explicitly constructed are $\eta$, $\chi_1$ and $\chi_2$ (with the spins treated in combinations as
described above). 

As ingredients in our model construction, we use hybrid waveforms, where the early inspiral is described by 
the un-calibrated SEOBv2 model (see Paper 1, and Sec.~\ref{sec:inspiral_choice}
below), and the late inspiral
and merger-ringdown by NR
waveforms. The mass and spin of the final \acp{BH}, $M_f$ and $a_f$, which are
key parts of the 
merger-ringdown model, are provided by fits to \NR data. The details of the hybrid construction, and of the 
final mass and final spin fits, are given in Paper 1. 

We model separately three frequency regimes of the waveform. The first region covers the inspiral, up to the
frequency $Mf = 0.018$. Here the information is predominantly from the analytical \EOB inspiral waveforms, 
although there is some information at higher frequencies from the early parts of the longer \NR waveforms;
the frequency at which each hybrid switches to an \NR waveform is provided in Tab.~\ref{tab:wftable}.
The second two regions are informed purely from \NR data. We note that in principle one could also construct the
individual inspiral and merger-ringdown models separately from \PN or \EOB models (for the inspiral) and
\NR data (for the merger-ringdown), without constructing any hybrid waveforms. In this work we chose to 
use hybrid waveforms, because they allow us to use the maximum \NR information (which influences to 
some extent our inspiral model), and allows for a consistent choice of calibration points in parameter space
for both the inspiral and merger-ringdown. 

The resulting model is modular: we are free to use a different inspiral model,
or a different merger-ringdown model, as we wish. This introduces a flexibility
that was not present in previous models. If in the future we have access to a
more accurate inspiral model (\EOB, \PN, or otherwise), or more accurate
merger-ringdown model (e.g., calibrated to waveforms over a larger region of
parameter space), then we can easily replace that part of the model without any
additional tuning. The model calculates appropriate time- and phase-shifts (a
linear correction to the frequency-domain phase) to ensure that the phase
connects smoothly between the inspiral and merger-ringdown, and the model of the
amplitude in the intermediate region between inspiral and merger-ringdown is
constructed such that the function is continuous.

\subsection{Matches} 
\label{sec:match} 

To assess the accuracy of our model and generally quantify the (dis)agreement
between two waveforms $h_1$ and $h_2$ (real-valued in the time domain), we use
the standard inner product weighted by the power spectral density of the
detector $S_n(f)$. It is defined
as \cite{Cutler:1994ys}, 
\begin{equation}
\langle h_1, h_2 \rangle = 4 \, \Re \int_{f_{\rm min}}^{f_{\rm max}}
\frac{\tilde h_1(f) \; \tilde h^\ast_2(f)}{S_n(f)} \, df. \label{eq:innerprod}
\end{equation}

The \emph{match} between two waveforms is defined as the
inner product between \emph{normalised} waveforms $(\hat{h}=h/\sqrt{\langle h, h
\rangle})$
maximised over relative time and phase shifts between the two waveforms,
\begin{equation}
M(h_1,h_2) = \max_{t_0,\phi_0} \, \langle \hat{h}_1 , \hat{h}_2 \rangle.
\end{equation} 
A time- and phase-shift has no significance for the physical fidelity of an
aligned-spin
waveform --- they correspond, respectively, to a change in the merger time of the binary, and of the initial
phase of the binary, i.e., an overall rotation. 

Results will be quoted in terms of the \emph{mismatch} $\mathcal{M}$, defined as,
\begin{equation}
\mathcal{M}(h_1,h_2)= 1 - M(h_1,h_2).
\end{equation} 

We use two noise spectra in this work: the ``early \aLIGO'' spectrum, which
approximates the
detector response during the first observing run, 
planned for late 2015, and the ``zero-detuned high-power'' (zdethp) spectrum,
which is the design goal of \aLIGO that is anticipated
by 2019-20~\cite{Aasi:2013wya}. Calculations with the early \aLIGO curve use
a lower cutoff frequency of $f_{\rm min} = 30\,{\rm Hz}$, and
zdethp calculations are carried out with $f_{\rm min} = 10\,{\rm Hz}$. In both
cases, we use $f_{\rm max} = 8000\,{\rm Hz}$ which is greater than the highest
frequencies contained in the signals we are considering.

In various steps of the model construction in this paper, we are interested in
analyzing the agreement of waveform sections that are only defined over a
certain frequency range. (A good example are \NR waveforms that are typically
too short to fill the entire \aLIGO frequency band.) In these cases, one could
reduce the integration limits in \eqref{eq:innerprod} to the frequency range
defined by the waveform sections, but the resulting matches would be difficult
to interpret as they have no direct application in \GW searches. Here instead,
we ask the question ``What influence does the difference in a certain signal
part have on the full waveform, assuming all other parts are perfectly
modeled?''
We address this question by aligning the signal parts that we wish to
compare as if they were hybridized with a common model of the remaining signal
and set the phase difference for this particular alignment to zero over all
frequencies that are not covered by the waveform sections we consider. To
construct the full integrand in \eqref{eq:innerprod}, we additionally need a
model of the amplitude, which we take from our final PhenomD model, although
this particular choice is far less important than the phase disagreement we wish
to quantify. We can then use a standard algorithm to calculate the mismatch
between both signals, and due to their simple form in the frequency domain,
time and phase shifts will be properly taken into account across the entire
signal. More details and a full discussion of this approach is given in
\cite{Ohme:2011zm}.

\section{Numerical-Relativity Waveforms}
\label{sec:NR}

We calibrated the PhenomD model with publicly available \NR waveforms from the
Simulating Extreme 
Spacetimes (SXS) collaboration \cite{SXS:catalog}, 
and a set of new simulations produced with the BAM code~\cite{Bruegmann2008, Husa2008}. 
Details of the new BAM simulations and their numerical accuracy
are presented in Paper 1. Here we summarize the 19 NR waveforms that we used to calibrate the model, and the 
additional waveforms that were used to further test its accuracy. 

Our two main goals are to extend the parameter-space coverage of aligned-spin phenomenological models to higher mass 
ratios, and to improve the overall accuracy to well within the requirements of GW detection and parameter estimation with 
Advanced LIGO and Virgo; in practice we consider a mismatch error of less than 1\% to be sufficient. 
The first goal dictated our choice of new \NR simulations. 

The previous aligned-spin phenomenological models, PhenomB \cite{Ajith2011} and PhenomC \cite{Santamaria2010}, 
were constructed from waveforms up to mass ratios of 1:4, and (equal) spins up to $\pm 0.75$
(with $\pm 0.85$ for equal-mass binaries), although spinning-binary waveforms were used only up to 
mass-ratio 1:3. We found in constructing those models
that it was sufficient to use only four or five \NR waveforms in each direction
of parameter space. This suggests that we can construct
a model across the entire $(\eta, \hat\chi)$ parameter space with only 25 waveforms. 

Five waveforms equally spaced in $\eta$ would be placed at $\eta = (0.25, 0.20, 0.15, 0.10, 0.05)$. 
(In the current model we do not include extreme-mass-ratio $\eta \rightarrow 0$ 
waveforms, e.g., Refs.~\cite{Harms:2014dqa,Taracchini:2014zpa}, but we plan to use these to 
complete our parameter-space mapping in future work).
We focus on simulations at
mass ratios $q = 1, 4, 8, 18$, which correspond to $\eta \approx (0.25, 0.16, 0.10, 0.05)$; we find that waveforms at 
$\eta \approx 0.2$
are not necessary to produce an accurate model, although the model is tested against waveforms at $q = 2, 3$ 
($\eta = 0.222, 0.1875$). 

\begin{figure}[tb]
\begin{center}
  \includegraphics[width=0.4\textwidth]{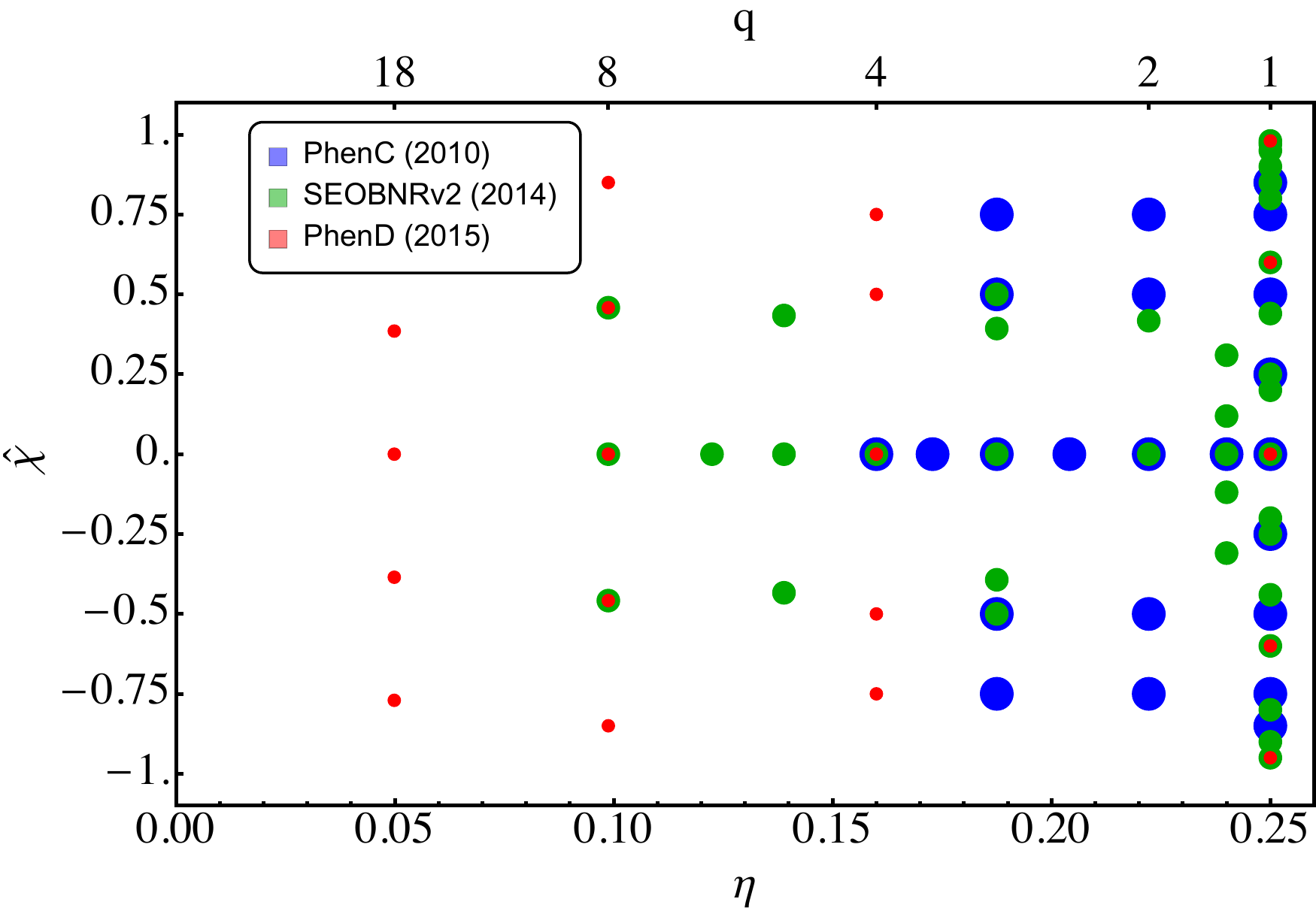}
  \caption{
  \label{fig:ParSpace} 
  Parameter space over which the PhenomD model has been calibrated. The locations in parameter space
  of the calibration waveforms are indicated by red points. Also shown are the calibration points for the 
  SEOBNRv2 (green)
  and PhenomC (blue) models.
  }
\end{center}
\end{figure}

We produced new waveforms with the BAM code up to mass ratio 1:18, and for a range of spins.  
At lower mass ratios we have also used publicly available waveforms, which were
produced by the SXS collaboration using the Spectral Einstein Code (SpEC). In
particular, their catalogue provides waveforms 
for equal-mass binaries with high \BH spins of $-0.95$ and $+0.98$. The
parameter space coverage of \NR 
waveforms used in previous models, and in our new
model, are shown in Fig.~\ref{fig:ParSpace}, and the details of the waveforms that we used are summarized in 
Tab.~\ref{tab:wftable}. We tested the model against an extended sent of waveforms, and this is described in more detail in 
Sec.~\ref{sec:modelval} and Tab.~\ref{tab:wftable2}.

The accuracy of the new BAM simulations was discussed in some detail in Paper 1. 
In this work we are interested in constructing accurate waveform models for GW astronomy with \aLIGO
and \AdV. In that context, an important accuracy measure is the mismatch between the waveforms with
respect to the \aLIGO noise spectrum. We calculate the mismatch between the numerical waveforms following
the procedure outlined in Sec.~\ref{sec:match}; in particular, we take into account the inspiral signal power, allowing us
to calculate mismatches for low-mass systems, and reliably infer the (typically larger) mismatches in these
systems due to any errors in the merger-ringdown waveforms. This procedure tends to estimate larger mismatches
than integrating Eq.~(\ref{eq:innerprod}) over only the frequency range of the \NR waveforms, as in, e.g., 
Ref.~\cite{Hinder2013}, and is a more conservative estimate of the mismatch error in the \NR waveforms.

We consider the effect of two sources of error on the mismatch: the errors due to (1) finite numerical resolution,
and (2) finite waveform extraction radius. In all cases we have found the overall mismatch error from these sources
to be $< 0.5$\%. Here we focus on two configurations, $q=4$, $\chi_1 = \chi_2 = \hat\chi= 0.75$ (A10), and 
nonspinning $q = 18$ (A18). 

\begin{figure}[tb]
\centering
\includegraphics[width=0.8\linewidth]{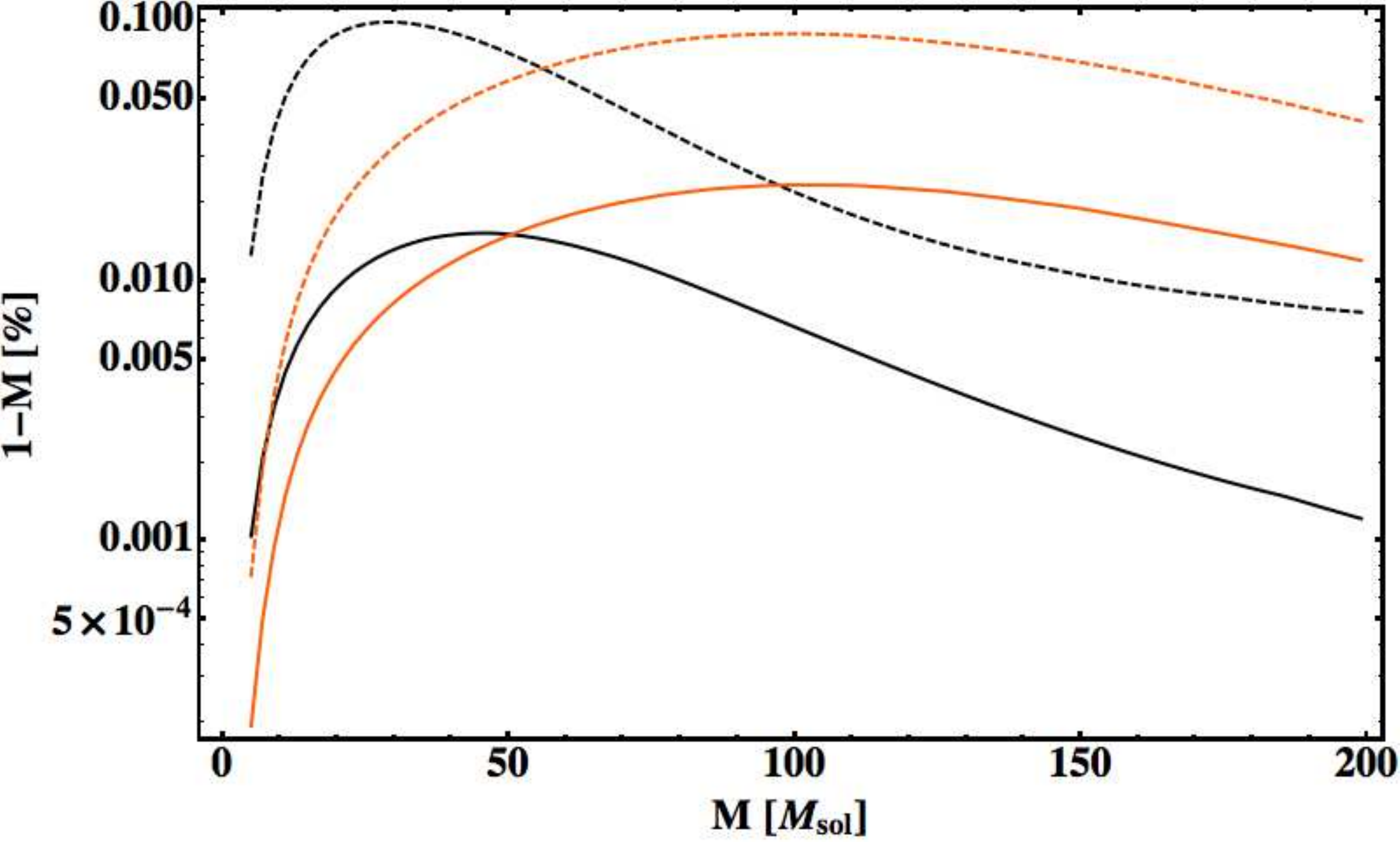}
\caption{
Mismatch error due to numerical resolution, for the $q=4$, $\chi_1 = \chi_2 =
\hat\chi = 0.75$
(black lines) and non-spinnning $q=18$ simulations (orange lines). 
The solid black line shows the mismatch between waveform $q=4$ 112- and 96-point
simuations,
and the dashed black line shows the mismatch between the 96- and 80-point
simulations. 
For the $q=18$ configuration, the solid orange line shows the mismatch between
the 
144- and 120-point simulations, and the dashed orange line shows the mismatch
between the 
144- and 96-point simulations (see text).
}
\label{fig:q4q18resolutions}
\end{figure}

\begin{figure}[tb]
\centering
\includegraphics[width=0.8\linewidth]{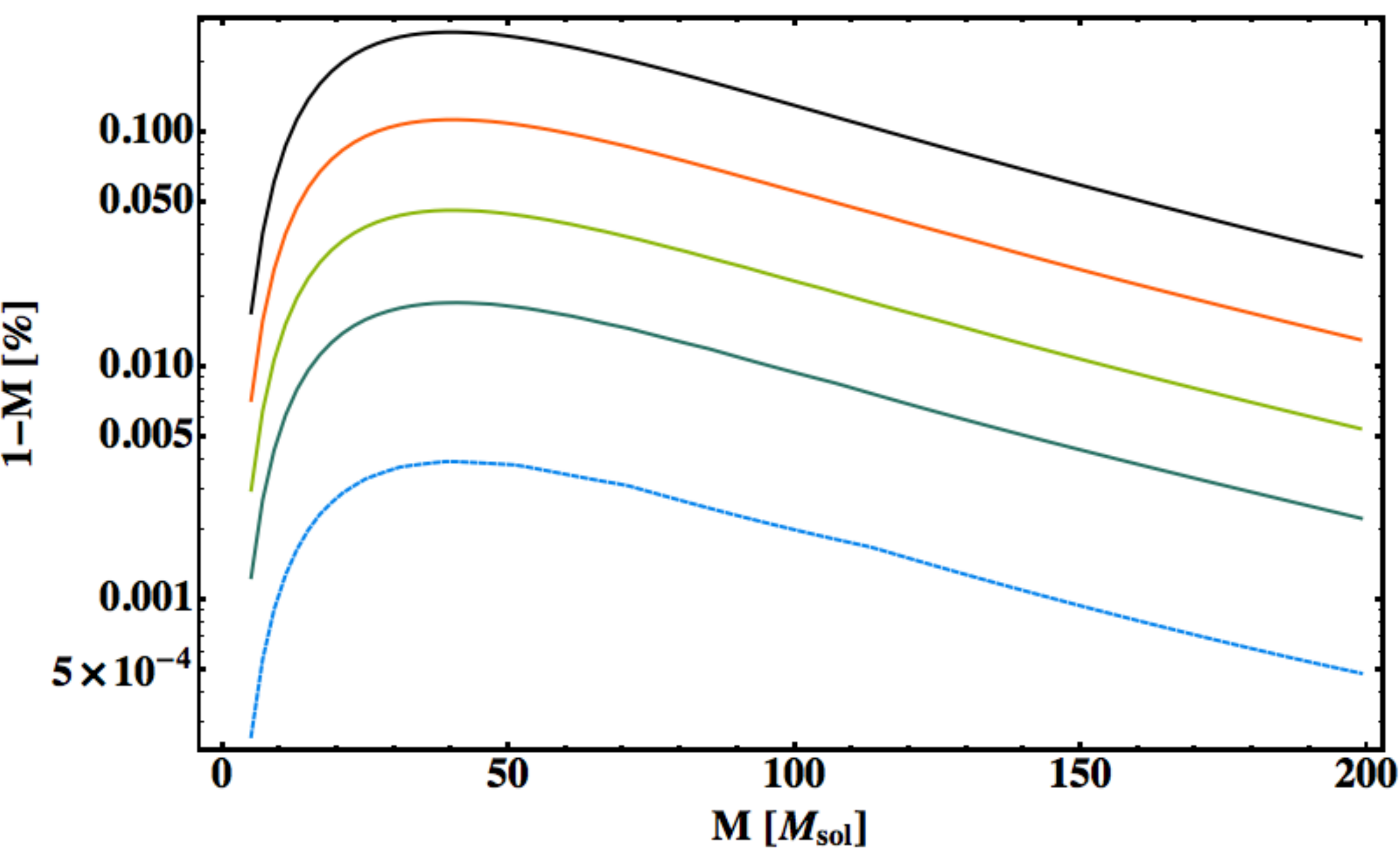}
\caption{
Mismatch errors due to finite-radius waveform extraction for the 120-point
simulations of
the same $q=4$ case as in Fig.~\ref{fig:q4q18resolutions}. Mismatches are
between the 
$R_{ex} = 100\,M$ waveform and those extracted at $R_{ex} = \{50, 60, 70, 80,
90\}\,M$
(from top to bottom). 
}
\label{fig:q4extrac}
\end{figure}

Fig.~\ref{fig:q4q18resolutions} shows the mismatch error due to numerical resolution. In the $q=4$ configuration,
the reference simulation uses a base grid size of $112^3$ points, with the finest grid spacing being 
$h_{\rm min} = M/230$. 
Comparisons are made against simulations with the same resolution, but a base grid size of $96^3$ points, 
and an $80^3$ simulation with the resolution scaled to give the same physical grid sizes as in the $96^3$ 
simulation. The solid black line shows the mismatch between the $112$-point and $96$-point simulations, i.e.,
simulations where only the physical grid sizes were changed. This change introduces a mismatch error of 
at most $\sim 0.01$\%. The dashed black line shows the mismatch between the $112$-point and $80$-point 
simulations, i.e., both the physical grid sizes and the numerical resolution have been reduced. Here the 
mismatch difference is at most $\sim 0.1$\%. 

The orange lines show the mismatch between the $q=18$ waveforms, with grid sizes of $96^3$, $120^3$
and $144^3$ points. These three simulations constitute a convergence series, and we have shown in  
Paper 1 that they exhibit evidence of sixth-order convergence. The solid orange line shows the mismatch 
between the $144^3$ and $120^3$ simulations, and the dashed orange line shows the mismatch between
the $144^3$ and $96^3$ simulations. The higher mismatches at high mass, compared to the $q=4$ configuration,
suggests that the merger-ringdown errors are larger in this case, although their effect on the mismatches at
lower masses is comparable. We again conclude that the waveforms are accurate to well within our 1\%
criterion.

Although the convergence of our simulations is in general unclear, we typically find that our 80-point 
simulations are not in the convergence regime, and are much less accurate than higher-resolution 
simulations. We therefore expect that if the mismatch between the 112-point and 80-point simulations
is no larger than 0.1\%, then the mismatch between the 96- or 112-point simulations and the continuum
limit will be lower than this; it will certainly be lower than the 1\% accuracy requirement that we place on 
our model. 

Fig.~\ref{fig:q4extrac} shows the mismatch between waveforms extracted at different radii. The waveforms
were extracted at $R_{ex} = \{50, 60, 70, 80, 90, 100\}\,M$, and the mismatch calculations are performed
against the $R_{ex} = 100\,M$ waveforms. We expect the error to fall of as $\sim 1/R_{ex}$, and in general
we observe this for our simulations, but only for $R_{ex} \gtrsim 60\,M$. Since even the $R_{ex} = 50\,M$ 
waveform has a mismatch of only $\sim 0.3\%$ with the $R_{ex} = 100\,M$ waveform, and assuming a
$1/R_{ex}$ fall-off in waveform extraction error, we expect that the contribution of this error to the 
$R_{ex} = 100\,M$ waveforms is less than 0.1\%. 

Based on this analysis, we conclude that our simulations are well within the accuracy requirements to construct
a waveform model with an overall mismatch error of $\lesssim 1$\%.

\begin{table*}
\begin{tabular}{lllllllllllc}
  \hline
  \hline
  \# & \text{Code/ID} & \text{q} & $\eta$  & \text{$\chi_1$} & \text{$\chi_2$} &
  $\hat\chi$ & $M_f$ & $a_f$ & $Mf_{\text{RD}}$ & $Mf_{\text{hyb}}$ & $N_{\rm GW,NR}$ \\
  \hline
 A1 & \text{SXS:BBH:0156} & 1. & 0.25 & -0.95 & -0.95 & -0.95 & 0.9681 & 0.3757
& 0.0713 & 0.00522 & 22 \\
 A2 & \text{SXS:BBH:0151} & 1. & 0.25 & -0.6 & -0.6 & -0.6 & 0.9638 & 0.4942 &
0.0764 & 0.00517  & 26 \\
 A3 & \text{SXS:BBH:0001} & 1. & 0.25 & 0. & 0. & 0. & 0.9516 & 0.6865 & 0.0881
 & 0.00398 & 54 \\
 A4 & \text{SXS:BBH:0152} & 1. & 0.25 & 0.6 & 0.6 & 0.6 & 0.9269 & 0.8578 &
0.1083 & 0.00501 & 42 \\
 A5 & \text{SXS:BBH:0172} & 1. & 0.25 & 0.98 & 0.98 & 0.98 & 0.8892 & 0.9470 &
0.1328 & 0.00497 & 48 \\
 A6 & \text{BAM} & 4. & 0.16 & -0.75 & -0.75 & -0.75 & 0.9846 & 0.0494 & 0.0614
 & 0.00713 & 15 \\
 A7 & \text{BAM} & 4. & 0.16 & -0.5 & -0.5 & -0.5 & 0.9831 & 0.1935 & 0.0649 &
 0.00716 &  18 \\
 A8 & \text{SXS:BBH:0167} & 4. & 0.16 & 0. & 0. & 0. & 0.9779 & 0.4715 & 0.0743
 & 0.00665 & 28 \\
 A9 & \text{BAM} & 4. & 0.16 & 0.5 & 0.5 & 0.5 & 0.9674 & 0.7377 & 0.0906 &
 0.00811 & 26 \\
 A10 & \text{BAM} & 4. & 0.16 & 0.75 & 0.75 & 0.75 & 0.9573 & 0.8628 & 0.1054 &
 0.00818 & 30 \\
 A11 & \text{BAM} & 8. & 0.099 & -0.85 & -0.85 & -0.85 & 0.9898 & -0.3200 &
0.0546 & 0.00918 & 8 \\
 A12 & \text{SXS:BBH:0064} & 8. & 0.099 & -0.5 & 0. & -0.458 & 0.9923 & -0.0526
& 0.0589 & 0.00632 & 36 \\
 A13 & \text{SXS:BBH:0063} & 8. & 0.099 & 0. & 0. & 0. & 0.9894 & 0.3067 &
0.0677 & 0.00623 & 49  \\
 A14 & \text{SXS:BBH:0065} & 8. & 0.099 & 0.5 & 0. & 0.458 & 0.9846 & 0.6574 &
0.0838 & 0.00615 & 66 \\
 A15 & \text{BAM} & 8. & 0.099 & 0.85 & 0.85 & 0.85 & 0.9746 & 0.8948 & 0.1087
 & 0.01580 & 15 \\
 A16 & \text{BAM} & 18. & 0.05 & -0.8 & 0. & -0.77 & 0.9966 & -0.5311 & 0.0514
 & 0.01035 & 14 \\
 A17 & \text{BAM} & 18. & 0.05 & -0.4 & 0. & -0.385 & 0.9966 & -0.1877 &
0.0563 & 0.01283 & 15 \\
 A18 & \text{BAM} & 18. & 0.05 & 0. & 0. & 0. & 0.9959 & 0.1633 & 0.0633 &
 0.01284 & 13 \\
 A19 & \text{BAM} & 18. & 0.05 & 0.4 & 0. & 0.385 & 0.9943 & 0.5046 & 0.0745
 & 0.00916 & 23 \\
\hline
 \hline
\end{tabular}
\caption{
    \label{tab:wftable}
        Hybrid waveform configurations used to calibrate the PhenomD model. For each
configuration we list both the 
    mass ratio $q$ and symmetric mass ratio $\eta$, along with the spins
$\chi_1$ and $\chi_2$ and the reduced-spin
    combination, $\hat\chi$, which follows from Eq.~(\ref{eqn:chihat}). The
final \BH has mass $M_f$ and 
    dimensionless spin $a_f$, and the ringdown signal has frequency $Mf_{\rm
RD}$. The frequency $Mf_{\text{hyb}}$ marks the midpoint of the transition region between
 SEOBv2 inspiral and \NR data. The approximate number of \NR \GW cycles in each hybrid
 is given by $N_{\rm GW,NR}$.
    }
    
\end{table*}

\section{Choice of inspiral approximant}
\label{sec:inspiral_choice}

The early, gradual inspiral of compact binaries and the \acp{GW}
they emit can be accurately modeled by expanding the energy and flux of the
system into a \PN series. Depending on how the underlying equations are formulated
and solved, there is a variety of \PN approximants, each consistent with the
others when truncated at the same expansion order. However, as every approximant
is formulated with different, mostly implicit, assumptions of how higher order
terms are treated, the \GW signals they predict can differ considerably,
especially towards higher mass ratios, increased spin magnitudes and for
increasing orbital frequencies
\cite{Hannam:2007wf,Buonanno2009,Hannam:2010ec,Ohme:2011zm,MacDonald:2011ne,
MacDonald:2012mp,Nitz:2013mxa}. There are sophisticated methods that aim to
improve the convergence and accuracy of \PN-based approximants, and one of the
most successful approaches is the mapping to an \EOB system
\cite{Buonanno:1998gg,Buonanno:2000ef,Damour:2001tu}.

In the construction of a complete waveform model we face the following
two issues. First, we need to pick one approximant that, to our
current knowledge, models the inspiral most accrately. Second, this inspiral
description has to be complemented by \NR-based information about the
merger and ringdown. We briefly summarize our strategy to address both issues
below and give references to the following sections that describe our reasoning
in more detail.

Recent studies have indicated that among the family of non-precessing inspiral
approximants, the \EOB approximant by Tarachini et al.\
\cite{Taracchini2014} shows the most consistent agreement with \NR
simulations within the calibration range of the model
\cite{Szilagyi:2015rwa,Kumar:2015tha}. In Paper~1, we
have performed an independent consistency test between inspiral approximants and
our set of \NR data and confirmed this conclusion.
(Note that the most recently
calibrated version of a non-precessing \EOB model \cite{Nagar:2015xqa} has not
yet been included in any of these tests.) Hence, we used the Tarachini et al.\
model (dubbed SEOBNRv2 in the publicly available LIGO software library
\cite{lalsuite}) as our target inspiral approximant, albeit in its original,
uncalibrated form that does not include \NR fitted corrections (we refer to
this form as as SEOBv2). Specifically, this involves calculating the SEOBNRv2 waveforms
with all of the \NR calibration terms set to zero, to provide an ``uncalibrated'' SEOBv2 
calculation of the inspiral waveform. 

We do so because our goal is to explore an alternative modeling approach that
is independent of previous \NR-informed \EOB tuning. In particular, we
performed dedicated \NR simulations outside the calibration range of
SEOBNRv2, and instead of inheriting higher-order corrections that were fitted
in a smaller parameter space region, we prefer to use the uncalibrated \EOB
model purely in the inspiral regime and hybridize it with \NR data of the
merger and ringdown.

We are naturally limited by the lengths of the \NR waveforms, which are
different for every simulation. Previous studies of \NR waveform length
requirements have suggested that \PN inspiral waveforms up to 5-10
orbits before merger are sufficiently accurate for detection purposes
\cite{Hannam:2010ky,Ohme:2011zm}; many more
orbits are needed to fulfil more stringent accuracy requirements
\cite{MacDonald:2011ne,
MacDonald:2012mp,Damour:2010zb,Boyle:2011dy}, especially in the high-mass-ratio
and high-spin regime that we are covering. Many
of our \NR waveforms are too short to allow that. However, previous studies
estimated the accuracy of \PN approximants based on the differences between
\emph{all} available approximants at 3.5PN order (with highest spin corrections
at 2.5PN order at that time). One might argue that the \EOB approach is more
accurate, and therefore comparisons between \PN waveforms exaggerate the
uncertainty in our best current models. On the other hand, without fully
general-relativistic results to compare to, one might be sceptical of good
agreements between alternative \EOB waveforms that are very similar by
construction. 

Nevertheless, given that we can join \EOB with our \NR data in a
much more robust manner than any of the \PN approximants (see Sec.~II of Paper 1
for our full analysis), we
trust that they provide a reasonably accurate description of the inspiral up to
the point where \NR data take over. At what frequency this switch from \EOB to
\NR happens depends on the length of the individual \NR simulations. We note
that the lowest common starting frequency of our \NR waveforms is $Mf \sim
0.018$, and this is where we begin our phenomenological merger-ringdown model.
Note, however, that our hybridization procedure ensures that the maximum amount
of \NR information is used in every point of the parameter space to inform both
the inspiral and merger-ringdown part of our model.

\section{Model of the \NR regime (Region II)}
\label{sec:MRmodel}

\begin{figure}[tb]
\centering
\includegraphics[width=0.8\linewidth]{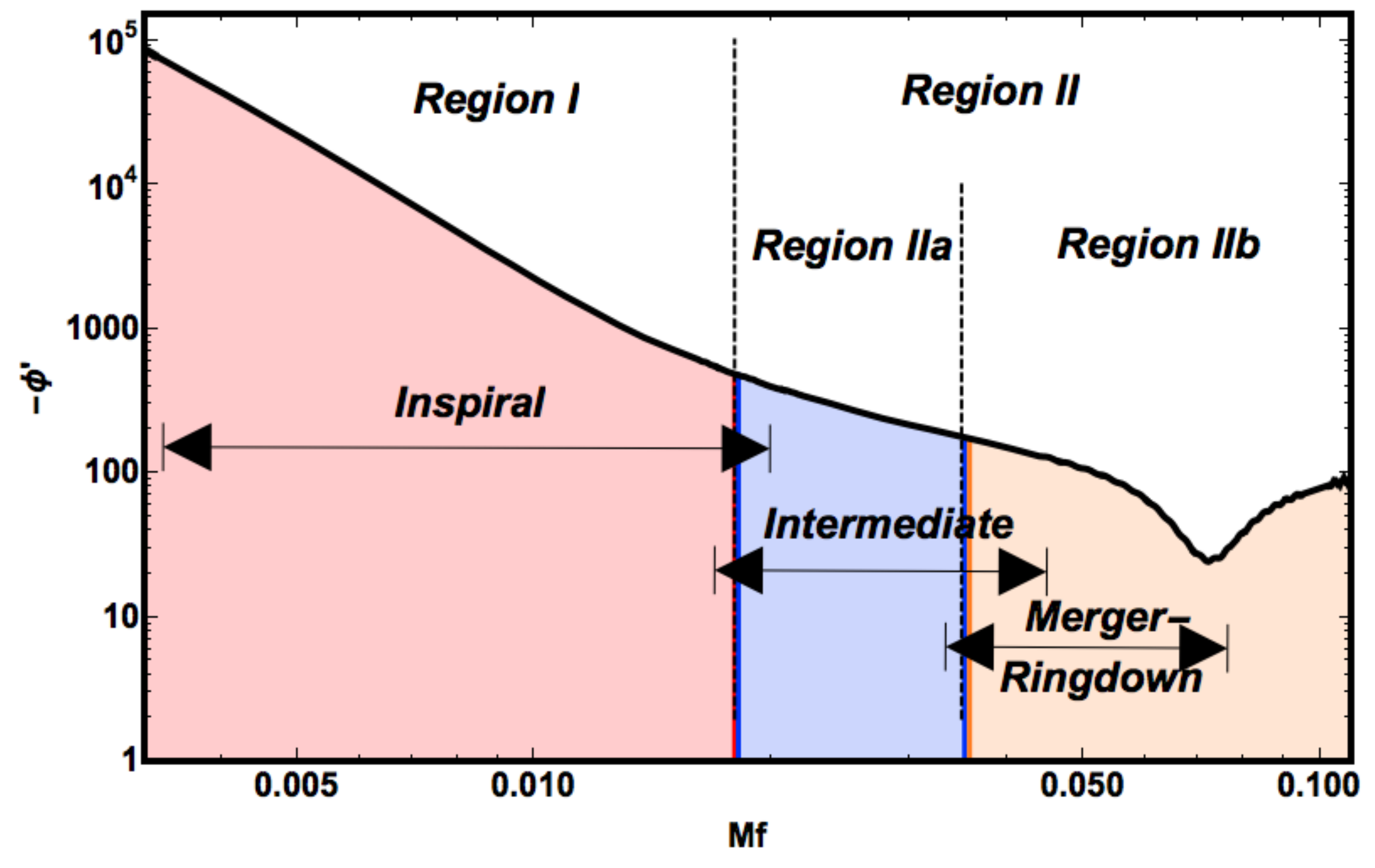}
\includegraphics[width=0.8\linewidth]{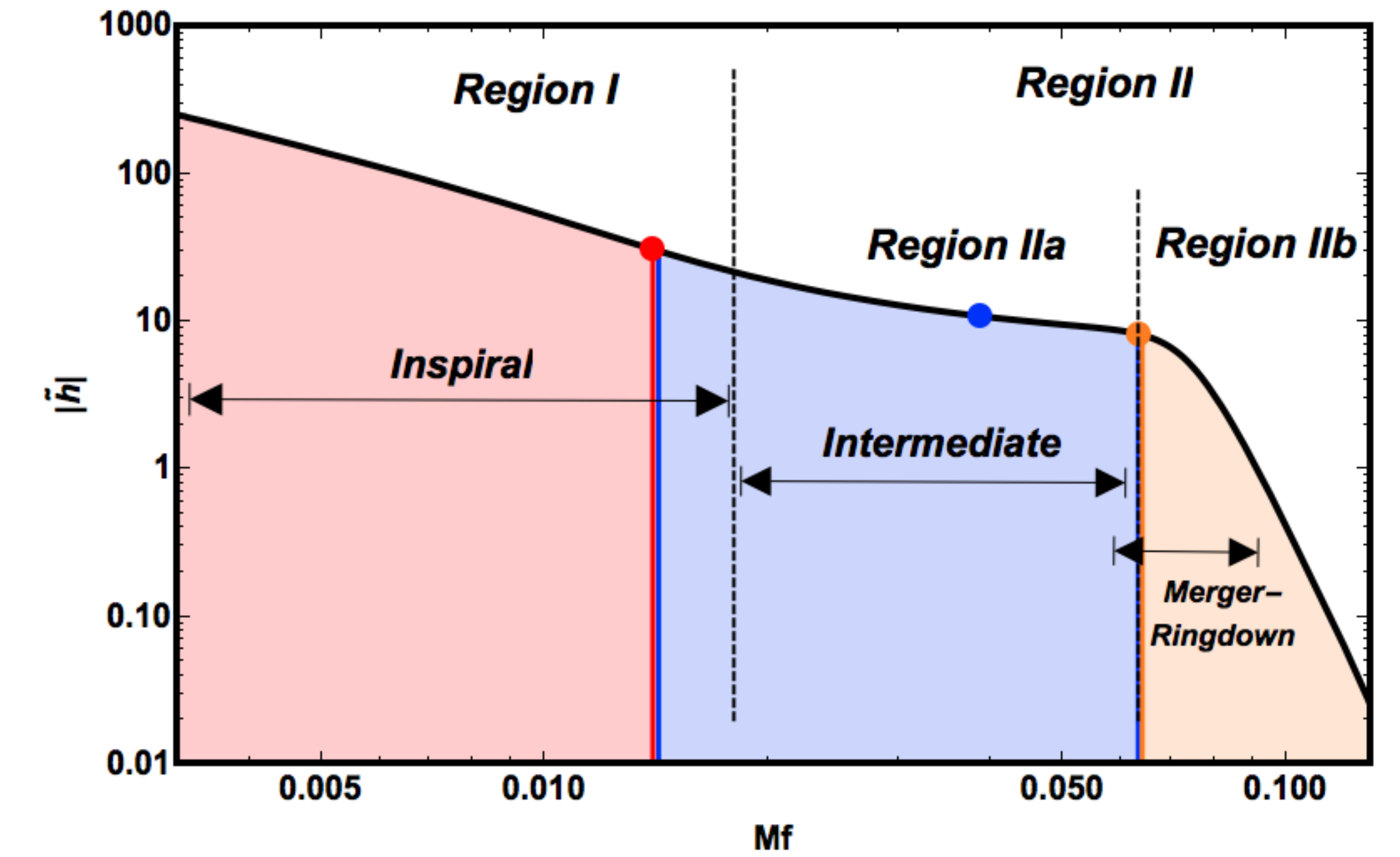}
\caption{
Phase derivative $-\phi'(f) \equiv - \partial \phi(f) / \partial f $ (upper panel) 
and amplitude (lower panel)
for the $q=1$, $\chi_1 = \chi_2 = -0.95$ configuration.
The frequency ranges that were used in the fits for each section are shown
as black double-ended arrows.
For reference, the frequency $Mf=0.018$ is marked with a black dashed line.
Shaded regions illustrate the
boundaries between the different regions when constructing the full
\IMR waveform.
The ringdown frequency for this case is $Mf=0.071$.
}
\label{fig:phasefittingwindows}
\end{figure}

We model separately three frequency regions of the waveforms. 
These are indicated in Fig.~\ref{fig:phasefittingwindows}.
\emph{Region I} is defined to be the portion of the hybrid
that contains the optimal blend of \NR and SEOBv2 data,
\emph{Region II} is the portion of the hybrid
that contains purely \NR data and corresponds to frequencies
$Mf \geq 0.018$.
This region is further sub-divided into two regions,
\emph{Regions IIa} and \emph{IIb}.
These divisions correspond to the \emph{intermediate} and
\emph{merger-ringdown} models for both the amplitude and phase.

The figures indicate both the frequency ranges over which the three parts are
connected,
but also the ranges that are used to calibrate the model's coefficients to the
hybrid data. These
regions are in general slightly larger than those used when piecing together the
final model. 

We will refer to other features of these figures in the forthcoming sections. 

\subsection{From PhenomC to PhenomD} 

The merger-ringdown portion of the phase was modelled in PhenomC
\cite{Santamaria2010} using the ansatz,

\begin{equation}
\begin{split}
\psi^{22}_{\rm PM}(f) & = \frac{1}{\eta} \left( \alpha_{1} f^{-5/3} + \alpha_{2}
f^{-1} \right. \\
					  & \left. + \alpha_{3} f^{-1/3} +
\alpha_{4} + \alpha_{5} f^{2/3} + \alpha_{6} f \right).
\end{split}
\label{eqn:PhCMRansatz}
\end{equation}

The phase was fit over the frequency range $[0.1,1] f_{\rm RD}$. The
reference phase and time of the fit are given by
the coefficients $\alpha_4$ and $\alpha_6$. At the ringdown frequency $f_{\rm
RD}$ the phase was smoothly connected to 
a linear function, $\psi^{22}_{\rm RD}(f)  = \beta_1 + \beta_2 f$, using a
$\tanh$ transition function.

\begin{figure}[tb]
\centering
\includegraphics[width=0.8\linewidth]{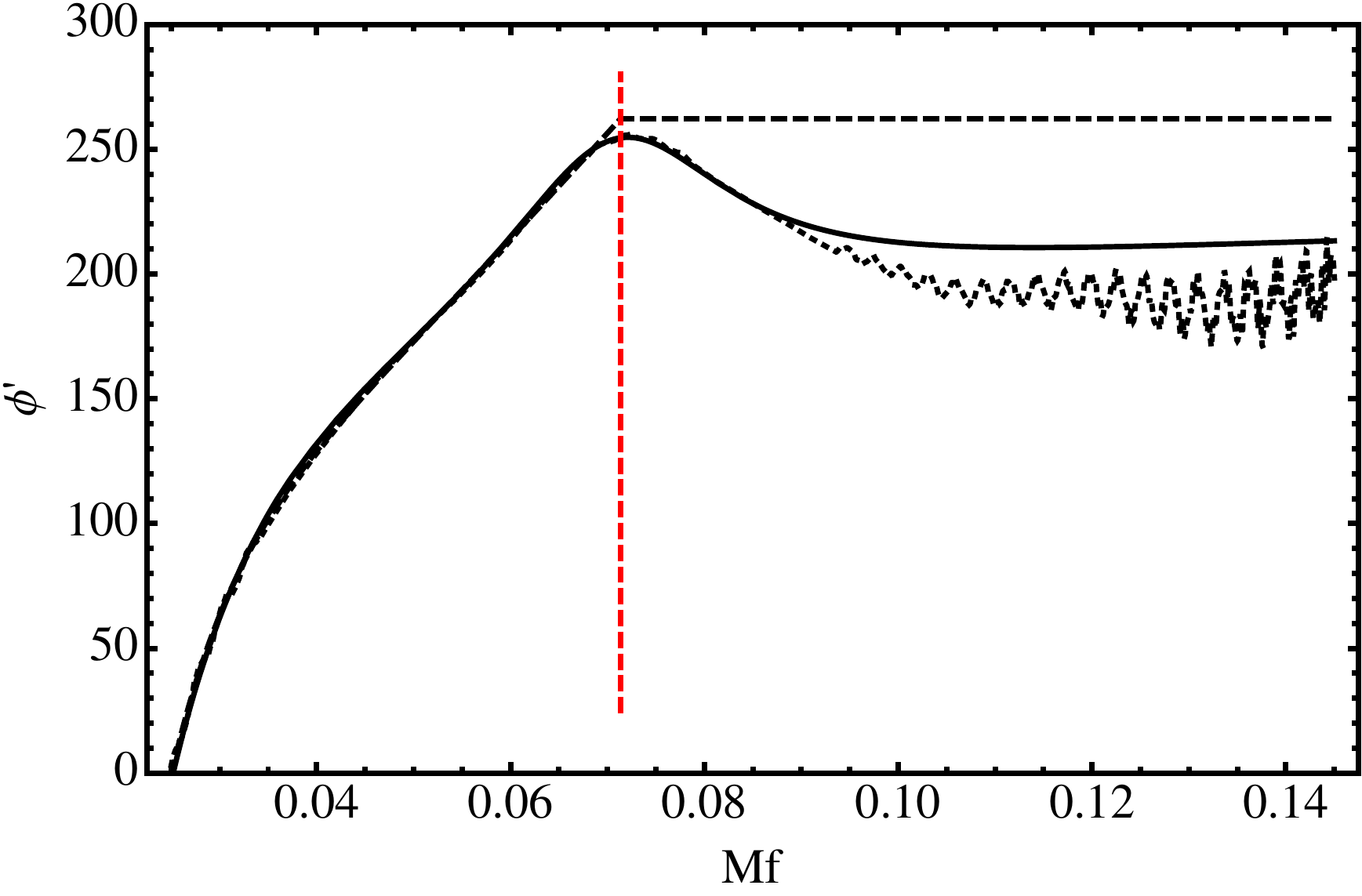}
\caption{Phase derivative $\phi'(f)$ for the $q=1$, $\chi_1 = \chi_2 = -0.95$
configuration. The numerical data (dotted) show
a distinctive extremum at the ringdown frequency, $M f_{\rm RD} = 0.071$,
indicated by a vertical dashed line. 
A fit that follows an approach similar to that used for PhenomC (dashed) 
is only a crude approximation to the phase for $f > f_{\rm RD}$, whereas the
approach used for the PhenomD model (solid) 
accurately models the phase into the ringdown.
}
\label{fig:dphiexample}
\end{figure}

We now aim to model the merger-ringdown phase of the \NR waveforms only from $Mf
= 0.018$, to ensure that we include only
\NR information in this part of the model. Fig.~\ref{fig:dphiexample} shows the
derivative of the frequency-domain phase for the configuration 
$q = 1$, $\chi_1 = \chi_2 = -0.95$. The dashed line shows a fit to the phase
using the procedure described above; beyond the 
ringdown frequency $M f_{\rm RD} = 0.071$ the derivative of the phase is
constant, and in this example the transition is only 
piecewise continuous. We see that, while Eq.~(\ref{eqn:PhCMRansatz}) is able to
accurately reproduce the phase up to the 
ringdown frequency, the linear approximation at higher frequencies is crude. 

The solid line in Fig.~\ref{fig:dphiexample} shows a fit to the phase following
the procedure we use to construct PhenomD,
which was motivated in detail in Paper 1, and is also described in
Sec.~\ref{Sec:MRDPhase} below. 
This accurately reproduces the main features of the phase 
derivative in the vicinity of the ringdown frequency. There is some disagreement
at higher frequencies, but we note that 
the accuracy of the \NR data typically degrades at these frequencies, and the
true behaviour of $\phi'(f)$ is not clear. 

In the next section we describe the methodology used to
produce models of the phase and amplitude
for the late inspiral, merger and ringdown parts of the
waveform, i.e., those frquencies for which we have \NR data.
These we have denoted Region II; see Fig.~\ref{fig:phasefittingwindows}.
We assume that we have a valid inspiral approximant that
we can join to our NR-based Region II model to construct a
full \IMR waveform model. The construction of a suitable inspiral
model (Region I) is given in Sec.~\ref{sec:Inspiral}.

Our current construction requires that the starting frequency of the
Region II model must be consistent for all waveforms.
This imposes the constraint that the starting frequency of the
NR-based Region II model
is the lowest common \GW frequency for
which we have \NR data, $Mf \sim 0.018$.
This is purely based on the available \NR data and could in principle be pushed
towards lower
frequencies given longer waveforms.

\subsection{Phase}
\label{sec:phaseNR}

To produce a robust model there are two key requirements:
(1) the ansatz must fit the data well, i.e.,
the fits have small residuals to the data, and (2)
the choice of ansatz should ideally be chosen to in such a way
that the coefficients vary smoothly across the parameter space,
to enable an accurate parameter-space fit in the final model.

We find that a simple approach is to split
Region II into an intermediate (Region IIa)
and merger-ringdown (Region IIb) part, and model them separately, as shown in
Fig.~\ref{fig:phasefittingwindows}.

The detailed features of the phase through Region II are most apparent when we
consider the derivative of the phase, $\partial \phi / \partial f \equiv \phi'(f)$. 
For this reason we first model 
$\phi'$, and then integrate the resulting expression to produce the final phase
model. We also note that the overall $1/\eta$ dependence in the inspiral, 
Eq.~(\ref{equ:phiTF2}),
also holds for the merger and ringdown, and so all of
our primary fits are to $\eta \phi'$.

\subsubsection{Region IIb - merger-ringdown}
\label{Sec:MRDPhase}

An example of the derivative of the phase, $\phi'$ is shown in
Fig.~\ref{fig:phasefittingwindows}
for a binary with $q = 1$, $\chi_1 = \chi_2 = -0.95$. As described in Paper 1,
we propose the following ansatz to 
model this functional form,
\begin{equation}
\phi'_{\text{MR}} = \alpha_{1}
                  + \alpha_{2} f^{-2}
                  + \alpha_{3} f^{-1/4}
                  + \frac{ a }
                  {b^2+(f - f_0)^2}.
\end{equation} 
The last term models the `dip'  in Fig.~\ref{fig:phasefittingwindows}.
The location of the minimum is given by $f_0$, while $a$ is the overall
amplitude of the dip and 
$b$ is the width. We find that the frequency location of the dip is very close
to the final \BH's
ringdown frequency, $f_{\rm RD}$ (they agree within our uncertainty in
calculating $f_{\rm RD}$), 
and that the ringdown damping frequency $f_{\rm damp}$ is a good approximation
to
our best fit of the width. These quantities are calculated from our final mass
and spin fits.
For these reasons the ansatz that we use in practice is,
\begin{equation}
\eta \, \phi'_{\text{MR}} = \alpha_{1}
                  + \alpha_{2} f^{-2}
                  + \alpha_{3} f^{-1/4}
                  + \frac{\alpha_{4} f_\text{damp}}
                  {f_\text{damp}^2+(f-\alpha_{5}f_\text{RD})^2}.
\label{equ:phiprimeMRD}
\end{equation}

We find that the parameter $\alpha_5$ is in the range $[0.98,1.04]$. 
The power law terms account for the overall trend of the data,
and its behaviour at lower frequencies.
The constant term translates into a time shift in the overall phase,
which will be determined by the continuity requirements of the
final \IMR phase; see Sec.~\ref{sec:IMR}.
The phase derivative data are fit to Eq.~(\ref{equ:phiprimeMRD}) 
over the frequency range $\left[ 0.45, 1.15 \right]f_{\text{RD}}$.
The upper frequency $1.15 f_{\text{RD}}$ approximates the highest
frequency for which we have clean \NR data.
This fitting window was chosen to have some overlap between the intermediate
phase
model, as indicated in Fig.~\ref{fig:phasefittingwindows}.

The merger-ringdown phase is given by the integral of
Eq.~(\ref{equ:phiprimeMRD}),
\begin{equation}
\begin{split}
\phi_{\text{MR}} & = \frac{1}{\eta} \left\{ \alpha_{0}
                 + \alpha_{1} f
                 - \alpha_{2} f^{-1}
                 + \frac{4}{3} \alpha_{3} f^{3/4} \right. \\
                 & \left. + \, \, \alpha_{4}
                   \tan^{-1}\left(\frac{f-\alpha_{5} f_\text{RD}}
                   {f_\text{damp}}\right)\right\} \,.
\end{split}
\label{eqn:MRDPhase}
\end{equation}

For the full \IMR phase we use the above fit for frequencies
larger than $0.5 \,f_{\text{RD}}$.
At lower frequencies we find that $\eta \, \phi'$ is fit better
by $\sim 1/f$ and we model this region (IIa) separately.

The phase offset that appears as a constant of integration $\alpha_0$, and the
time-shift
term $\alpha_1$, will both be determined in the final 
model by requiring a smooth connection with the phase from Region IIa. 

\begin{figure}[tb]
    \centering
    \includegraphics[width=1\linewidth]{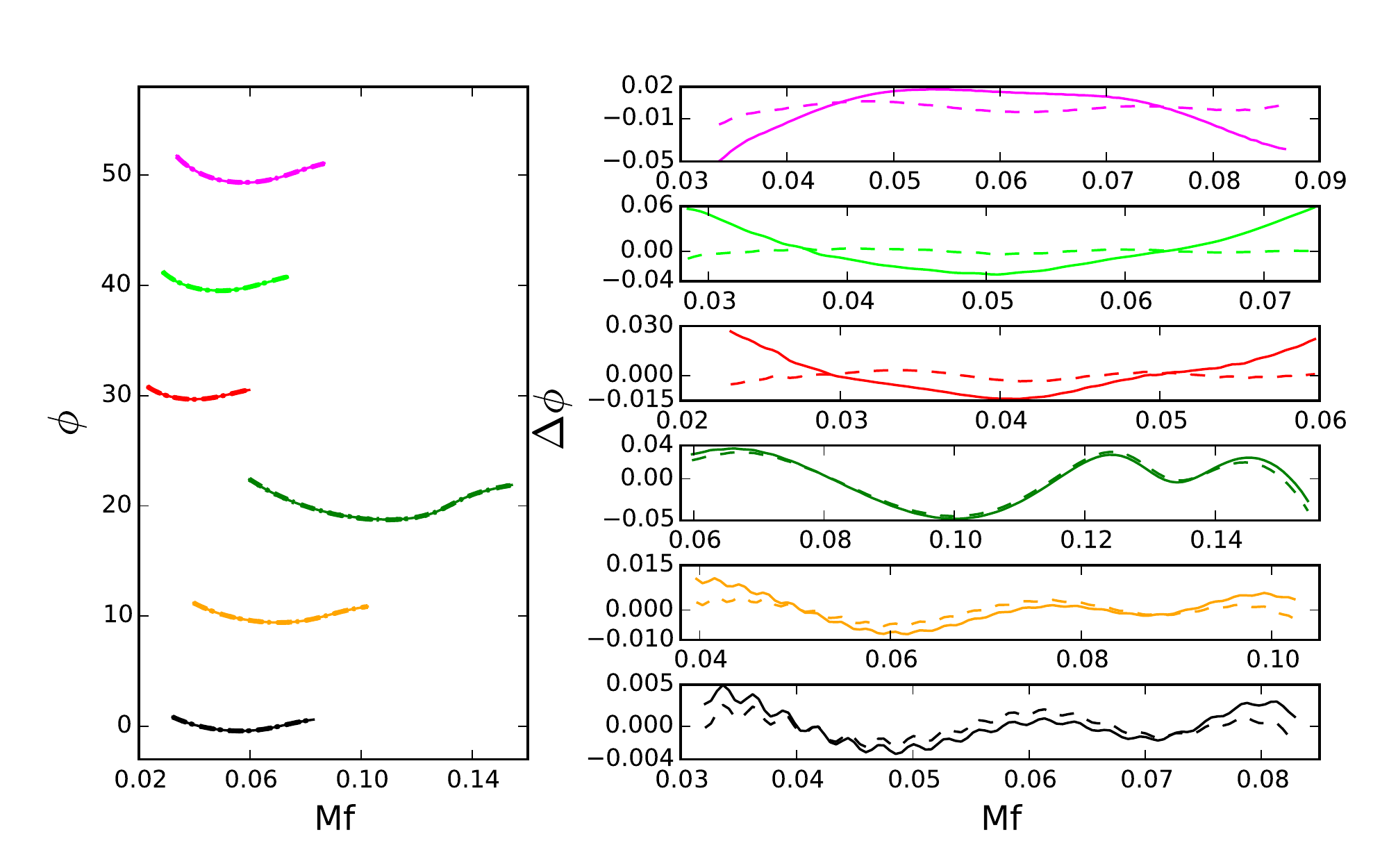}
    \caption{
    Examples of the merger-ringdown (Region IIb) model for 
     three $q=18$ configurations where the spin on the large \BH is $\chi_1 = 
    \{+0.4, 0, -0.8\}$ and 
    three equal-spin $q=1$ configurations ($\chi_{1,2} = +0.98, 0, -0.95$). 
    The configurations are ordered top to bottom in the figure. 
    The left panel shows the hybrid data, best-fit and final-model
    predictions over Region IIb. 
    The right panel shows the difference between 
    the hybrid data and the best-fit (dashed line) and between
    the hybrid data and the final model (solid line).
    }
    \label{fig:pltmrd}    
\end{figure}

Examples of the results are shown in 
Fig.~\ref{fig:pltmrd} for six
configurations at the edges of
our calibration parameter space. These are equal-spin $q=1$ waveforms with
spins 
$\hat\chi = \{-0.95, 0, 0.98\}$ and $q=18$ waveforms with spins on the larger
\BH of 
$\chi_1 = \{-0.8, 0, 0.4\}$ (the second \BH has no spin). In addition to demonstrating that both the
ansatz and the final model capture the data well, the figure also illustrates
the large differences in the frequency range of the merger-rigndown at different points in the
parameter space.

\subsubsection{Region IIa - intermediate}
\label{sec:phiint}

To bridge the gap between the lowest common frequency of the \NR data 
and the Region IIb merger-ringdown model, i.e., over the
frequency range $Mf \in \left[0.018, 0.5 f_{\text{RD}} \right]$,
we use the following ansatz,
\begin{equation}
\eta \, \phi'_{\text{Int}} = \beta_{1}
				   + \beta_{2} f^{-1}
				   + \beta_{3} f^{-4} \, .
\label{eqn:IntAnsatz}
\end{equation}

The behaviour of the data over this frequency range is predominately
proportional to $1/f$.
This is not sufficient at higher mass ratios 
and high anti-aligned spins, where $f_{\text{RD}}$ can be approximately half 
that of the equal mass non-spinning case.
We find that the additional $f^{-4}$ term fits the data
well across the entire parameter space.
The intermediate (Region IIa) ansatz is used over the frequency interval
$[0.018, 0.5 f_{\text{RD}}]$,
but we found that the best results were obtained if the data were fit over
$[0.017, 0.75 f_{\text{RD}}]$.

\begin{figure}[tb]
    \centering
    \includegraphics[width=1\linewidth]{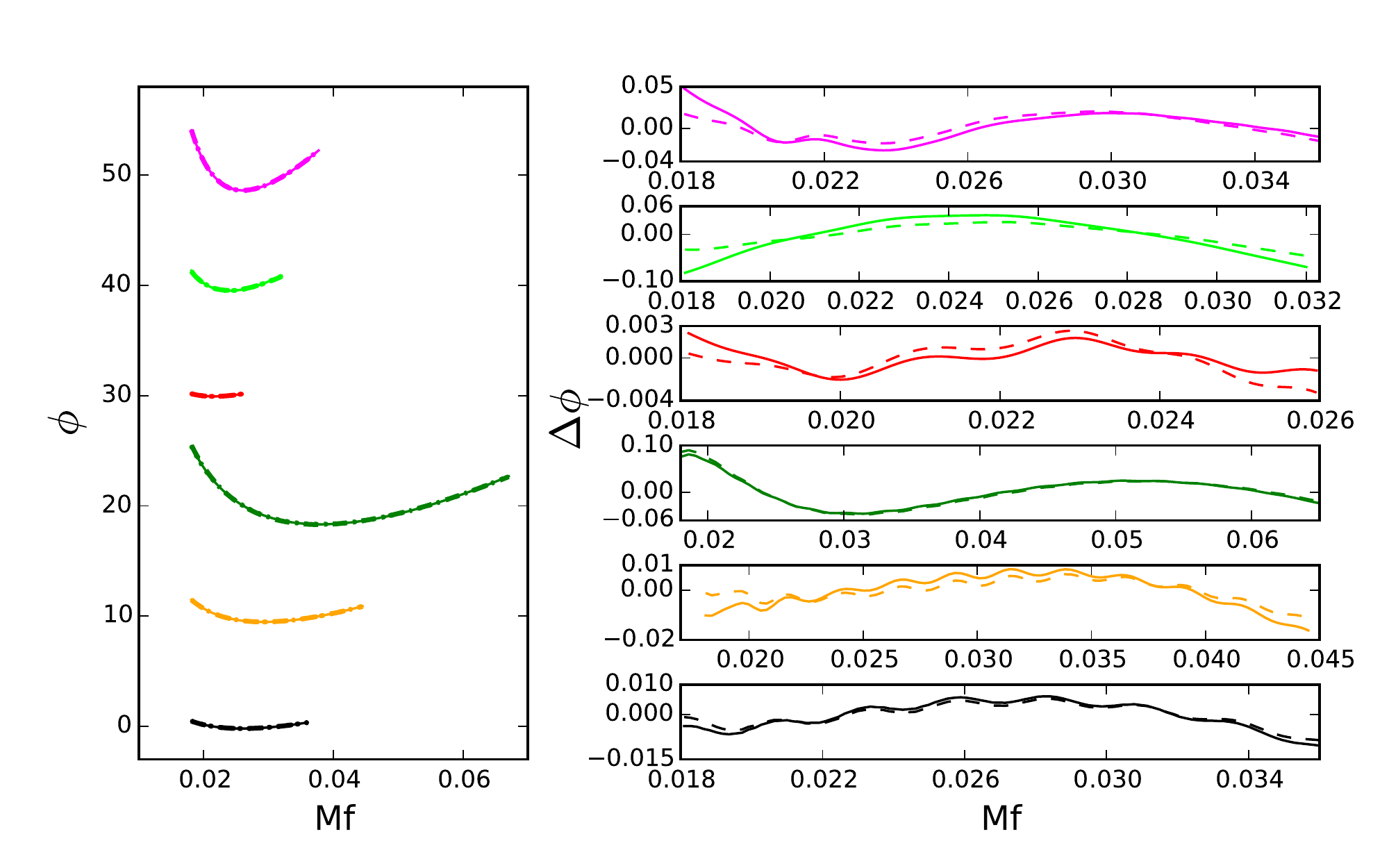}
    \caption{
    The same configurations and layout as in Fig.~\ref{fig:pltmrd}, but now
    showing phase over the intermediate region (IIa). 
        }
    \label{fig:pltint}    
\end{figure}

Once again the phase is obtained by integrating Eq.~(\ref{eqn:IntAnsatz}),
\begin{equation}
\phi_{\text{Int}} = \frac{1}{\eta} \left(\beta_{0} + \beta_{1} f
				  + \beta_{2} \, \text{Log}(f)
				  - \frac{\beta_{3}}{3}f^{-3}\right) \, .
				  \label{eqn:IntPhase}
\end{equation}
As in Region IIb, the phase-shift due to the constant of integration $\beta_0$,
and the 
time-shift term $\beta_1$, will be fixed by requiring a smooth connection to the
Region I phase. The results for the corner cases are shown in Fig.~\ref{fig:pltint}.

This completes the modelling of the phase over the frequencies for which we have
\NR data, Region II. 
We will now consider the signal amplitude over the same region, before moving on
to the 
inspiral, Region I.

\subsection{Amplitude}

When we perform the fits to the amplitude
across Region I and Region II, we first factor out the leading order \PN
$f^{-7/6}$ behaviour. The
resulting data tend to unity as the frequency tends to zero, and as with the use
of the phase
derivative, allows us to identify and model detailed features of the amplitude
behaviour;
see Fig.~\ref{fig:ampscaleInsp}, which shows both amplitude for \PN
inspiral waveforms,
and for the full hybrids.

The normalisation is given by,
\begin{equation}
\lim_{f\to 0} \left[ f^{7/6} \, A_{\text{PN}}(f) \right] \rightarrow
\sqrt{\frac{2 \, \eta}{3 \, \pi^{1/3}}},
\end{equation}
and our normalisation factor is therefore,
\begin{equation}
A_0 \equiv \sqrt{\frac{2 \, \eta}{3 \, \pi^{1/3}}} f^{-7/6}
\label{equ:A0}.
\end{equation}

\subsubsection{Region IIb - merger-ringdown}
\label{sec:IIb-MRD}

In all previous phenomenological models \cite{Ajith2007, Ajith2011,
Santamaria2010},
the ringdown amplitude has
been modelled with a Lorentzian function, which is the Fourier transform of the 
(two-sided) exponential decay function. The Fourier transform of the full \IMR
data
instead exhibit an exponential decay, as discussed in Paper 1. 
The amplitude in Region IIb is fit over the frequency range
$Mf \in \left[1/1.15, 1.2 \right]f_{\text{RD}}$ using the following ansatz,
\begin{equation}
\frac{A_{\text{MR}}}{A_0} = \gamma_{1}
                \frac{\gamma_{3} f_{\text{damp}}}
                {(f-f_\text{RD})^2+(\gamma_{3}f_{\text{damp}})^2}               
e^{-\frac{\gamma_{2} (f-f_{\text{RD}})}{\gamma_{3}f_{\text{damp}}}}
\, .
\label{equ:AmpMR}
\end{equation}
The coefficient $\gamma_1 \in [0.0024,0.0169]$ determines the overall amplitude
of the 
ringdown. We expect that the frequency width and location of the amplitude peak
can be 
inferred from the remnant \BH
parameters, which motivates the appearance of the ringdown damping frequency
$f_{\rm damp}$ 
in Eq.~(\ref{equ:AmpMR}). In practice we find that the width is increased by the
factor
$\gamma_3 \in [1.25,1.36]$, and the decay rate $1/(f_{\text{damp}}\gamma_{3})$
is modified by the factor $\gamma_2 \in [0.54,1.0339]$.

If we used only the Lorentzian part of Eq.~(\ref{equ:AmpMR}), the amplitude peak
would be located at 
$f_{\rm RD}$. With the additional exponential factor, the peak is located at
\begin{equation}
f_{\rm peak} = \left|  f_{\text{RD}} + \frac{f_{\text{damp}} \gamma_3 \left(
\sqrt{1 - \gamma_2^2} -1  \right)}{\gamma_2}  \right|. \label{eq:fpeak}
\end{equation}

\subsubsection{Region IIa - intermediate}
\label{sec:AmpRegIIa}

\begin{figure*}[tb]
    \begin{minipage}[l]{1.0\columnwidth}
        \centering
        \includegraphics[width=1\linewidth]{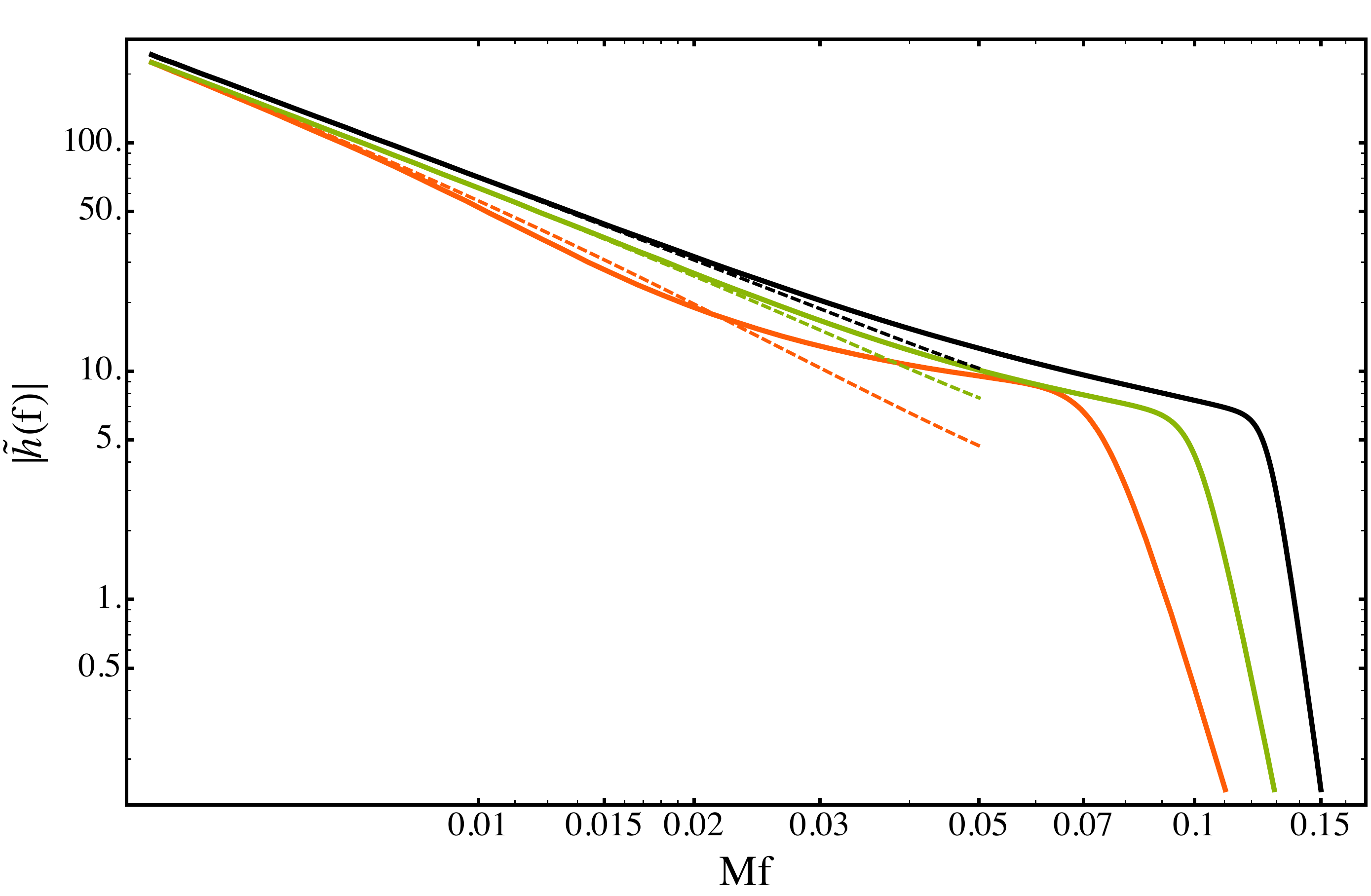}
    \end{minipage}
    \hfill{}
    \begin{minipage}[r]{1.0\columnwidth}
        \centering
        \includegraphics[width=1\linewidth]{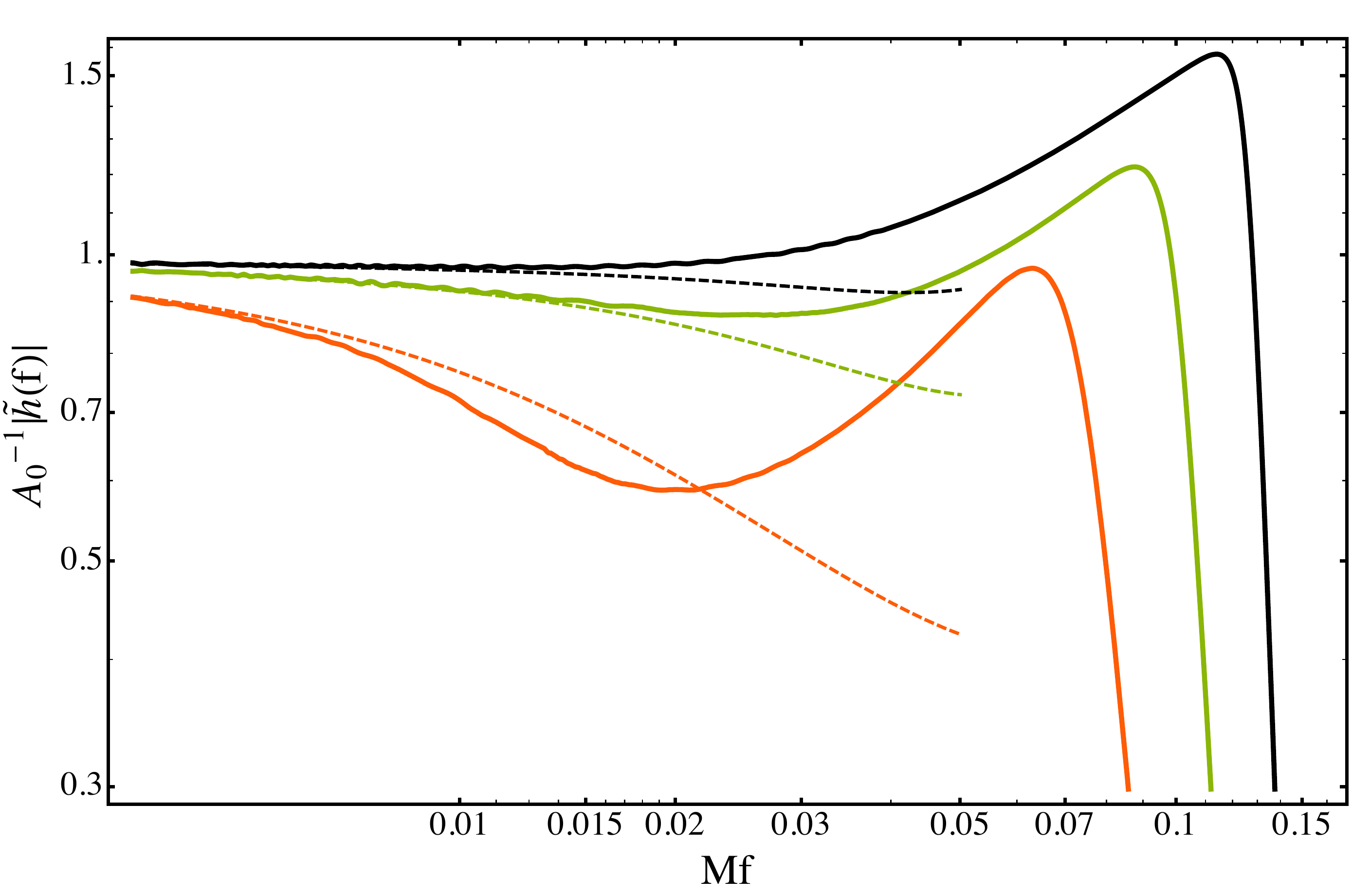}
    \end{minipage}
    \caption{
    Hybrid Fourier domain amplitude for three equal mass cases
    $q=1$, $\chi_1 = \chi_2 = 0.98$, $\chi_1 = \chi_2 = 0$ and $\chi_1 = \chi_2
= -0.95$, indicated by black, orange and green lines respectively. The \PN prediction is shown
    as dashed lines. The left panel shows the full fourier domain amplitude, while the
    right panel shows the fourier domain amplitude but rescaled by $A_0^{-1}$, Eq.~(\ref{equ:A0}).
	}
	\label{fig:ampscaleInsp}
\end{figure*}

We now consider the intermediate region (IIa) between the end of the
inspiral region (I) and the start of the merger-ringdown region (IIb).

Fig.~\ref{fig:ampscaleInsp} shows the TaylorF2 inspiral amplitude in comparison
to the amplitude in 
the hybrid data. In some cases, we see that we can model the intermediate
(Region IIa) amplitude
by simply smoothly connecting regions I and IIb. For example, we could fit the
four coefficients of a third-order polynomial by 
matching the value of the amplitude and its derivative at the end of the Region
I (nominally $Mf=0.018$) and at the beginning
of Region IIb, $f_{\rm peak}$. 

In other cases, however, we see that the rescaled amplitude will have a minimum in the 
intermediate region, and a naive connection of the inspiral and merger-ringdown
regions would not in general locate this 
minimum correctly. 

For this reason, we model the intermediate amplitude with a fourth-order
polynomial. Four of the coefficients are fixed (as
above), by matching the value and derivative of the amplitude at the endpoints
of our intermediate fit. The lower frequency 
is chosen as $M f_1  = 0.014$, i.e., slightly before the end of the inspiral at
$M f=0.018$, and the upper frequency is 
$f_3 = f_{\rm peak}$. The fifth coefficient is determined by the value of
amplitude of the \NR waveform at the frequency mid-way
between the two, $f_2 = (f_1 + f_3)/2$. 

In practice, the amplitude values and derivatives at the endpoints are given by
the models for Region I and Region IIB.
The only additional piece of information that needs to be modelled from the \NR
data is the value of the amplitude at 
$f_2$. We find that this can be accurately modelled across the parameter space
by a polynomial ansatz in ($\eta, \hat{\chi}$), 
as will be described in Sec.~\ref{Sec:mapping}. 

This collocation method is similar to that used in spectral methods. Given an
ansatz with $n$ free coefficients we require $n$ 
pieces of information from the data to constrain the ansatz and solve the
system. In this case we use the value of the function at
three points, and the derivative at two points. The intermediate ansatz is given
by
\begin{equation}
A_{\text{Int}} = A_0 \left( \delta_{0} + \delta_{1} f + \delta_{2} f^2 +
\delta_{3} f^3 + \delta_{4} f^4 \right),
\label{eqn:AInt}
\end{equation} and the $\delta_i$ coefficients are the solution to the 
system of equations,
\begin{eqnarray}
A_{\text{Int}}(f_1) & = & v_1, \\
A_{\text{Int}}(f_2) & = & v_2, \\
A_{\text{Int}}(f_3) & = & v_3, \\
A'_{\text{Int}}(f_1) & = & d_1, \\
A'_{\text{Int}}(f_3) & = & d_3.
\end{eqnarray} The frequencies and values are given in Tab.~\ref{tab:coldef}.

\begin{table}[tb]
\begin{tabular}{lll}
\hline \hline
\multicolumn{1}{c}{Collocation Point (Mf)} & \multicolumn{1}{c}{Value} &
\multicolumn{1}{c}{Derivative} \\ \hline
$f_1 = 0.014$               & $v_1 = A_{\text{Ins}}(f_1)$     & $d_1 =
A'_{\text{Ins}}(f_1)$  \\ 
$f_2 = (f_1 + f_3)/2$ & $v_2 = A_{\text{Hyb}}(f_2)$ &         \\ 
$f_3 = f_{\text{peak}}$   & $v_3 = A_{\text{MR}}(f_3)$      & $d_3 =
A'_{\text{MR}}(f_3)$   \\ \hline \hline
\end{tabular}
\caption{
    \label{tab:coldef} 
    Locations of the collocation points, $f_1, f_2, f_3$, and the corresponding
values of the amplitude $A(f)$ and its
    derivative $A'(f)$. All information comes from either the inspiral or
merger-ringodwn models, except for the value 
    $v_2$, which is read off the input waveform data. 
    }
\end{table}

\begin{figure*}[tb]
    \begin{minipage}[l]{1.0\columnwidth}
        \centering
        \includegraphics[width=1\linewidth]{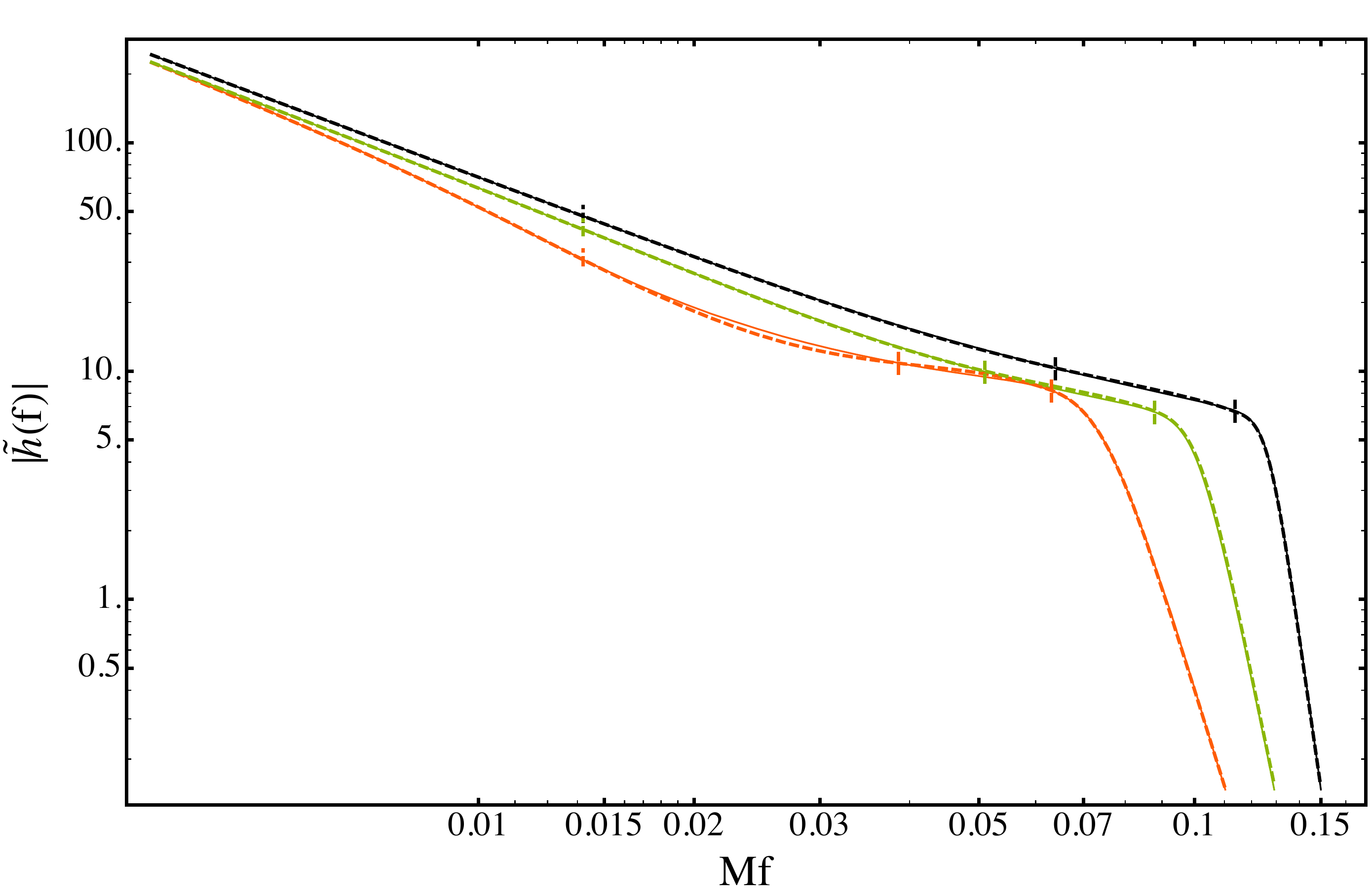}
    \end{minipage}
    \hfill{}
    \begin{minipage}[r]{1.0\columnwidth}
        \centering
        \includegraphics[width=1\linewidth]{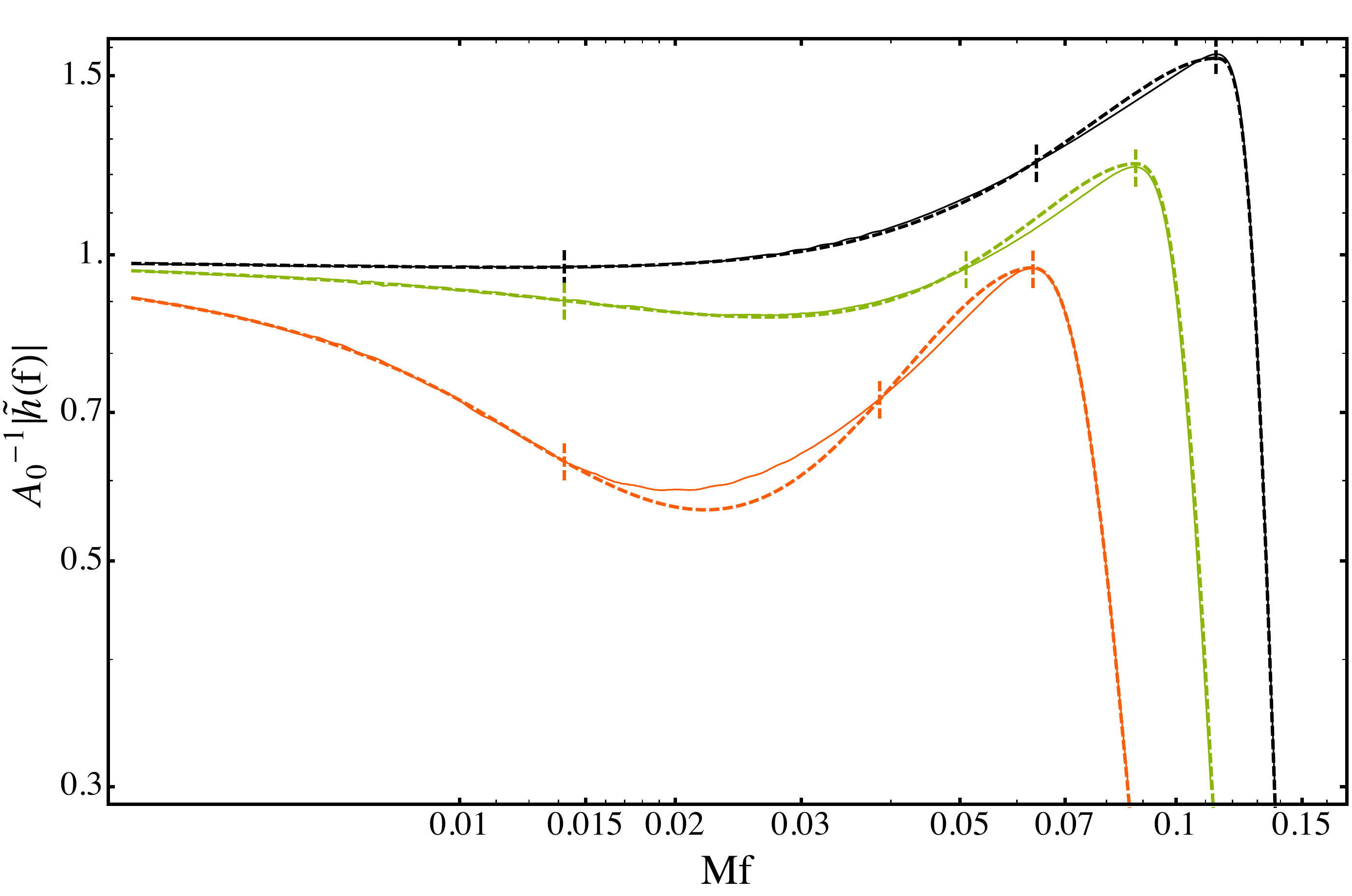}
    \end{minipage}
    \caption{
    Hybrid and model Fourier-domain amplitude for three equal-mass
configurations,
    $\chi_1 = \chi_2 = 0.98$, $\chi_1 = \chi_2 = 0$ and $\chi_1 = \chi_2 =
-0.95$, indicated by
    black, orange and green lines respectively. The hybrid data are shown by
solid lines, and the 
    PhenomD model by dashed lines.
    The left panel shows the full Fourier-domain amplitude, while the
    right panel shows the Fourier-domain amplitude but rescaled by $A_0^{-1}$, Eq.~(\ref{equ:A0}).
The short vertical dashed lines mark
the three frequency points in Tab~(\ref{tab:coldef}),
while the lines at lower and higher frequency coincide with the transition
points between regions I and IIa and between regions IIa and IIb respectively.
    }
    \label{fig:ampscale}
\end{figure*}

\begin{figure*}[tb]
    \begin{minipage}[l]{1.0\columnwidth}
        \centering
        \includegraphics[width=1\linewidth]{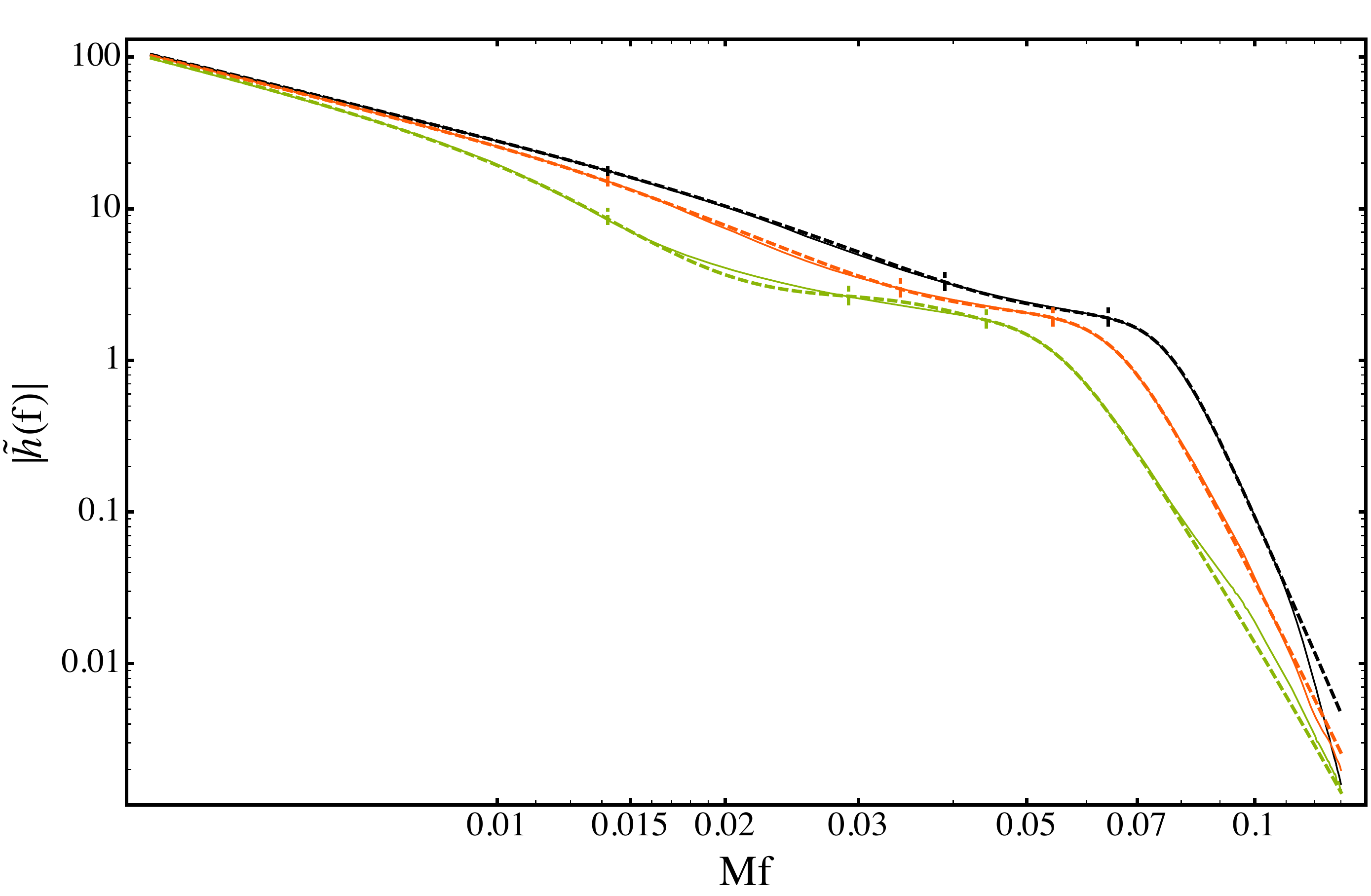}
    \end{minipage}
    \hfill{}
    \begin{minipage}[r]{1.0\columnwidth}
        \centering
        \includegraphics[width=1\linewidth]{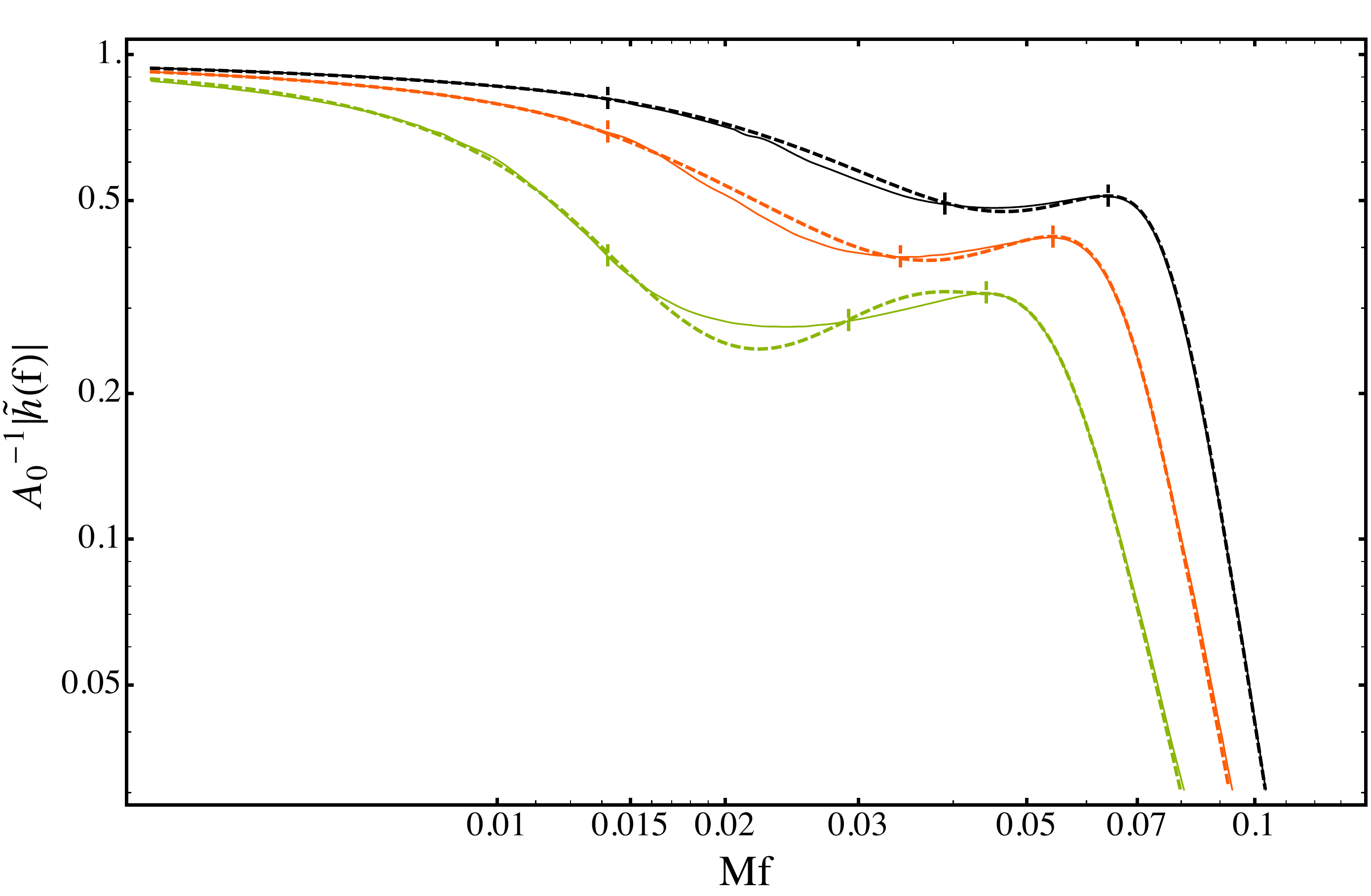}
    \end{minipage}
    \caption{
    The same quantities as in Fig.~\ref{fig:ampscale}, but now for three $q=18$
configurations,
    $\chi_1 = 0.4, \chi_2 = 0$, $\chi_1 = \chi_2 = 0$ and $\chi_1 = -0.8, \chi_2
= 0$.    }
    \label{fig:ampscale2}
\end{figure*}

The results of our amplitude model are shown in Figs.~\ref{fig:ampscale} and
\ref{fig:ampscale2}, which
show the same equal-mass and $q=18$ cases as in Fig.~\ref{fig:pltmrd}. The left
panels show the full signal
amplitude, while the right panels show the amplitude scaled by the $f^{7/6}$
factor, Eq.~\ref{equ:A0}.

The scaled plots indicate that the weakest part of the model is that which
describes the intermediate
Region IIa amplitude. This is because the minimum that we see in the scaled
figures (those in the right
panels) is captured only through the value of the amplitude at the frequency in
the middle of Region IIa.
If we were in addition to model the frequency at which the minimum occurs, and
prescribe the amplitude
value there, the model may perform better. We could also, of course, add further
collocation points. 
However, we can see from the full unscaled amplitude (the left panels) that the
amplitude is nonetheless
very accurately represented, and in addition, small variations in the amplitude
play a far smaller role in 
\GW applications (both searches and parameter estimation) than the \GW phase. 

\section{Inspiral model (Region I)}
\label{sec:Inspiral}

We now turn our attention to modelling Region I, i.e., the inspiral portion
of the waveform, below the frequency $Mf=0.018$;
see Fig.~\ref{fig:phasefittingwindows}.

The non-spinning \cite{Ajith2007} and the first aligned-spin \cite{Ajith2011}
phenomenological
models used a PN-like ansatz for the inspiral phase, calibrated against PN+NR
hybrids. In the PhenomC model~\cite{Santamaria2010}, the TaylorF2 phase 
was used for the equivalent of Region I; in that model the
inspiral region ended at $0.1 f_{\text{RD}}$. For the parameter space covered by
our new 
model, this would corresponds to frequencies between $Mf \sim 0.005$ and $Mf
\sim 0.012$.

In Paper 1 we presented evidence
that the uncalibrated SEOBv2 model is currently the inspiral approximant that is
most consistent
with \NR data for the inspiral.
In this section we construct a frequency domain model of
the SEOBv2 inspiral, up to $Mf = 0.018$, using our
SEOBv2+\NR hybrids. As discussed
previously, we
expect that the SEOBv2 model is sufficiently accurate up to this frequency, and
very likely to
higher frequencies, allowing us to match to our merger-ringdown model at
significantly higher
frequencies than was considered reasonable with the TaylorF2 approximant used
for PhenomC.

Note that it is possible, in principle, to cover the parameter space with an
arbitrarily high
density of SEOBv2 waveforms, and use those to calibrate an inspiral model. In
this paper,
however, we use hybrid SEOBv2+\NR waveforms and therefore calibrate the inspiral
model to the
same points in parameter space as used for the Region II merger-ringdown
models.

\subsection{Phase}
\label{sec:inspiral_phase}

The inspiral portion $Mf \in [0.0035,0.018]$ of the hybrids
can be accurately modelled with an ansatz consisting of
the known TaylorF2 terms for the phase,
augmented with the next four higher order \PN terms, with their 
 coefficients fit to the SEOBv2+\NR hybrid data.
We find that these higher order terms are enough to capture the \EOB and \NR
data over this
frequency range to a very high level of accuracy.

The full TaylorF2 phase is, 
\begin{eqnarray}
\phi_{\text{TF2}} &=  &\, 2 \pi f t_c - \varphi_c - \pi/4 \nonumber \\
                     &   & + \frac{3}{128 \, \eta}(\pi f M)^{-5/3} \sum_{i=0}^{7}
\varphi_i(\Xi) (\pi f M)^{i/3},
\label{equ:phiTF2}
\end{eqnarray}
where $\varphi_i(\Xi)$ are the \PN expansion coefficients that are functions of
the intrinsic binary parameters. Explicit expressions are
given in Appendix~\ref{sec:app_pncoeffs}. We incorporate
spin-independent corrections up to 3.5PN order ($i=7$)
\cite{Buonanno2009,Blanchet2014}, linear spin-orbit corrections up to 3.5PN
order \cite{Bohe:2013cla} and quadratic spin corrections up to 2PN order
\cite{Poisson:1997ha,Arun:2008kb,Mikoczi:2005dn}. In re-expanding the \PN energy
and flux to obtain the TaylorF2 phase, we drop all quadratic and
higher-order spin corrections beyond 2PN order as they would
constitute incomplete terms in our description.
With these choices, we are entirely consistent with the current state of the
LIGO software library~\cite{lalsuite}. 
We note that we also constructed a full model that incorporated recently
calculated higher-order terms, specifically quadratic spin terms at 3\PN
order \cite{Bohe:2015ana} and cubic spin terms at 3.5\PN order
\cite{Marsat2015}, but we found no significant difference between both
constructions.

Equation \eqref{equ:phiTF2}  includes both spins, $\chi_1$ and $\chi_2$, while
our 
fit for the coefficients of  additional terms will be parameterized only by
$\hat{\chi}$.
This means that the final phase expression will incorporate some effects from
the spins of 
each \BH, but, although the model is sufficiently accurate for use in GW
astronomy 
applications across a wide range of the two-spin parameter space, it should not be considered
an accurate 
representation of two-spin effects. We expect the model to be more than
sufficient
for searching for \BH binaries with any \BH spins within the
calibration parameter 
space, or for estimation of the parameters $(M, \eta, \hat{\chi})$, but we
\emph{do not} 
recommend its use in, for example, theoretical studies of detailed double-spin
effects in binaries.   

The phase ansatz is given by,

\begin{align}
\begin{split}
\phi_{\text{Ins}}  = &  \phi_{\text{TF2}}(Mf; \Xi) 
                  \\ + &  \frac{1}{\eta}\left(\sigma_{0} + \sigma_{1} f
                  + \frac{3}{4} \sigma_{2} f^{4/3}
                  + \frac{3}{5} \sigma_{3} f^{5/3} + \frac{1}{2} 
                  \sigma_{4}
f^{2} \right).
\label{equ:phiins}
\end{split}
\end{align}

Note that to compute the phenomenological coefficients the fit is
performed over the frequency range $Mf \in [0.0035,0.019]$
to achieve an optimal balance between goodness of fit and
accuracy in reproducing phenomenological coefficients
and to reduce boundary effects at the interface between Region I and Region IIa
(i.e., $Mf = 0.018$).
In practice the fits were performed over the $\phi'$ data,
as with Region II above. We will see in Sec.~\ref{sec:mismatches} that this
model
also sufficiently accurately represents SEOBv2+\NR hybrids down to much lower 
frequencies.

\begin{figure}[tb]
    \centering
    \includegraphics[width=1\linewidth]{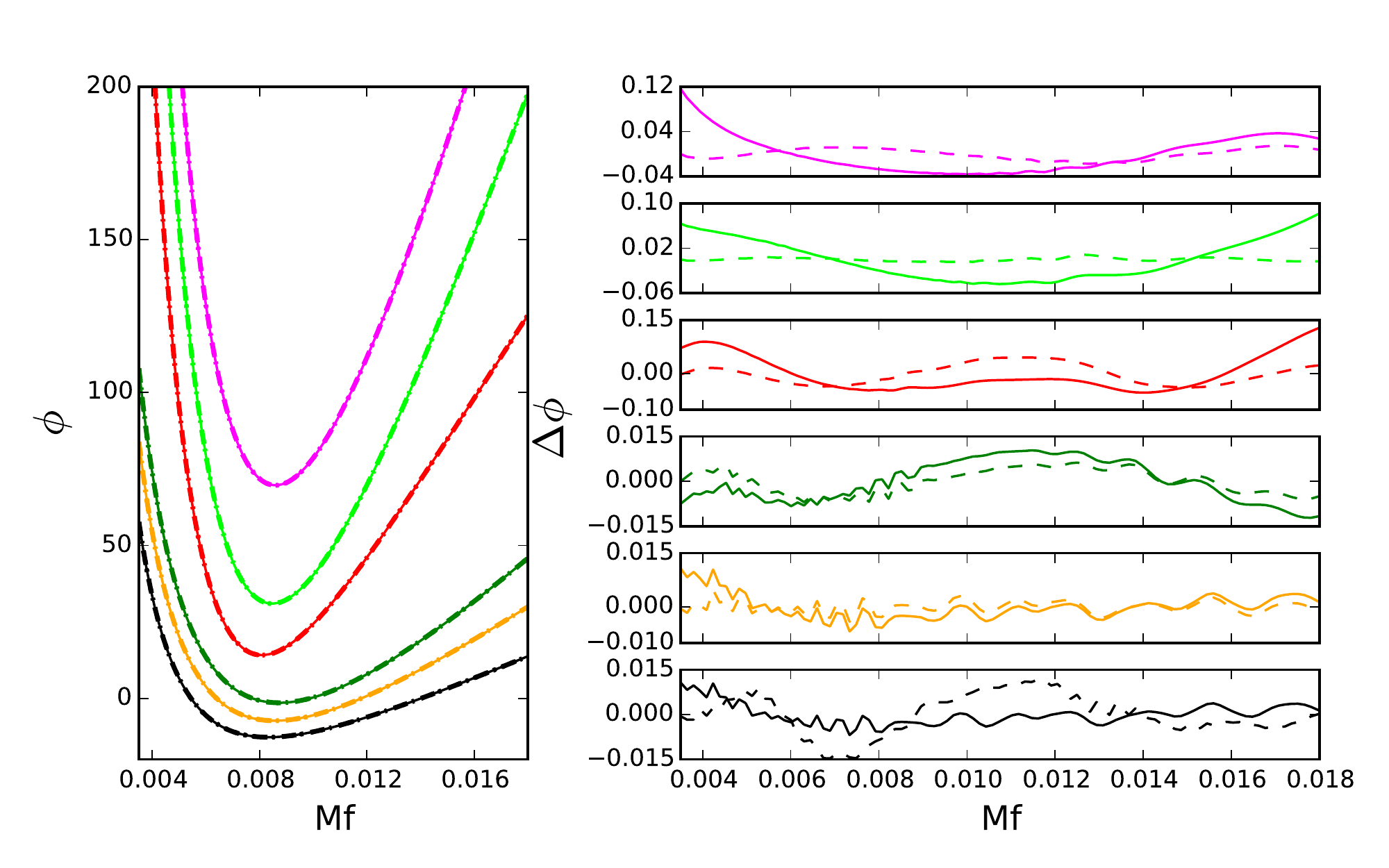}
    \caption{
    The same analysis as in Figs.~\ref{fig:pltmrd} and \ref{fig:pltint}, but now for the 
    inspiral model. 
    }
    \label{fig:pltins}    
\end{figure}

The results for the three example $q=1$ and $q=18$ configurations are shown in Fig.~\ref{fig:pltins}.
We see that once again our ansatz accurately models the data, and that the Fourier-domain phase 
error is below 0.15\,rad for the entire inspiral for the high-mass ratio configurations, while for the 
equal-mass configurations the phase error is typically an order of magnitude smaller.

\subsection{Amplitude}
\label{sec:InsAmp}

Our model of the inspiral amplitude is based on a re-expanded \PN
amplitude, as discussed in Sec.~IV of Paper 1. The base amplitude is given by,
\begin{equation}
A_{\text{PN}}(f) = A_0 \sum_{i=0}^{6} \mathcal{A}_i (\pi f)^{i/3} \, ,
\label{equ:AmpPN}
\end{equation}
where $A_0$ is the leading order $f^{-7/6}$ behaviour in Eq.~(\ref{equ:A0}).
The higher order terms that we calibrate are the next natural terms in the \PN
expansion,
\begin{equation}
A_{\text{Ins}} = A_{\text{PN}}
                  + A_0  \sum_{i=1}^{3} \rho_{i} \, f^{(6+i)/3}.
\label{equ:Ains}
\end{equation}

\section{Mapping the phenomenological coefficients to physical parameters}
\label{Sec:mapping}

Our model has 11 amplitude and 14 phase coefficients.
However four of the amplitude coefficients in Region IIa
(Sec.~\ref{sec:AmpRegIIa})
and four of the phase coefficients
$(\alpha_{0},\alpha_{1},\beta_{0},\beta_{1})$
across Region II
(See Sec.~\ref{sec:IMR}) are constrained analytically;
there is only one time and phase-shift freedom
for the full waveform.
This leaves a total of 17 phenomenological
parameters which need to be mapped on to the physical parameter space.
We parametrise the phenomenological coefficients by
two physical parameters, 
$(\eta, \chi_{\rm PN})$.
Our model is also dependent on the total mass $M$ of the system
through a trivial rescaling.

As in previous phenomenological models \cite{Ajith2007, Ajith2011,
Santamaria2010}
we map the phenomenological coefficients in terms
of polynomials of the physical parameters, up to second
order in $\eta$ and third order in 
$\chi_{\rm PN}$, although 
 in this work our polynomial
ansatz is expanded around 
$\chi_{\rm PN} = 1$. Note that in the fit across the parameter space we use
the unscaled reduced-spin parameter $\chi_{\rm PN}$, 
\begin{equation}
\begin{split}
\Lambda^i = & \, \lambda^i_{00}+\lambda^i_{10} \eta \\
		  & (\chi_{\rm PN}-1) \left(\lambda^i_{01}+\lambda^i_{11} \eta +
\lambda^i_{21} \eta ^2\right) \\
		  & +(\chi_{\rm PN}-1)^2
\left(\lambda^i_{02}+\lambda^i_{12}\eta +
\lambda^i_{22}\eta ^2\right) \\ 
		  & +(\chi_{\rm PN}-1)^3
\left(\lambda^i_{03}+\lambda^i_{13} \eta +
\lambda^i_{23}\eta ^2\right) \, ,
\end{split}
\label{eqn:mapping}
\end{equation}
where $\Lambda^i$ indexes both the amplitude and phase coefficients given by,

\begin{equation}
\Lambda^i = \big\{\hspace{-0.25cm} \underbrace{\{\rho_j\} , \{v_2\} ,
\{\gamma_j\}}_{\text{Amplitude Coefficients}}
			,
			\underbrace{\{\sigma_j\} , \{\beta_j\} ,
\{\alpha_j\}}_{\text{Phase Coefficients}}
			\big\} \, .
\end{equation}

The index $i \in \{1,2,3,4,5,6\}$ selects the coefficient vector for either
amplitude or phase and
model for Region I, IIa or IIb.
Tab.~\ref{tab:coefftable} in Appendix~\ref{app:phencoeff} contains the values of
all the mapping coefficients for each phenomenological parameter.

\section{Full \IMR waveforms}
\label{sec:IMR}

By construction, all the regions
of the amplitude and phase models
are joined by $C(1)$-continuous conditions.
This ensures the first derivative of the
amplitude and phase at the boundary between
the various regions, which are used in analytic calculations,
are smooth. We assume that this is sufficient and
simply join together the piecewise regions
with step functions.
Our step function is defined as
\begin{eqnarray}
\label{eq:theta}
\theta(f-f_0) = 
\left \{ \begin{array}{lllll}
 -1,  & & f < f_0, \\
 1,   &  & f \ge f_0, \\
 \end{array}
 \right .
\end{eqnarray}
and,
\begin{equation}
\theta_{f_0}^{\pm} = \frac{1}{2}\left[1 \pm \theta(f-f_0)\right].
\end{equation}

The full \IMR phase is determined up to an arbitrary time- and phase shift.
These shifts are absorbed into the constant and linear coefficients of the
inspiral part
$(\sigma_0, \sigma_1)$. 
The constant and linear coefficients of the Region IIa
$(\alpha_0, \alpha_1)$
and IIb models
$(\beta_0, \beta_1)$
are fixed by the requirement of $C(1)$ continuity.

The full \IMR phase is given by the following equation
\begin{equation}
\Phi_{\text{IMR}}(f) = \phi_{\text{Ins}}(f) \, \theta_{f_1}^{-} 
                     +  \theta_{f_1}^{+} \, \phi_{\text{Int}}(f) \,
\theta_{f_2}^{-}
                     + \theta_{f_2}^{+} \, \phi_{\text{MR}}(f) \, ,
\end{equation}
where $\phi_{\text{Ins}}$ is given by Eq.~(\ref{equ:phiins}),
$\phi_{\text{Int}}$ by 
Eq.~(\ref{eqn:IntPhase}), and 
$\phi_{\text{MR}}$ by Eq.~(\ref{eqn:MRDPhase}), and the transition frequencies
are 
$f_1 =0.018$ and $f_2=0.5 f_{\text{RD}}$.
As noted previously, when evaluating the known \PN part of $\phi_{\text{Ins}}$,
given in 
Eq.~(\ref{equ:phiins}), we use the full two spin dependence.

The full \IMR amplitude is given by
\begin{equation}
A_{\text{IMR}}(f)  = A_{\text{Ins}}(f) \, \theta_{f_1}^{-} 
                     +  \theta_{f_1}^{+} \, A_{\text{Int}}(f) \,
\theta_{f_2}^{-}
                     + \theta_{f_2}^{+} \, A_{\text{MR}}(f) \, ,
\end{equation}
where $A_{\text{Ins}}$ is given by Eq.~(\ref{equ:Ains}), $A_{\text{Int}}$ by 
Eq.~(\ref{eqn:AInt}),
and $A_{\text{MR}}$ by Eq.~(\ref{equ:AmpMR}), and where the transition
frequencies are 
$f_1 =0.014$ and $f_2=f_{\rm peak}$, Eq.~\eqref{eq:fpeak}.
The amplitude is $C(1)$-continuous by construction.
Once again, note that the base inspiral \PN amplitude includes both spin
contributions.

\begin{figure*}[h!]
\centering
\includegraphics[width=1\linewidth]{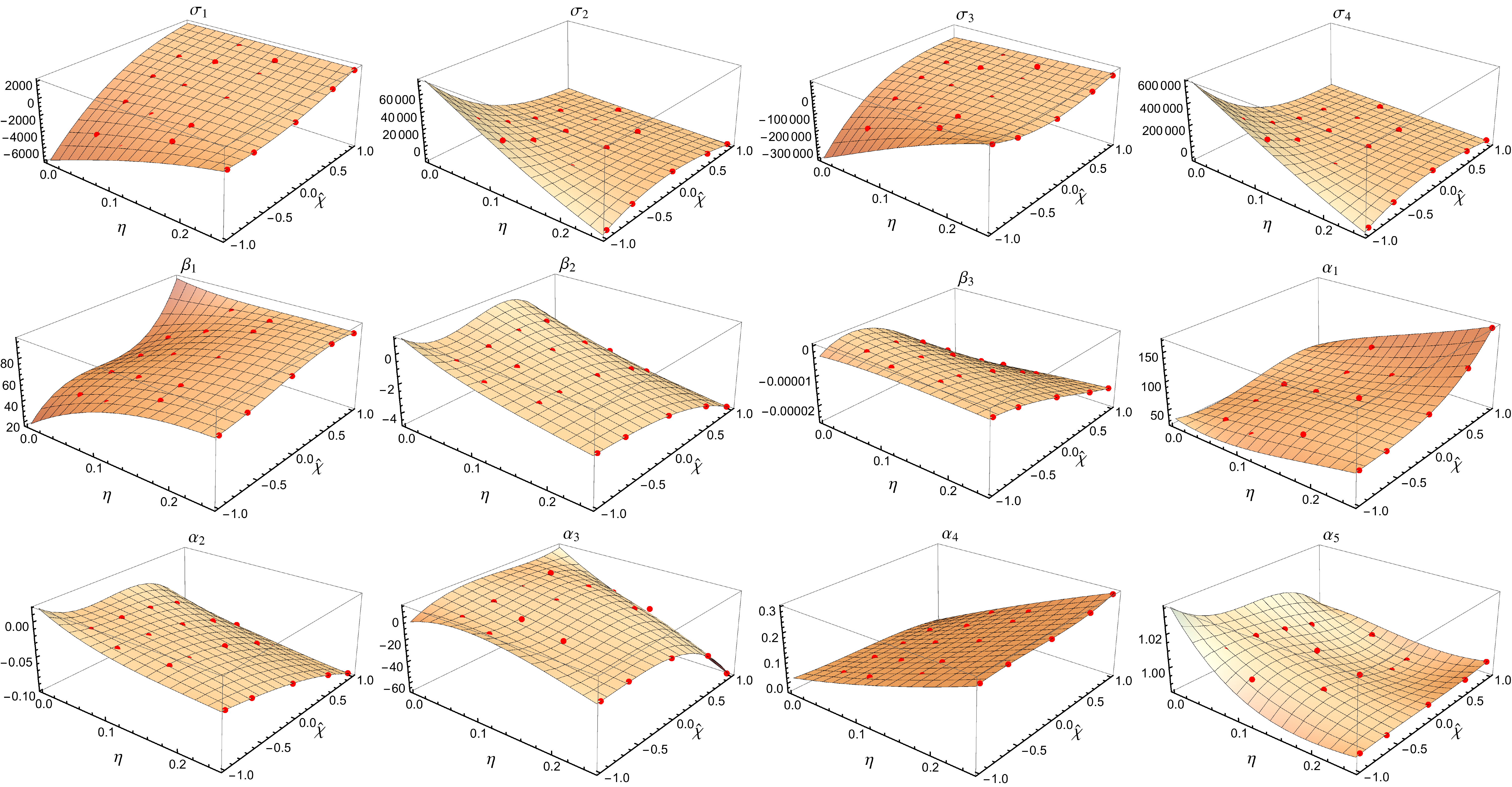}
\caption{
    Phase coefficients for region I and II.
    The calibration points and the model,
    extrapolated to the boundary of the physical
    parameter space are shown.
}
\label{fig:phasecoeffs}
\end{figure*}

\begin{figure*}[h!]
\centering
\includegraphics[width=1\linewidth]{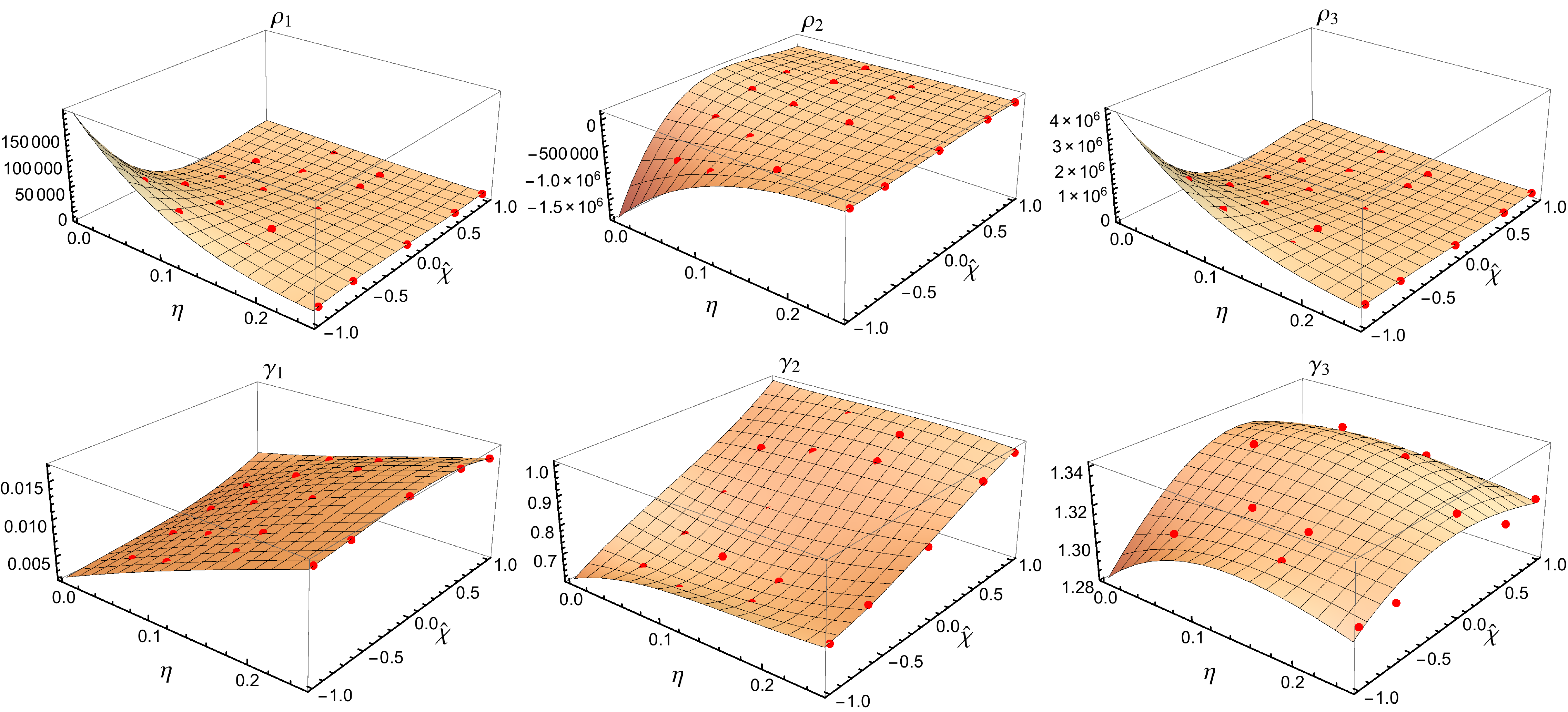}
\caption{
    Amplitude coefficients for region I and IIb.
    The calibration points and the model,
    extrapolated to the boundary of the physical
    parameter space are shown.
}
\label{fig:ampcoeffs}
\end{figure*}

\begin{figure}[tb]
\centering
\includegraphics[width=0.75\linewidth]{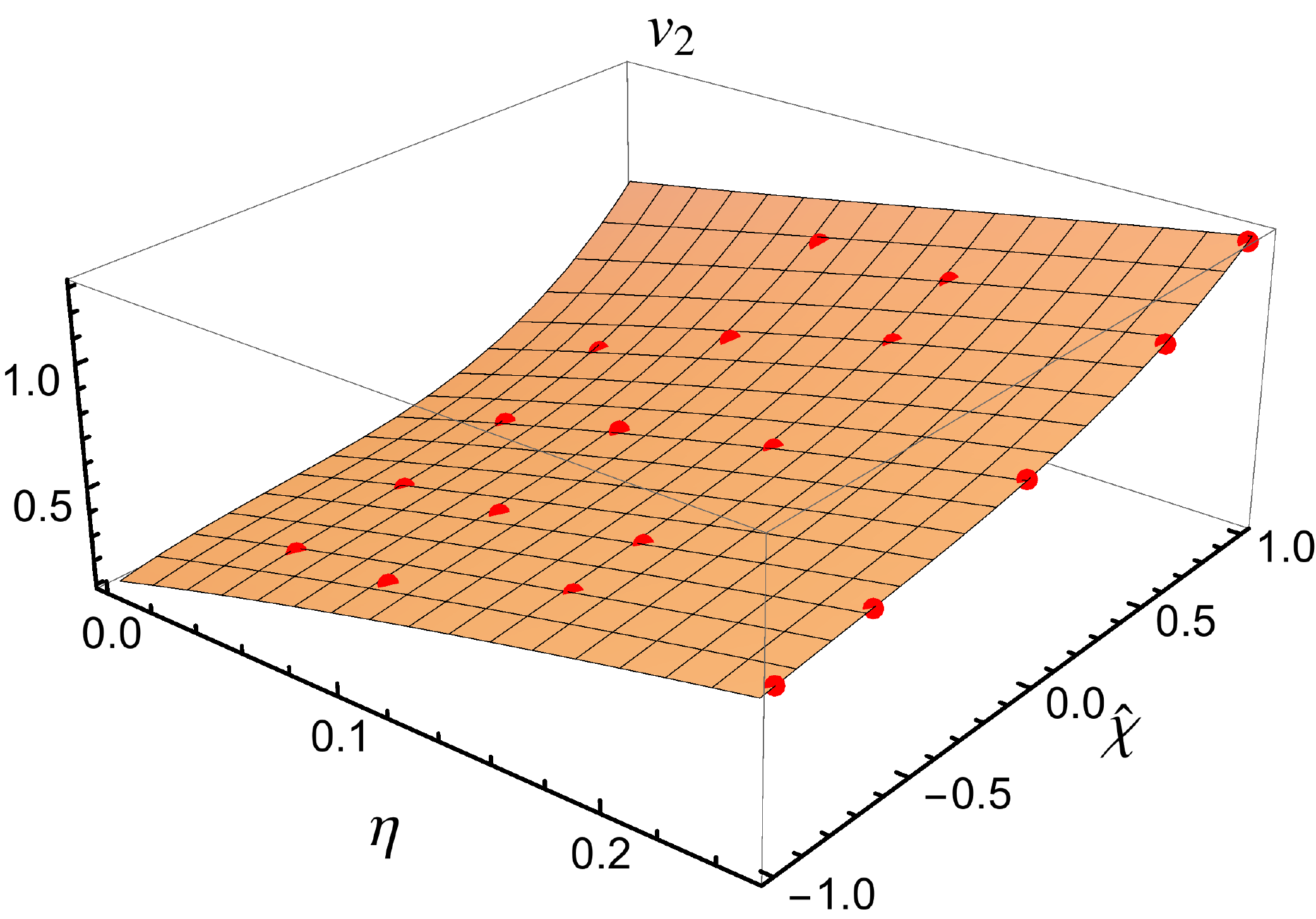}
\caption{
    Intermediate (Region IIb) amplitude coefficient.
    The calibration points and the model,
    extrapolated to the boundary of the physical
    parameter space are shown.    
}
\label{fig:ampcoeffsv2}
\end{figure}

The phase and amplitude coefficients across the $(\eta,\hat\chi)$ parameter
space are
shown in Figs.~\ref{fig:phasecoeffs}, \ref{fig:ampcoeffs} and
\ref{fig:ampcoeffsv2}. We see that
in general the coefficients vary smoothly across the parameter space, and are
captured well
by our fits.

\section{Model Validation}
\label{sec:modelval}

To evaluate the accuracy of our model we
compute the mismatch, defined in Sec.~\ref{sec:match}, between the model
and a set of hybrid waveforms,
including the 19 waveforms used to calibrate the model (Tab.~\ref{tab:wftable}),
and an additional 28 waveforms, listed in Tab.~\ref{tab:wftable2}.
The additional SpEC \NR waveforms comprise most of the remaining
aligned spin simulations in the public SXS catalogue~\cite{SXS:catalog}.
The remaining \NR waveforms were produced with BAM.

In this section we quantify the agreement for each
of these waveforms against the PhenomD model.
We also show (Sec.~\ref{sec:minset})
that using additional waveforms
in the calibration does not significantly change our model, and provide
evidence that the set of waveforms we have chosen may be close to the
minimal set necessary to accurately calibrate our model. 

A further, complementary
validation  based on time-domain transformations is presented in
Appendix~\ref{app:td}.

\begin{table}[tb]
\begin{tabular}{llllllllllll}
  \hline
  \hline
  \# & \text{Code/ID} & $q$  & \text{$\chi_1$} & \text{$\chi_2$} \\
  \hline
 B1 & \text{SXS:BBH:0159} & 1. & -0.9 & -0.9 \\
 B2 & \text{SXS:BBH:0154} & 1. & -0.8 & -0.8 \\
 B3 & \text{SXS:BBH:0148} & 1. & -0.438 & -0.438 \\
 B4 & \text{SXS:BBH:0149} & 1. & -0.2 & -0.2 \\
 B5 & \text{SXS:BBH:0150} & 1. & 0.2 & 0.2 \\
 B6 & \text{SXS:BBH:0170} & 1. & 0.437 & 0.437 \\
 B7 & \text{SXS:BBH:0155} & 1. & 0.8 & 0.8 \\
 B8 & \text{SXS:BBH:0153} & 1. & 0.85 & 0.85 \\
 B9 & \text{SXS:BBH:0160} & 1. & 0.9 & 0.9 \\
 B10 & \text{SXS:BBH:0157} & 1. & 0.95 & 0.95 \\
 B11 & \text{SXS:BBH:0158} & 1. & 0.97 & 0.97 \\
 B12 & \text{SXS:BBH:0014} & 1.5 & -0.5 & 0. \\
 B13 & \text{SXS:BBH:0008} & 1.5 & 0. & 0. \\
 B14 & \text{SXS:BBH:0013} & 1.5 & 0.5 & 0. \\
 B15 & \text{SXS:BBH:0169} & 2. & 0. & 0. \\
 B16 & \text{BAM} & 2. & 0.5 & 0.5 \\
 B17 & \text{BAM} & 2. & 0.75 & 0.75 \\
 B18 & \text{BAM} & 3. & -0.5 & -0.5 \\
 B19 & \text{SXS:BBH:0036} & 3. & -0.5 & 0. \\
 B20 & \text{SXS:BBH:0168} & 3. & 0. & 0. \\
 B21 & \text{SXS:BBH:0045} & 3. & 0.5 & -0.5 \\
 B22 & \text{SXS:BBH:0031} & 3. & 0.5 & 0. \\
 B23 & \text{SXS:BBH:0047} & 3. & 0.5 & 0.5 \\
 B24 & \text{BAM} & 4. & -0.25 & -0.25 \\
 B25 & \text{BAM} & 4. & 0.25 & 0.25 \\
 B26 & \text{SXS:BBH:0060} & 5. & -0.5 & 0. \\
 B27 & \text{SXS:BBH:0056} & 5. & 0. & 0. \\
 B28 & \text{SXS:BBH:0166} & 6. & 0. & 0. \\
 B29 & \text{BAM} & 10. & 0. & 0. \\
\hline
 \hline
\end{tabular}
\caption{
    \label{tab:wftable2} 
   Additional waveforms used to verify the model, but \emph{not} used in its
calibration.
   }
\end{table}

\subsection{Mismatches}
\label{sec:mismatches}

In this section we compute the mismatch between PhenomD and
all of the hybrid waveforms in Tabs.~\ref{tab:wftable} and
\ref{tab:wftable2}.

The model was calibrated to hybrid waveforms with a starting frequency of 
$Mf = 0.0035$, but the waveforms from many astrophysical compact binaries will
be detectable 
by \aLIGO and \AdV from much lower frequencies. We 
assume that the minimum mass
for one of the compact objects is given by the typical
mass of a neutron star i.e., $M_{\rm{NS}} \sim 1.4 M_\odot$.
The total mass of the binary can then be no lower than 
$M_{\rm{min}} = (q + 1) M_{\rm{NS}}$ for configurations with mass-ratio $q$. 
Our goal is to produce a model that is accurate for binaries that can be
detected from 10\,Hz down to either $12\,M_\odot$~\cite{Buonanno2009}, 
or $M_{\rm min}$, if this
exceeds
$12\,M_\odot$, which is the case for systems with $q \gtrsim 8$. 
At 10\,Hz, the waveform frequency of a $12\,M_\odot$ binary is $Mf \approx
0.0006$,
and so in this section we compare our model to much longer hybrids that extend
down 
to $Mf = 0.0006$. 

\begin{figure*}[tb]
    \begin{minipage}[l]{1.0\columnwidth}
        \centering
        \includegraphics[width=1\linewidth]{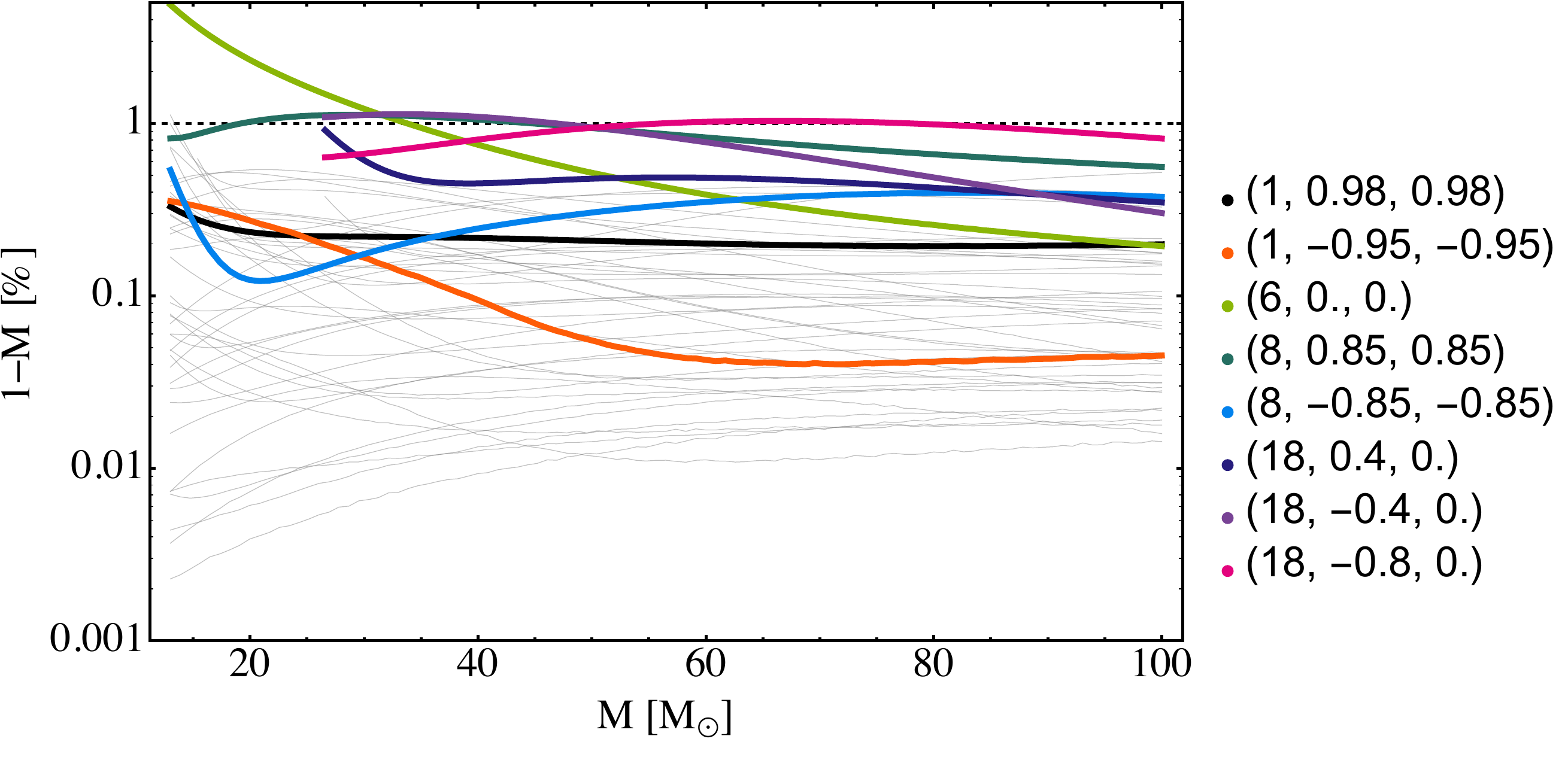}
    \end{minipage}
    \hfill{}
    \begin{minipage}[r]{1.0\columnwidth}
        \centering
        \includegraphics[width=1\linewidth]{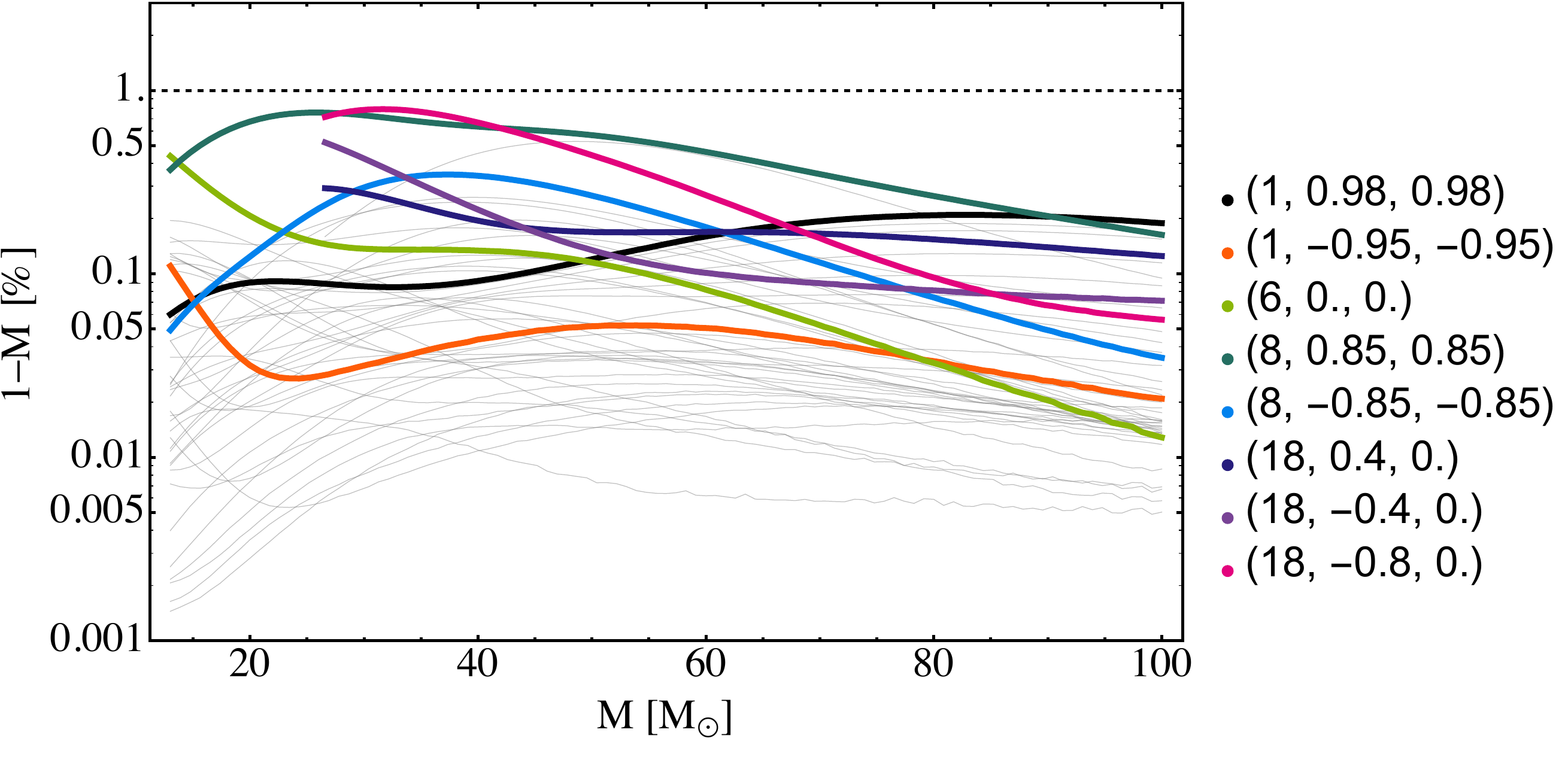}
    \end{minipage}
    \caption{
    Mismatches of the PhenomD model against all 48 available hybrid waveforms.
    The highlighted configurations are those closest to the edge of
    the $(\eta,\hat{\chi})$ parameter space as well as
    the case with the worst mismatch $(q,\chi_1,\chi_2)=(6,0.,0).$
    The majority of cases show mismatches well below 1\%.
    Left: Mismatches using the \aLIGO design sensitivity noise curve
    (zdethp) with a lower frequency cut off of 10 Hz.
    Right: Early \aLIGO noise curve with a 30 Hz cut off.
    }
    \label{fig:PhenDvsHybIMR}
\end{figure*}

The results are presented in Fig.~\ref{fig:PhenDvsHybIMR}.
The left panel uses the \aLIGO design sensitivity zero-detuned,
high-laser-power noise curve with $[f_{\rm min}, f_{\rm max}] = [10, 8000]$\,Hz
           \cite{advLIGOcurves}.
The worst mismatch is for the $\{q, \chi_1, \chi_2\} = \{6, 0, 0\}$ at
low masses which tends towards a mismatch of 3\% at $12\,M_\odot$.
All other mismatches fall below 1\%
with the majority distributed around 0.1\%.
We note, however, that the \emph{fitting factors} for the
waveforms in the model (i.e., matches optimised over
binary parameters, which is the relevant quantity for
searches) are better than 0.999 in all cases we have considered.
In particular, at
low masses the mismatch between different options of inspiral approximant
will be much larger than the mismatch between PhenomD and our hybrid
waveforms; the dominant error is in our uncertainty of the true inspiral
waveform,
and not in our model; this will be made clearer in Sec.~\ref{sec:modelcomparisons}.

The right panel in Fig~\ref{fig:PhenDvsHybIMR} shows the same
calculation but using the
predicted noise curve for early aLIGO science runs~\cite{Aasi:2013wya},
with a lower frequency cut-off of 30 Hz.
Due to the change in shape of the noise curve and lower frequency
cut-off the mismatches improve such that \emph{all} mismatches
are comfortably below 1\%. This gives a more realistic idea
of the performance of our model during the initial
science run of the advanced detectors.

In both panels, the highlighted cases are those at the edges of the calibration
region of 
parameter space. We note that the worst mismatches are for high mass ratios
and large spins. This suggests the region of parameter space that requires the
most improvement in future models --- although it is clear that for all of these
configurations
the model is well within the accuracy requirements for the second-generation
detectors.

\subsection{The effective spin approximation}
\label{sec:chiapprox}

The phenomenological fits to the waveform phase and amplitude are parameterised
by the weighted reduced spin, $\hat\chi$, Eq.~(\ref{eqn:chihat}). This is an
approximation,
based on the observation that the dominant spin effect on the inspiral phase is
due to 
this combination of the two spins, $\chi_1$ and $\chi_2$. This approximation is
not 
expected to be valid through the merger and ringdown; in the ringdown the
waveforms will
be characterized by the \emph{final spin}. The model was produced using mostly
equal-spin
$\chi_1 = \chi_2$ waveforms, and in general may not be accurate for systems with
unequal spins.

However, we have seen in the previous 
Sec.~\ref{sec:mismatches} that our model agrees well with all available hybrid
waveforms,
\emph{including} several with unequal spins. This included only four
unequal-spin configurations
that were not included in the calibration, and none were high-aligned-spin
systems. 

We expect that the reduced-spin approximation will perform worst for high mass
ratios and high
aligned spins. If we consider pure \PN inspiral waveforms, we find, for example,
that a system with mass-ratio
1:3 and total mass of 12\,$M_\odot$, with $\chi_1 = 1$ and $\chi_2 = -1$, that
the match against the 
corresponding reduced-spin waveform (with $\hat\chi = 0.655$) is less than 0.8.
However, if we consider
a configuration where the larger \BH has an anti-aligned spin, $\chi_1 =
-1, \chi_2 = 1$, then the 
match with the corresponding reduced-spin waveform ($\hat\chi = -0.655$) is much
better, 0.955.

This example was only an illustration. The performance of the reduced-spin
approximation at low masses 
does not concern us in the PhenomD model, where we use both spins $\chi_1$ and
$\chi_2$ to generate the 
base TaylorF2 phase. What we wish to know is how well the approximation holds
for high-mass systems,
where the late inspiral, merger and ringdown dominate the SNR. Those systems are
described by our 
merger-ringdown Region II model, for which the spin dependence is parameterized
only with $\hat\chi$.

\begin{figure}[tb]
\centering
\includegraphics[width=0.8\linewidth]{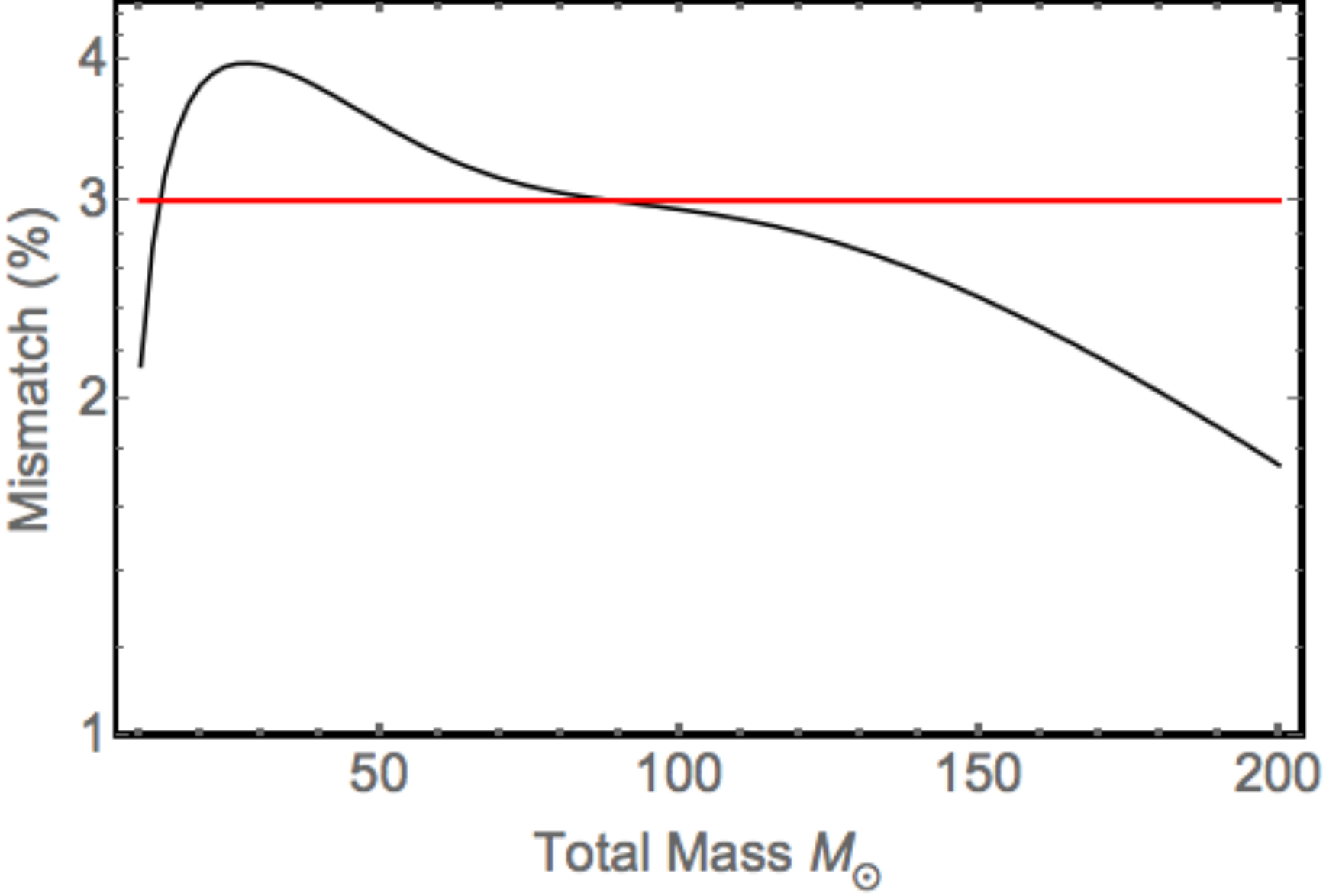}
\caption{Mismatch between a $q=8$, $\chi_1 = 0.8$, $\chi_2 = 0$ SEOBv2+NR
hybrid, and the PhenomD model. We see that the mismatch exceeds our 1\%
threshold
everywhere. However, the fitting factor is everywhere better than 0.9995, with
negligible
parameter biases (see text).
}
\label{fig:q8a0a08match}
\end{figure}

We have produced one high mass-ratio, high-spin \NR simulation to compare with,
$q=8$ and $\chi_1 = 0.8$,
$\chi_2 = 0$. Fig.~\ref{fig:q8a0a08match} shows the mismatch between this hybrid
waveform and the 
PhenomD model. As we expect in this region of the parameter space, the poor
quality of the reduced-spin
approximation causes a mismatch that exceeds our 1\% threshold for all masses.
However, if we 
calculate \emph{fitting factors} (i.e., minimise the mismatch with respect to
the model parameters $(\eta,
\hat\chi)$, as done in a GW search and, effectively, in parameter estimation),
then we find deviations from unity
of below 0.05\% 
for all masses. We also find biasses of less than 1\% in the total mass, less
than 2\% in the symmetric 
mass ratio, and less than 0.005 in the reduced spin, $\hat\chi$. We expect these
biasses to be far less than 
the statistical uncertainties in these quantities for observations with
second-generation detectors, and so we
conclude that the reduced-spin approximation will not impose any limit on the
science potential of these
detectors. 

Studies with the SEOBNRv2 model support this conclusion. Although we do not
expect that model to 
be accurate through the merger-ringdown for high spins, as we will see in 
Sec.~\ref{sec:modelcomparisons}, it is likely that its qualitative behaviour
with respect to parameter
variations is approximately correct, and the model allows us to study the
behaviour of the reduced-spin
approximation over the entire calibration parameter space of our model. 

Although the reduced-spin approximation will not limit our ability to measure
$\hat\chi$, one could argue
that it nonetheless prevents any measurement of individual spins. We have argued
in previous 
work~\cite{Puerrer2013} that it may be difficult to measure both \BH
spins even if we have a double-spin 
model. A further study, which provides much stronger evidence for this claim,
will be published in the 
near future~\cite{Puerrer2015}. In practice the measurable intrinsic parameters
of the binary will be 
$(M, \eta, \hat\chi)$, and these are the parameters of our model.

\subsection{Calibration Set of waveforms}
\label{sec:minset}

The construction of previous phenomenological 
models~\cite{Ajith2007,Ajith:2007kx,Ajith2011,Santamaria2010}
suggested that the parameter dependence of the coefficients in our models
depend sufficiently smoothly across the parameter space that each
coefficient can be presented by a low-order polynomial in each
parameter, and therefore we require only 4-5 waveforms for each
direction in parameter space. This expectation is borne out in the current
model, where we use four values of the mass ratio (1, 4, 8 and 18) and
four or five values of the spin at each mass ratio. 

In this section we consider versions of the model constructed with more
(or less) calibration waveforms. We find that our small set
of 19 calibration waveforms is just as accurate as a model
that is calibrated against a much larger set of 48 waveforms.
To quantify this test we compute the maximum mismatch of
four distinct models
against all hybrid waveforms used in this paper, i.e., the 48 waveforms in 
Tabs.~\ref{tab:wftable} and \ref{tab:wftable2}.

\begin{figure}[tb]
\centering
\includegraphics[width=1\linewidth]{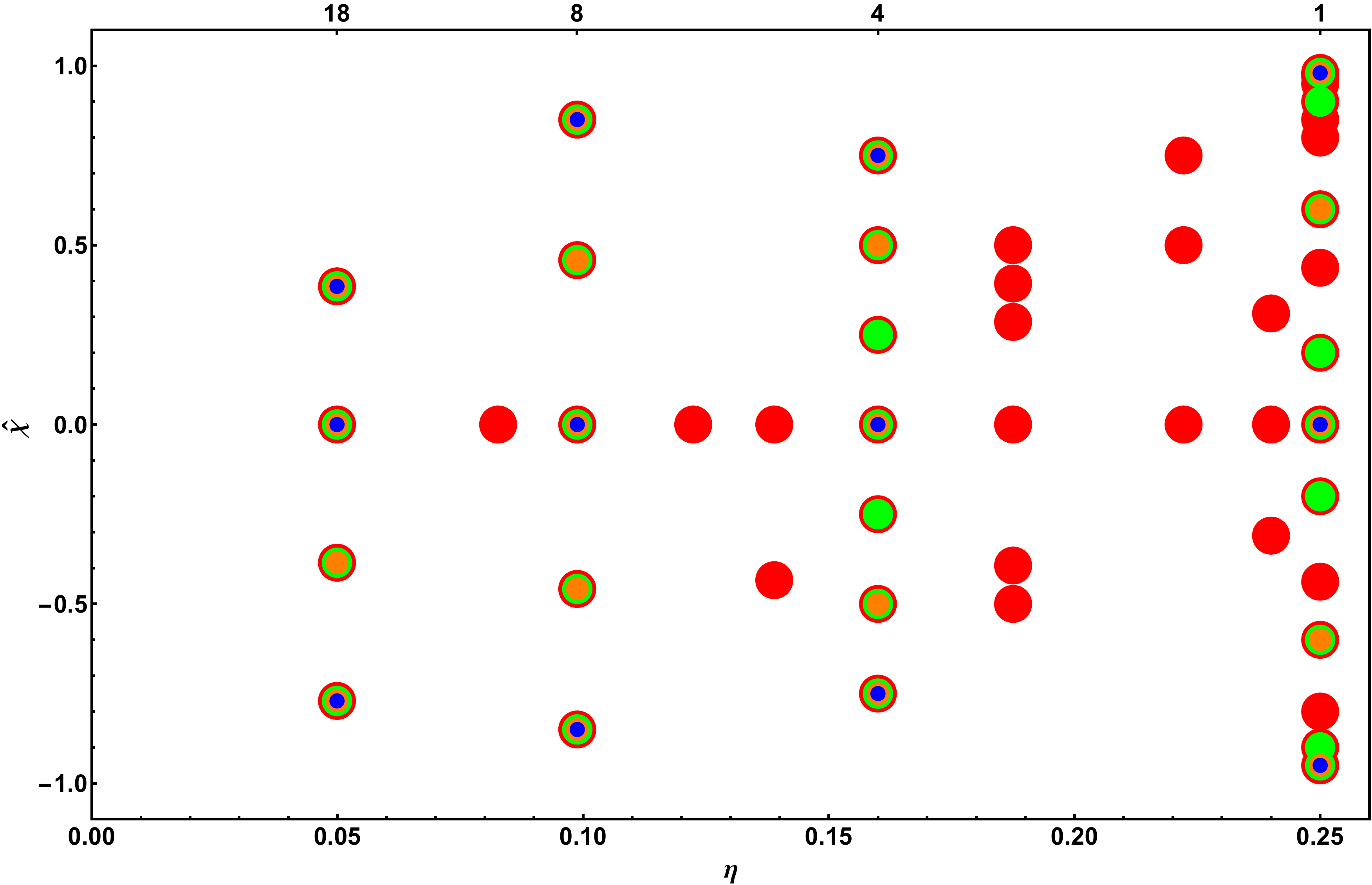}
\caption{
Four sets of calibration waveforms. Set 1 (48 waveforms) is indicated in red,
Set 2 (25
waveforms) in green, Set 3 (19 waveforms, used for the final PhenomD model) in
orange,
and Set 4 (12 waveforms) in blue.
}
\label{fig:ParSpaceMinSet}
\end{figure}

Fig~\ref{fig:ParSpaceMinSet} indicates four choices of parameter-space
coverage. 
The first set is the largest, and includes all 48 configurations indicated in
the figure. The
second set includes 25 waveforms, but only at mass ratios 1, 4 8 and 18, and
does not 
include all available spin values at mass ratios 1 and 8. The third set consists
of the 19
waveforms that we use for our final model. The fourth set is more sparsely
sampled in 
spins, with only three spin values at each mass ratio, and only 12 waveforms in
total. 

Four models were constructed, each using the same prescription, except for the
Set-4
model, for which we used a lower-order fit in the $\hat\chi$ direction, since in
general 
we cannot expect to fit four coefficients with only three spin values. 

The results are summarized in Tab.~\ref{tab:minset2}. We calculate the mismatch
between 
each of the models and all 48 hybrids, over the same mass range used in
Sec.~\ref{sec:mismatches} using the early aLIGO noise curve with a 30 Hz
cut off.
For each hybrid we calculate the largest mismatch in that mass range. The table
indicates the number of configurations for which we find mismatches larger than
0.1\%, 1\% and 3\% for each model. As we have already seen in
Fig.~\ref{fig:PhenDvsHybIMR},
the fiducial Set-3 model has mismatches of less than 1\% for all configurations.
We find that increasing the number of calibration waveforms
does not 
significantly improve the model's performance.

We also see that if we further \emph{reduce} the number of calibration
waveforms, as in 
the Set-4 model, then the accuracy of the model drops significantly. For this
model there
are now three configuration with mismatches worse than 1\%, and one
configuration with 
a mismatch worse than 7\%. We therefore conclude that, in the sense of the
simple comparison
that has been performed here, the Set-3 model represents the optimal choice of
calibration
waveforms.

\begin{table}
    \begin{tabular}{llllll}
    \hline \hline
    Model                  & \# waveforms &  $> 0.1$\% & $> 1$\% & $>3$\%    &
$\max \mathcal M$ (\%) \\ \hline
    Set 1                    & 48           & 19              & 0            & 0
              & 0.94       \\
    Set 2                    & 25           & 27              & 0            & 0
              & 0.83       \\
    Set 3 (*)                & 19           & 29              & 0             & 0
              & 0.87       \\
    Set 4                     & 12          & 37              & 3            & 1
              & 7.82      \\ \hline \hline
    \end{tabular}
\caption{Comparison of models constructed with different sets of calibration
waveforms. The table shows,
for each calibration set (see Fig.~\ref{fig:ParSpaceMinSet}), the number of
verifications waveforms (out of 48)
for which there is a mismatch $\mathcal M$ above $0.1$\%, 1\%, or 3\%,
over the same mass range used in
Sec.~\ref{sec:mismatches} using the early aLIGO noise curve with a 30 Hz
cut off.
We see that with a small set of 19 waveforms we achieve comparable
mismatches to models which used larger sets of calibration waveforms,
and that using \emph{less} waveforms significantly degrades
the quality of the model. Set 3
is used for the final model.
}
\label{tab:minset2}
\end{table}

\section{Model vs Model Comparisons}
\label{sec:modelcomparisons}

We have demonstrated the high degree of fidelity of PhenomD
to both the waveforms that were used in calibrating the model and
to those that were not. Without further comparisons to \NR waveforms
we cannot rigorously quantify the accuracy of our, or indeed any,
waveform model.
However, it is reasonable to assume that if two independent waveform models
agree over a portion of the parameter space then we can gain some
well-founded confidence in their accuracy.

The computational cost of the SEOBNRv2
model makes it difficult to make detailed comparisons across the entire
parameter
space with high resolution in $(\eta,\hat\chi)$. However, based upon the recent
work
in Ref.~\cite{Puerrer2014}, a \emph{reduced order model} (ROM) of SEOBNRv2,
called SEOBNRv2\_ROM, has been developed~\cite{Puerrer2015inprep}. 
This is a fast, frequency-domain approximation to the SEOBNRv2 model 
that has a worst mismatch against SEOBNRv2 of $1 \%$, but in general mismatches are
better than $\sim 0.1 \%$.
SEOBNRv2\_ROM is a two spin model which can be used to estimate
SEOBNRv2 waveforms with symmetric mass-ratios $\eta \in [0.01,0.25]$ and spins
$\chi_i \in [-1,0.99]$.
The ROM can be used over the frequency range $Mf \in [0.0001,0.3]$
Note that the underlying SEOBNRv2 model was calibrated to \NR waveforms
up to mass-ratios 1:8 and spins up to 0.5 (except along the equal
mass line where spins in the range $[-0.95,0.98]$ were used).
The merger-ringdown parts of the PhenomD and SEOBNRv2 models are almost completely
independent of one another with the only common features being that
they share some of the same calibration waveforms, i.e., the ones from
the public SXS catalogue and also the same underlying EOB Hamiltonian.

During the following comparison we restrict the computation
of the mismatch to the frequencies of the
SEOBNRv2\_ROM, namely
$[0.0006,0.135]$, using the design sensitivity noise curve
with a lower frequency cut-off of 10 Hz as in previous sections.

\begin{figure*}
    \centering
        {\includegraphics[width=0.3\textwidth]{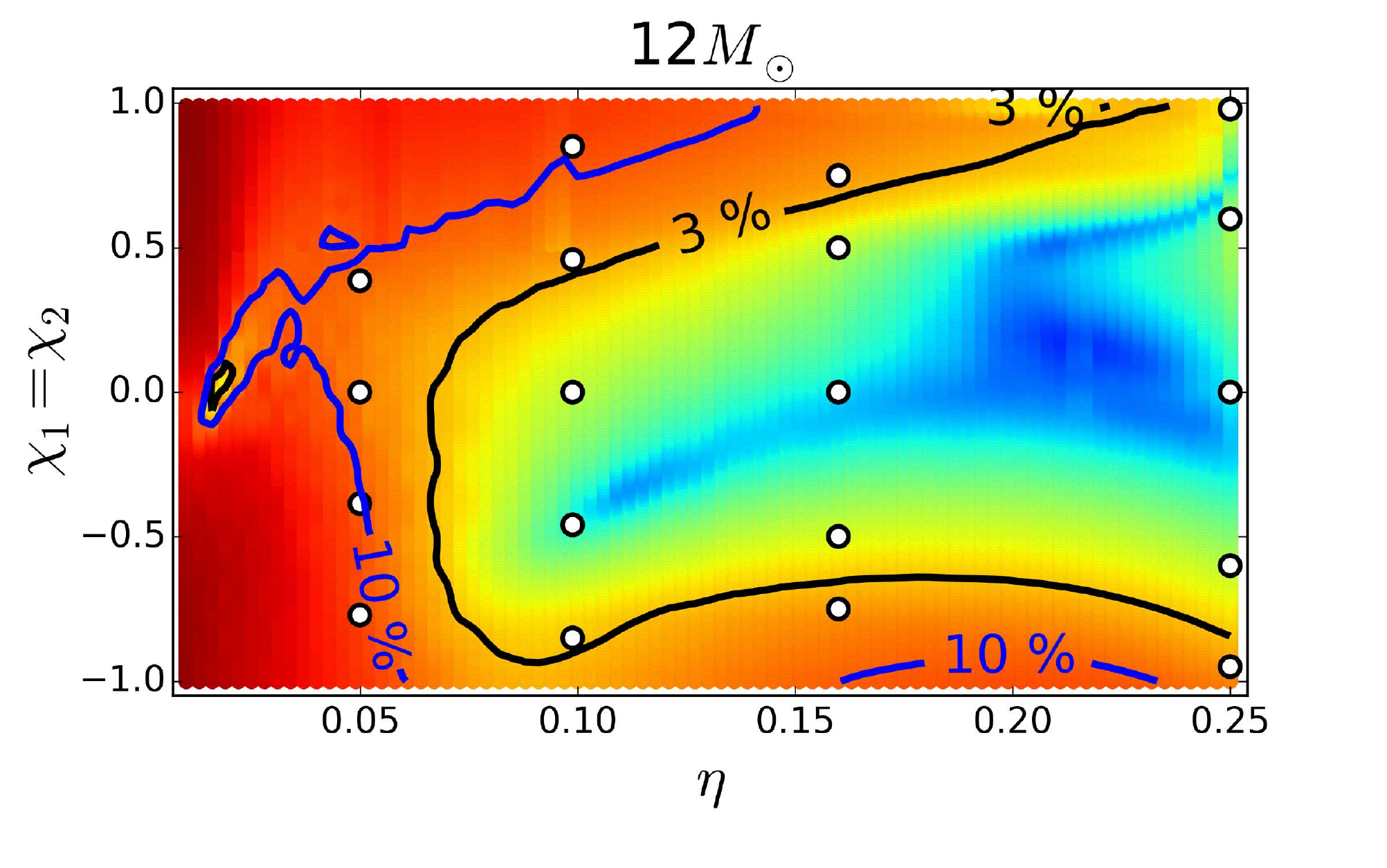}} \quad
        {\includegraphics[width=0.3\textwidth]{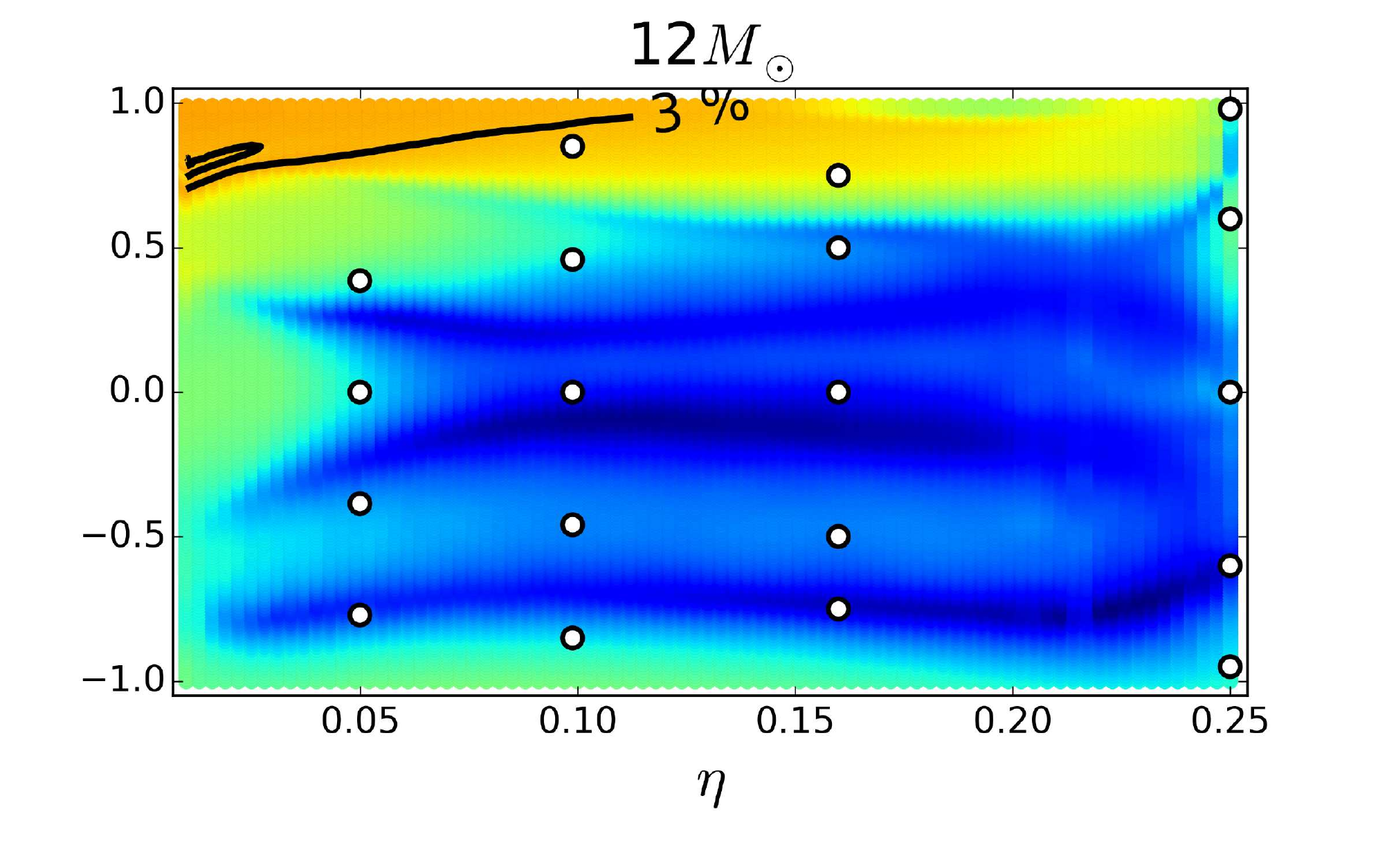}} \quad
        {\includegraphics[width=0.35\textwidth]{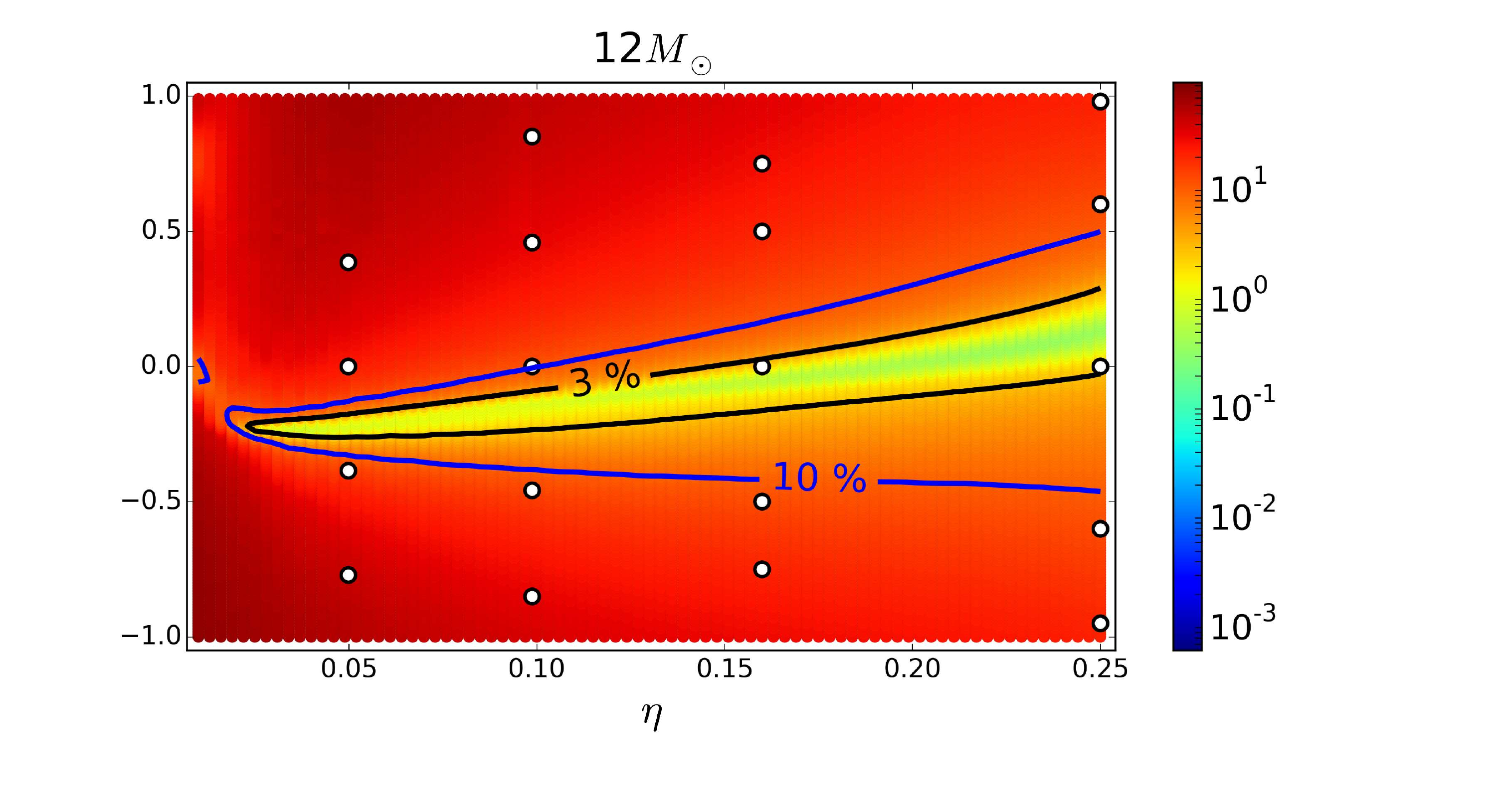}} \\
        {\includegraphics[width=0.3\textwidth]{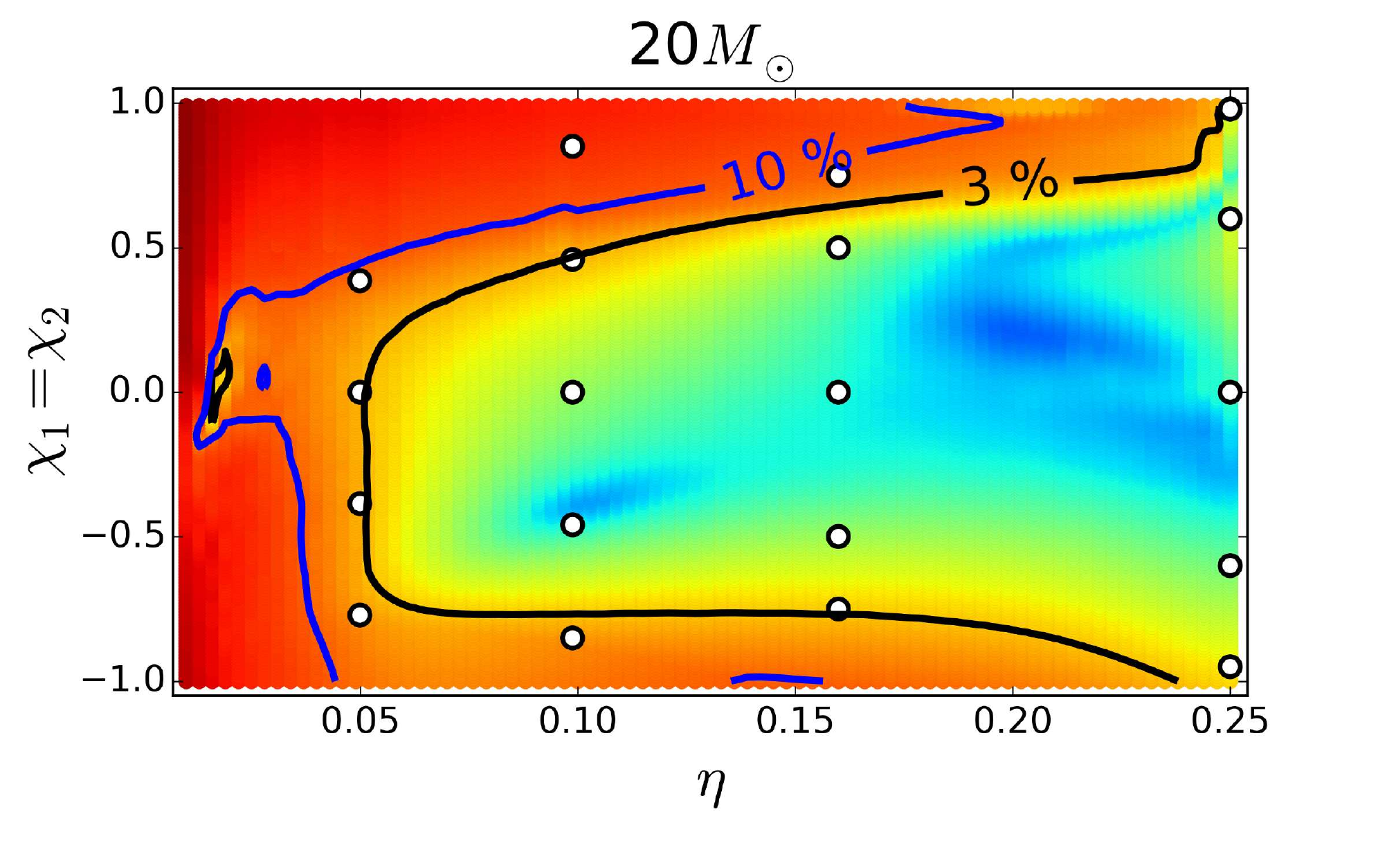}} \quad
        {\includegraphics[width=0.3\textwidth]{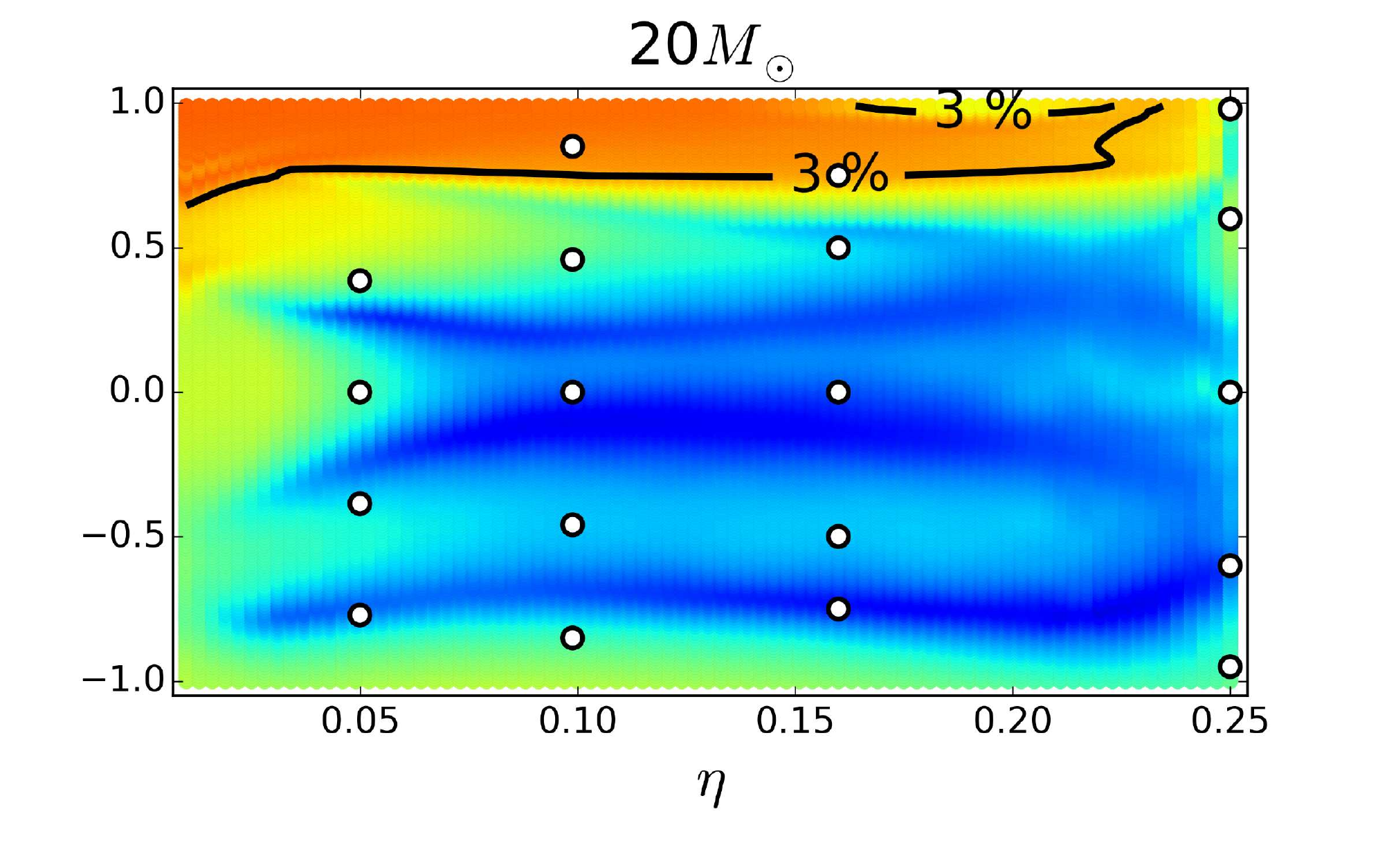}} \quad
        {\includegraphics[width=0.35\textwidth]{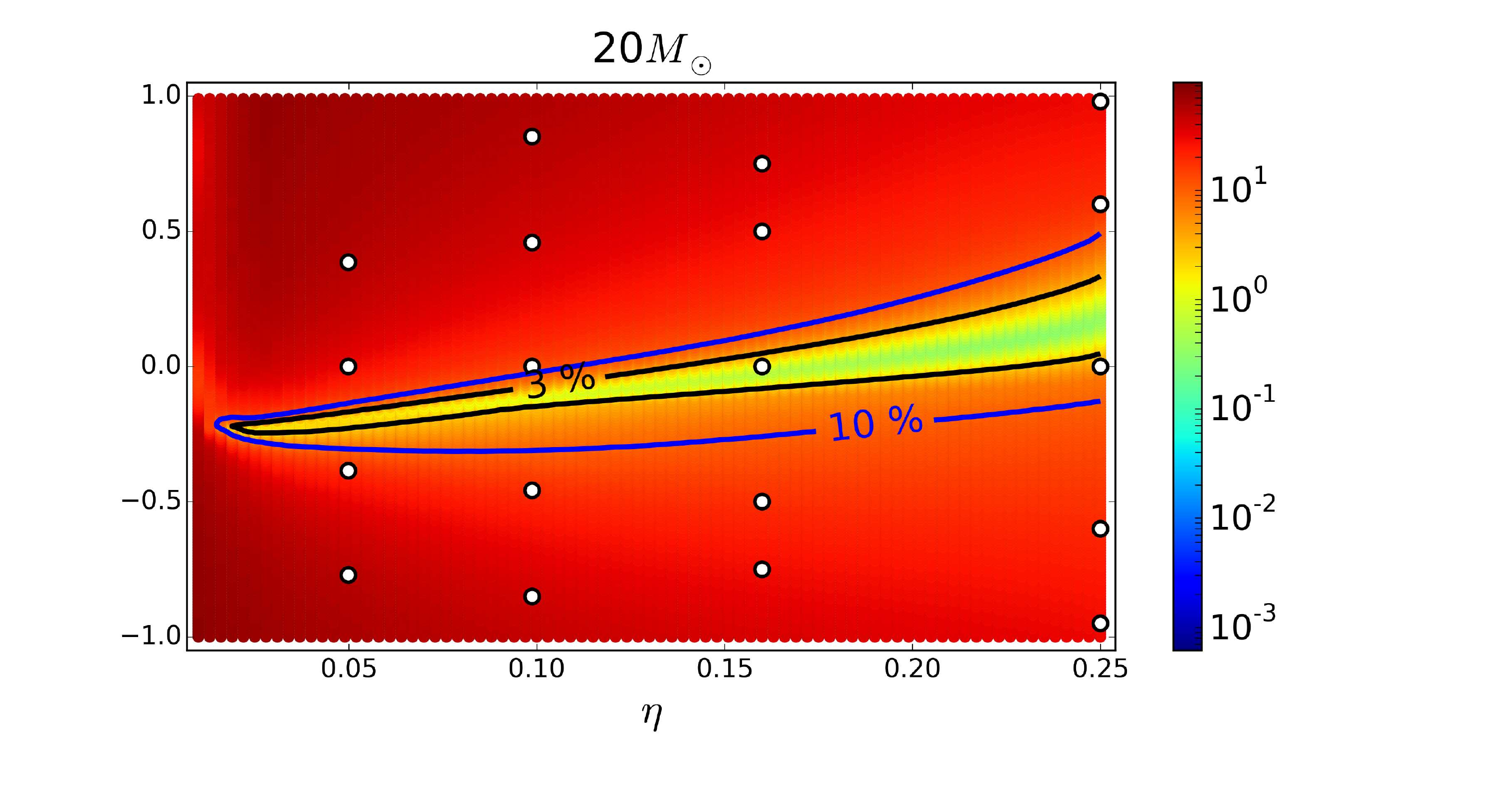}} \\
        {\includegraphics[width=0.3\textwidth]{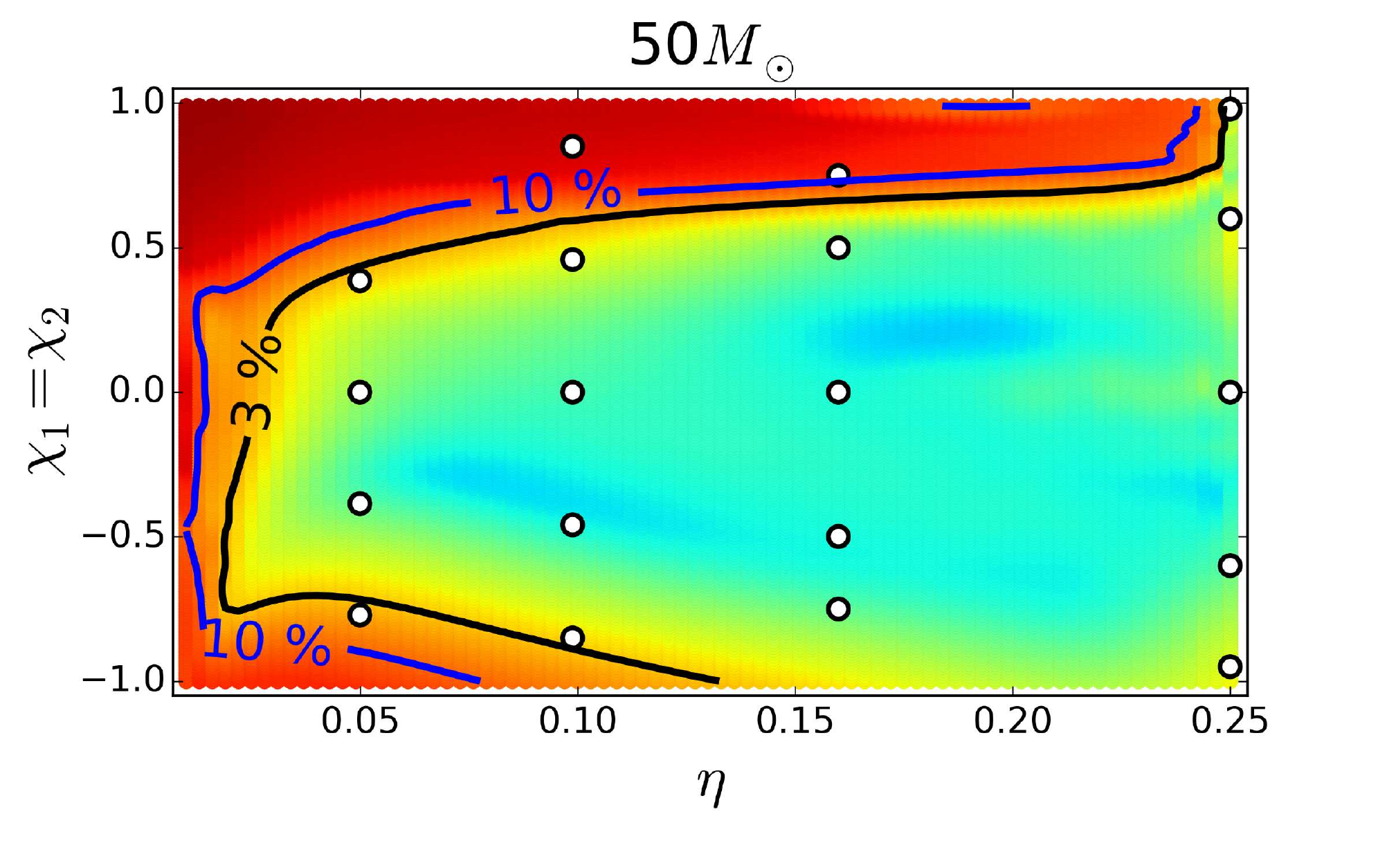}} \quad
        {\includegraphics[width=0.3\textwidth]{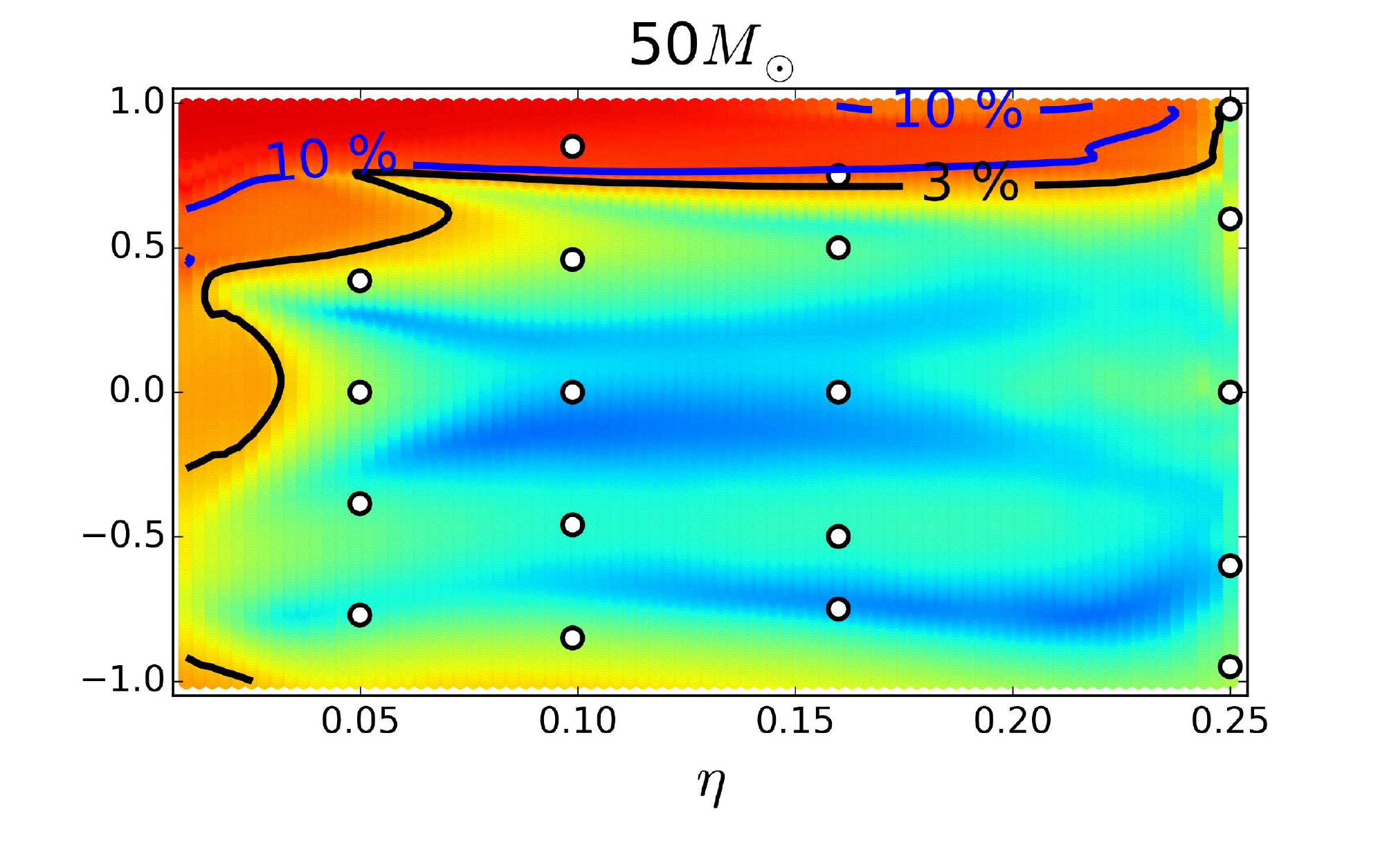}} \quad
        {\includegraphics[width=0.35\textwidth]{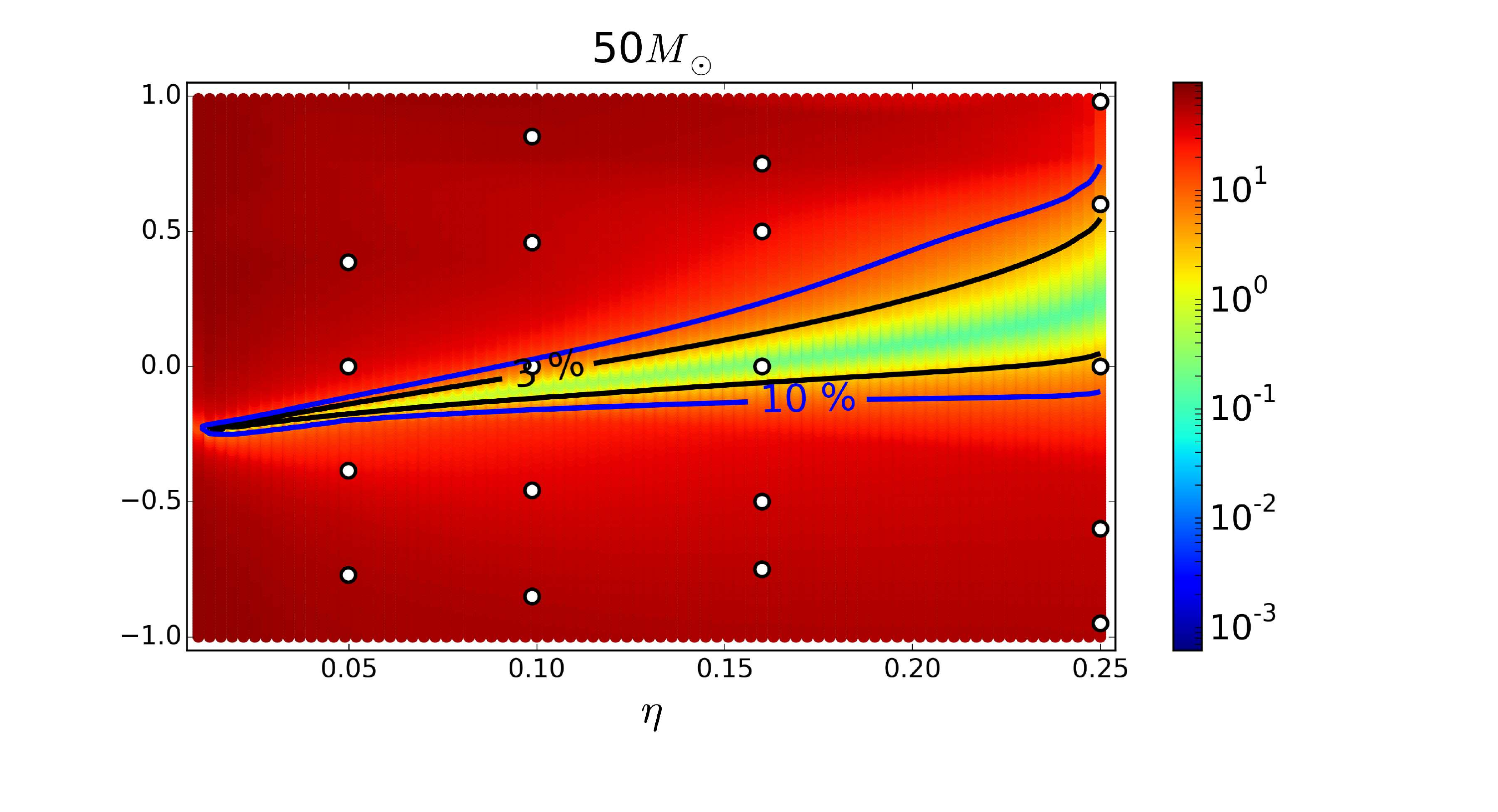}} \\        
        {\includegraphics[width=0.3\textwidth]{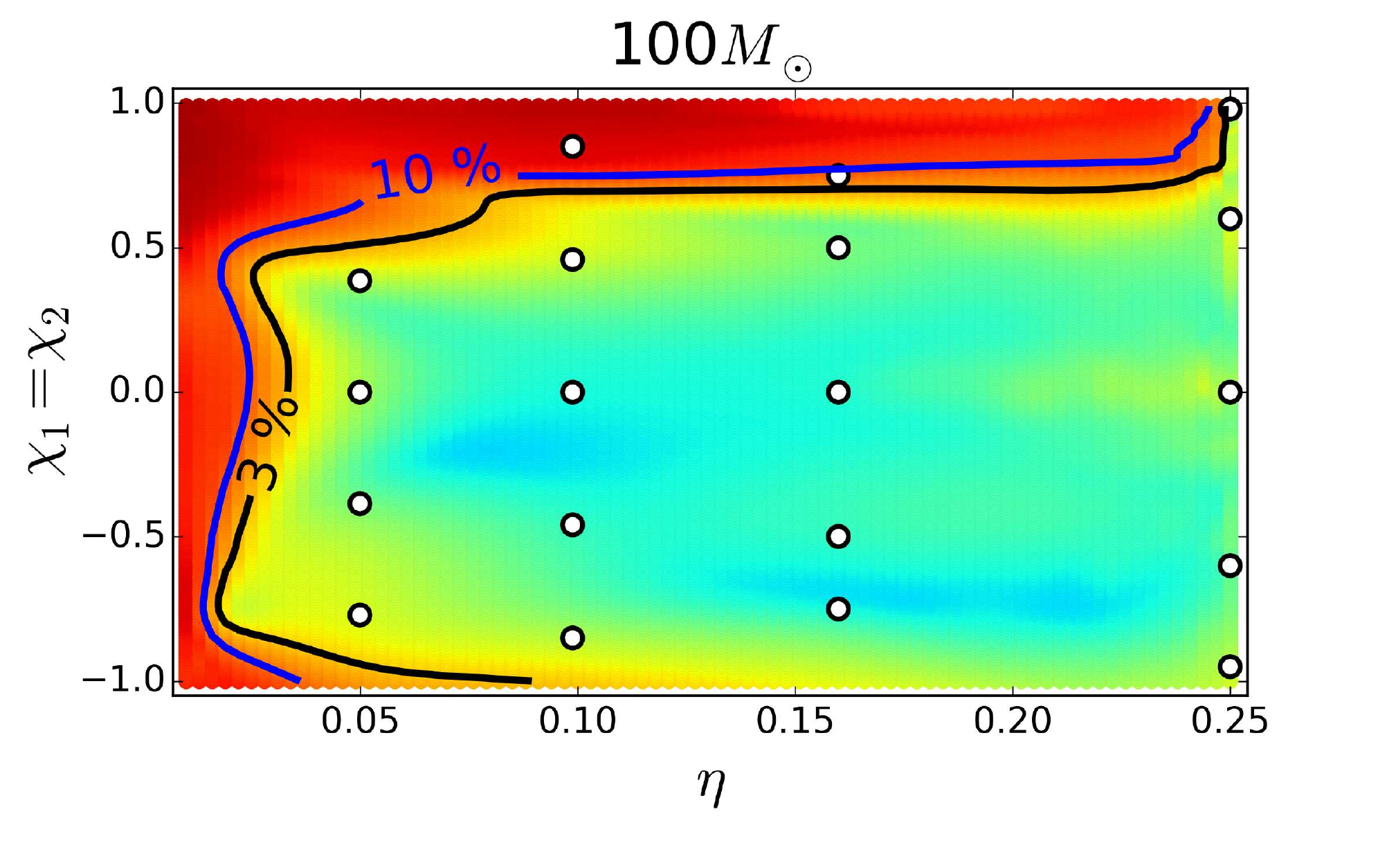}} \quad
        {\includegraphics[width=0.3\textwidth]{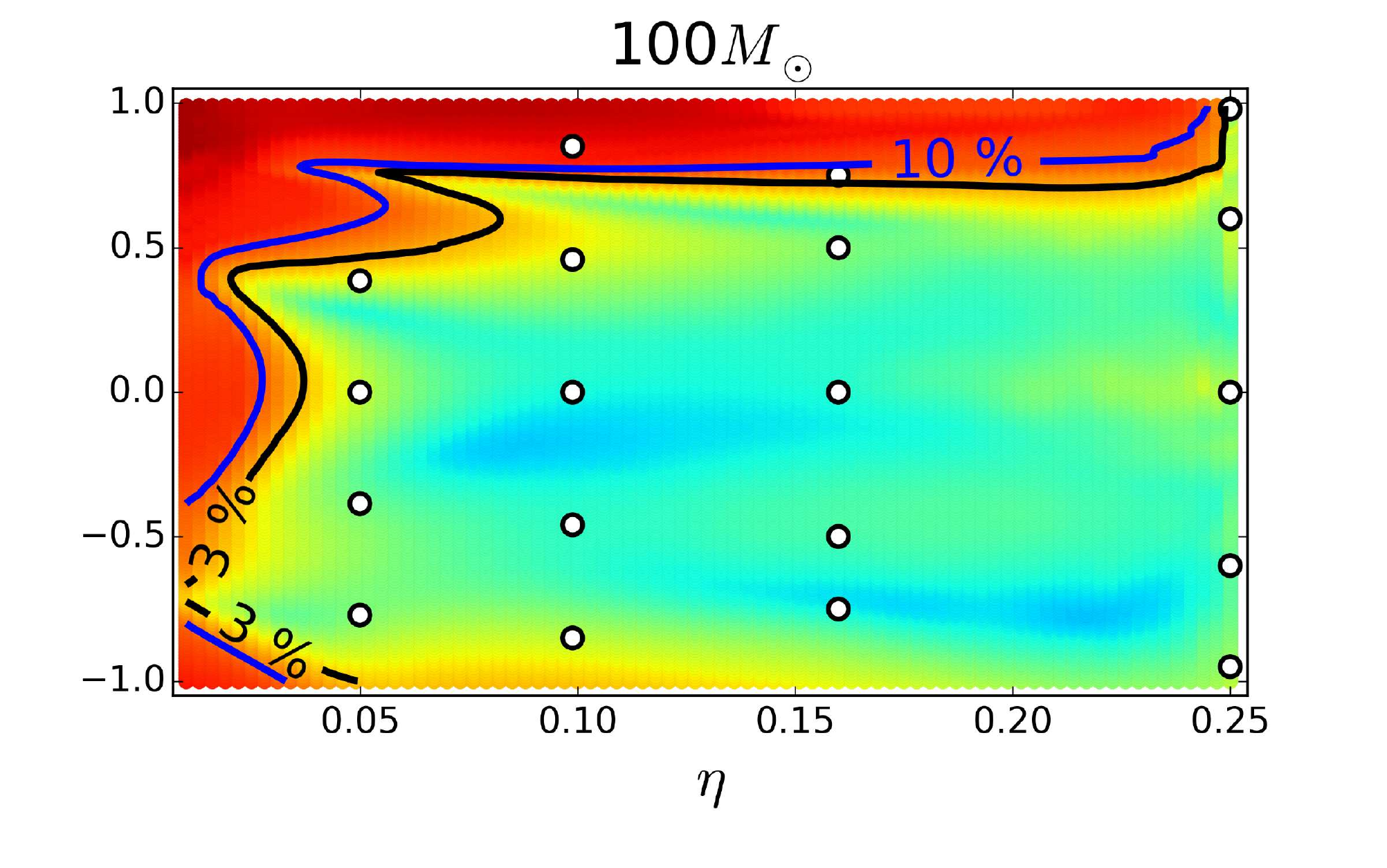}} \quad
        {\includegraphics[width=0.35\textwidth]{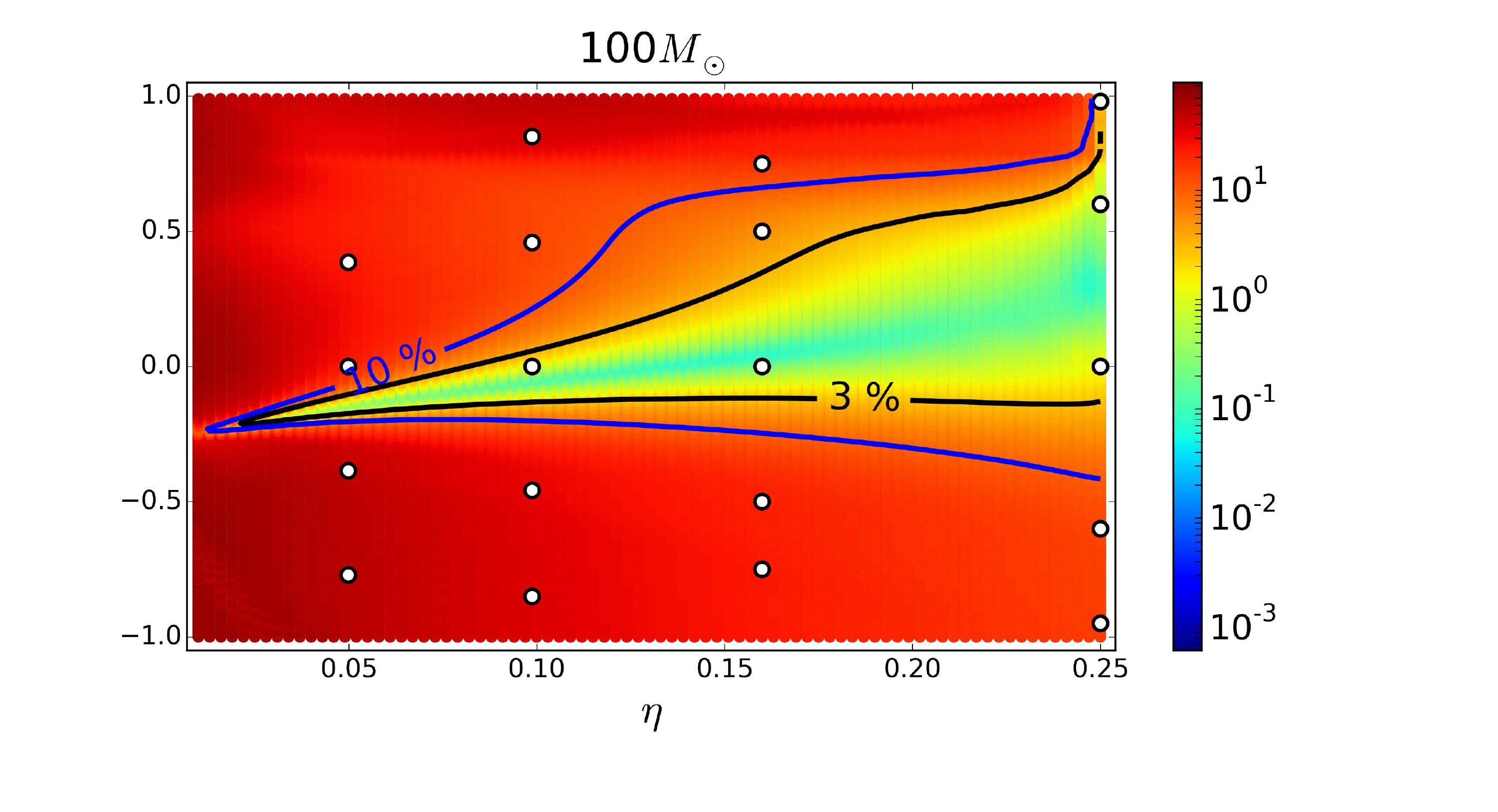}} \\                
        {\includegraphics[width=0.3\textwidth]{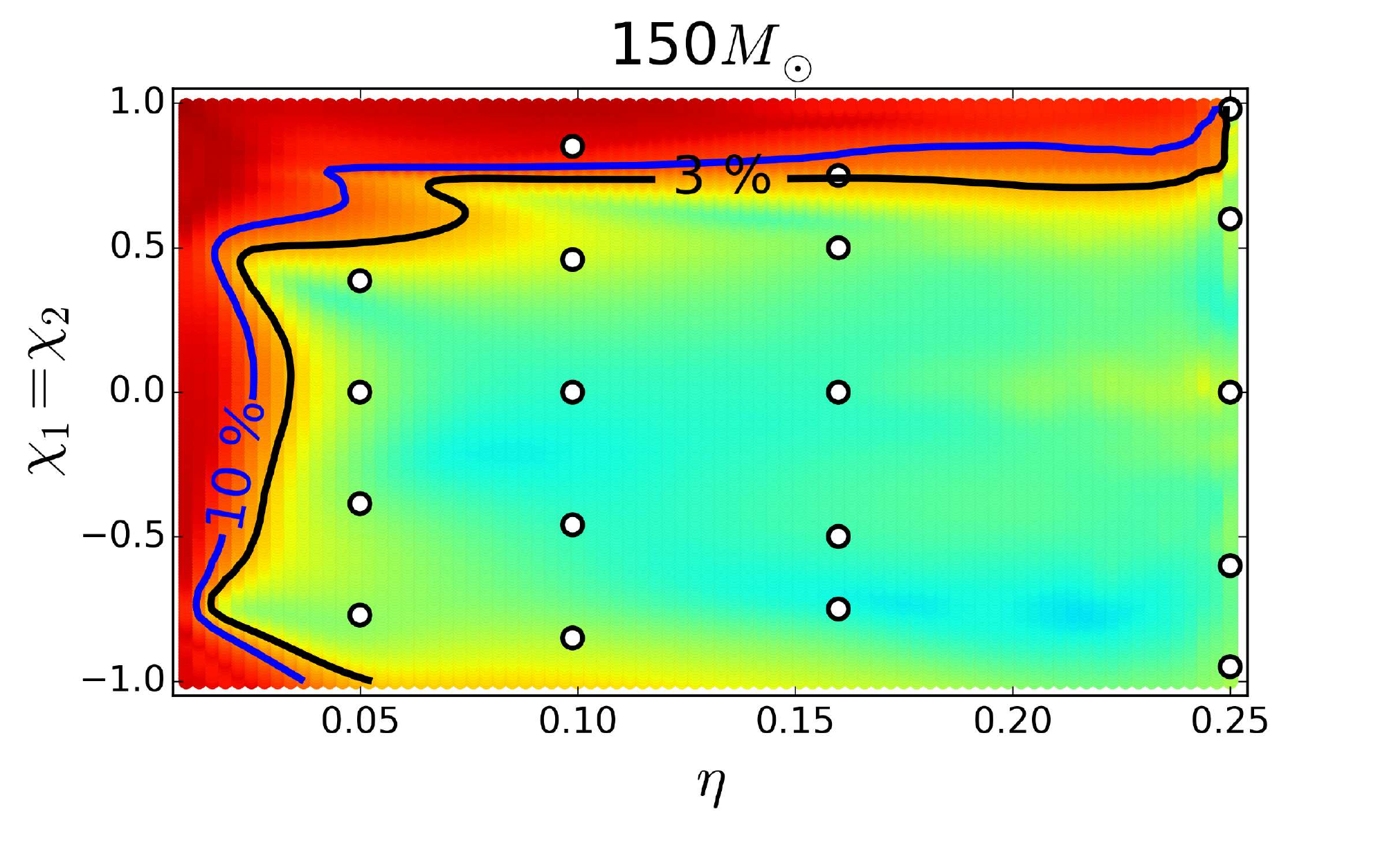}} \quad
        {\includegraphics[width=0.3\textwidth]{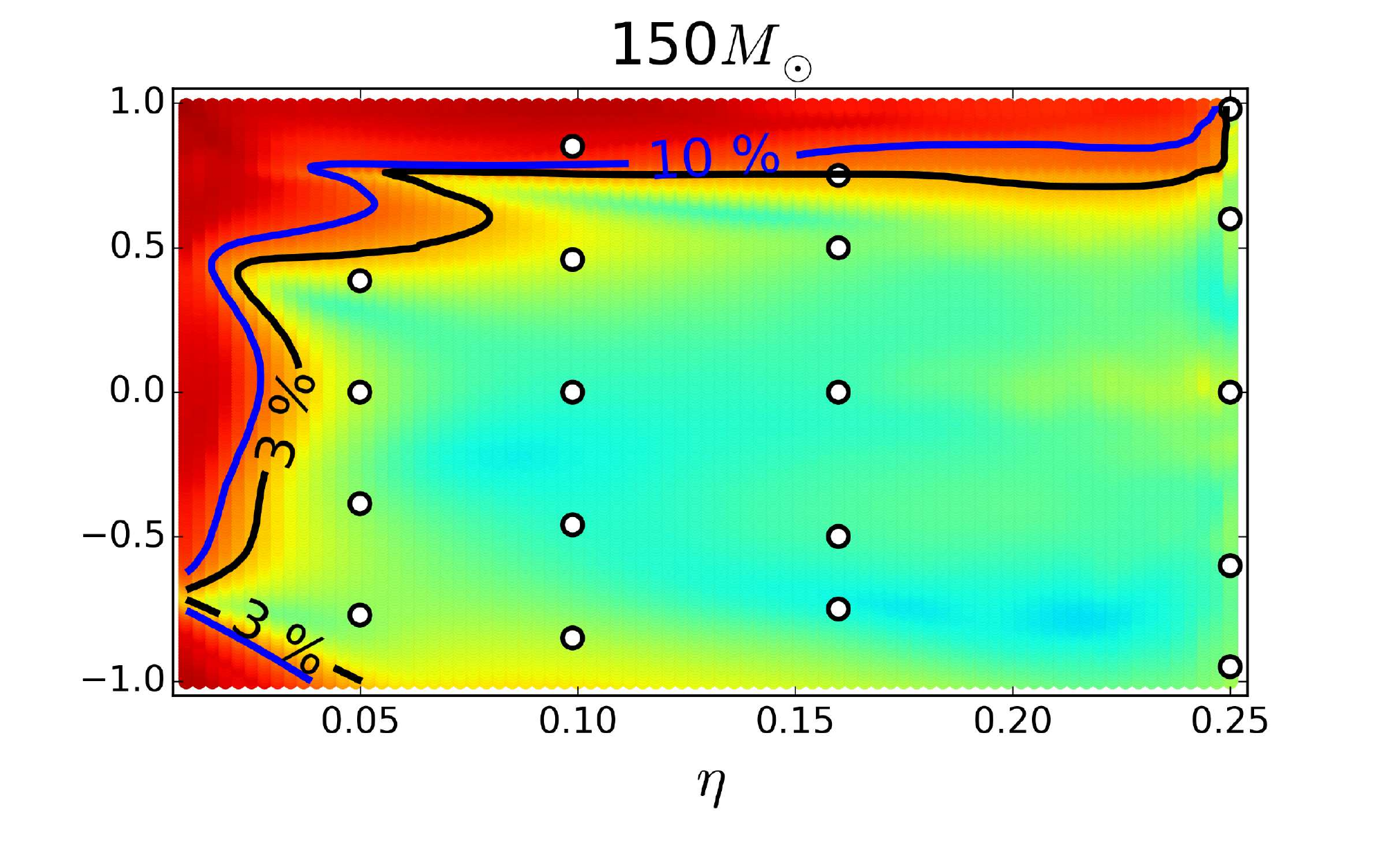}} \quad
        {\includegraphics[width=0.35\textwidth]{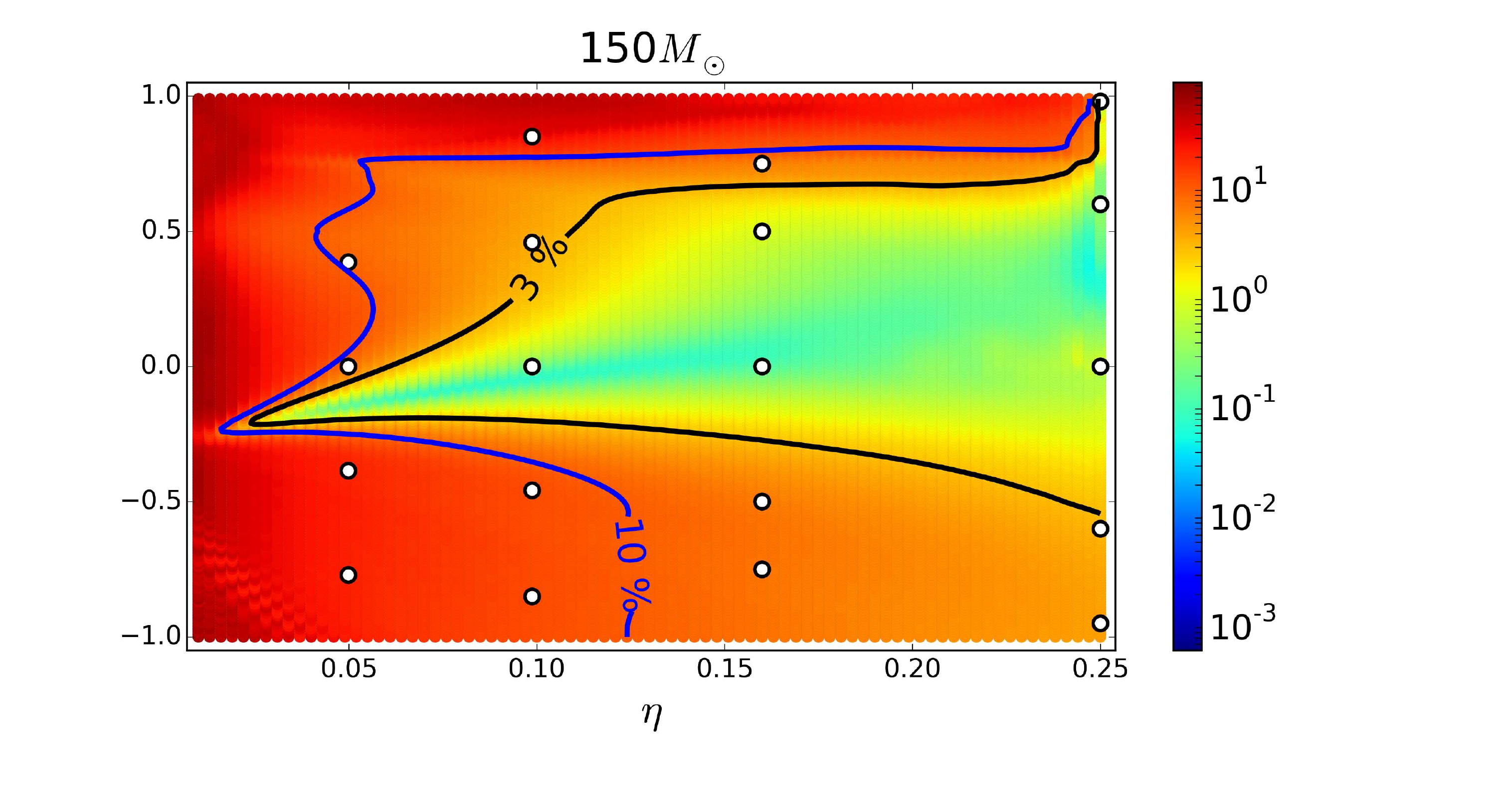}}      
    \caption{Mismatch comparisons between the SEOBNRv2\_ROM model, and three versions of PhenomD. Left: the final
    PhenomD model. Middle: SEOBNRv2\_ROM is used for the inspiral part of PhenomD, i.e., up to $Mf = 0.018$.
    Right: TaylorF2 is used for the inspiral part of PhenomD. See text for discussion.}
    \label{fig:PhenDvsv2ROM}
\end{figure*}

We noted earlier that the PhenomD model is modular, and we can use
alternative models of either the inspiral or merger-ringdown regions as
we wish. In the following comparisons we consider three versions of the model.
One is the full PhenomD model that we have presented in the previous 
sections. In comparisons with SEOBNRv2\_ROM at low masses, the 
mismatch is dominated by differences between the uncalibrated SEOBv2 model
that we used to calibrate the inspiral of PhenomD, and the calibrated SEOBNRv2
model; it is a reflection of a different choice of inspiral approximant, and not
the 
inherent accuracy of either model. For this reason we also perform a second
set of comparisons, where we use \emph{SEOBNRv2\_ROM} for the inspiral
(Region I) part of PhenomD; the merger-ringdown (Region II) remains unchanged. 
This allows us to compare PhenomD and SEOBNRv2\_ROM over only the 
merger-ringdown, and also illustrates the flexibility of the PhenomD model in
using alternative inspiral approximants. Finally, we replace the inspiral part of the 
PhenomD with TaylorF2.

The results of our comparions are shown in Fig.~\ref{fig:PhenDvsv2ROM}. 
Each panel shows the mismatch in percentage
between the PhenomD model and SEOBNRv2\_ROM (left column)
and between [SEOBNRv2\_ROM-inspiral + PhenomD-merger-ringdown] and
SEOBNRv2\_ROM (middle column), and between [TaylorF2-inspiral + PhenomD-merger-ringdown]
and SEOBNRv2\_ROM (right column). The calculations were performed 
over mass ratios $[1,100]$, spins in the range $[-1,0.99]$
and for the total masses $[12, 20, 50, 100, 150]M_\odot$.
Overlaid in white dots are the calibration points of
the PhenomD model. It is instructive when studying these plots to recall that 
the common region of parameter space calibration is up to mass-ratios 1:8 ($\eta
\sim 0.01$)
and spin $[-0.5, 0.5]$, except along the equal mass line where the
spins range from $[-0.95, 0.98]$. 

We focus first on the low-mass configurations ($M < 50\,M_\odot$). We see that
the 
agreement between PhenomD and SEOBNRv2\_ROM is in general quite poor --- some
parts of the common calibration region of both models show mismatches greater
than 3\%,
e.g, for anti-aligned spins. 
This is not necessarily due to the innaccuracy of either model. We have seen in 
Fig.~\ref{fig:PhenDvsHybIMR} that PhenomD typically has matches of better than 1\%
against 
our hybrid waveforms, which demonstrates that the model accurately reproduces the
\emph{uncalibrated} SEOBv2 model at low frequencies.
Therefore, we expect that
the poor mismatches
between PhenomD and SEOBNRv2\_ROM at low masses are due to differences between 
SEOBv2 and the calibrated SEOBNRv2 inspiral. This expectation is borne out in the middle
column,
where the SEOBv2-based PhenomD inspiral is replaced with the SEOBNRv2\_ROM 
inspiral.
Now the modified PhenomD and SEOBNRv2\_ROM models differ only in their
description of the merger-ringdown, and should agree well at very low masses,
where the 
merger-ringdown contributes little \SNR. This is what we find: at $12\,M_\odot$
the mismatches
are better than 1\% for most of the parameter space. The merger-ringdown still
has some influence,
increasing the mismatches for high-spin and high-mass-ratio systems, but in
general the agreement
is extremely good. 

Although the uncalibrated SEOBv2 and the calibrated SEOBNRv2 inspirals show poor 
matches at low masses, we note that both are still consistent with our full \NR data at
higher frequencies, and both are adequate options for an inspiral description, as we discussed
in detail in Paper 1, and also in Sec.~\ref{sec:inspiral_choice} above. The
right panel illustrates how
the model would change if we instead used TaylorF2 for the inspiral. At the matching frequency
with the merger-ringdown model ($Mf = 0.018$) the TaylorF2 phase disagree (in the sense of the 
time-shift analysis in Paper 1) at a level that makes it difficult to smoothly connect them over large
regions of the parameter space.
This, in addition to the differences
between TaylorF2 and SEOBv2(NR) at low frequencies, introduces high overlaps over all but a 
small strip of parameter space. 

As we progress down the table of plots to higher masses, the merger-ringdown
contributes more
power to the \SNR, and the results of the left and middle comparisons agree more. At
$150\,M_\odot$, where
the contribution from the inspiral (taken here as $Mf < 0.018$) is negligible,
we see that the 
two comparisons are almost identical. The poor agreement between TaylorF2 and our 
merger-ringdown model at $Mf = 0.018$ continues to lead to large mismatches. 

We now focus on the high-mass configurations ($M \geq 50\,M_\odot$), and the
left
panels that directly compare PhenomD and SEOBNRv2\_ROM.
It is evident that the region of agreement between the two models follows
closely
the region of common calibration points. Indeed, it is
very encouraging that there is a high level of agreement between
these two independent models even up to high mass-ratios of
1:18 and towards large negative spin values.

The positive spin section shows a different behaviour. At high masses (i.e.,
where the merger and ringdown are in the detector's most sensitive frequency 
range), there is a sudden drop in the agreement between the two models at
mass-ratios
larger than equal mass and spin greater than $\sim 0.75$.

PhenomD is calibrated to two high-spin unequal-mass cases, 
$(q, \chi_1, \chi_2)=\{(4,0.75,0.75),(8,0.85,0.85)\}$, 
and we have one additional case for verification, $(2, 0.75, 0.75)$.
These are the waveforms A10 and A15 from Table~\ref{tab:wftable} and
B17 from Table~\ref{tab:wftable2} respectively.
As we have already seen, PhenomD has better than 1\% mismatch
to all theses cases and therefore the poor mismatches are unlikely due to errors
in PhenomD. 
We note that these cases are well outside the calibration
region of the SEOBNRv2 model, and we therefore suspect that the accuracy of
its description of the merger-ringdown degrades significantly for high spins.

Our results also suggest that, despite the lack of calibration 
waveforms at high anti-aligned spins, the SEOBNRv2 model remains accurate in
that
region of parameter space, and the relatively good agreement between the two
models even
for nearly extreme anti-aligned spins suggests that additional calibration
waveforms, while
they would be valuable, are less crucial in those cases. 

We also observe poor mismatches for very high mass ratios. However, since this
is 
outside the calibration region of both models, we cannot conclude which (if
either) is
correct.

\begin{figure}[tb]
    \centering
    \includegraphics[width=1\linewidth]{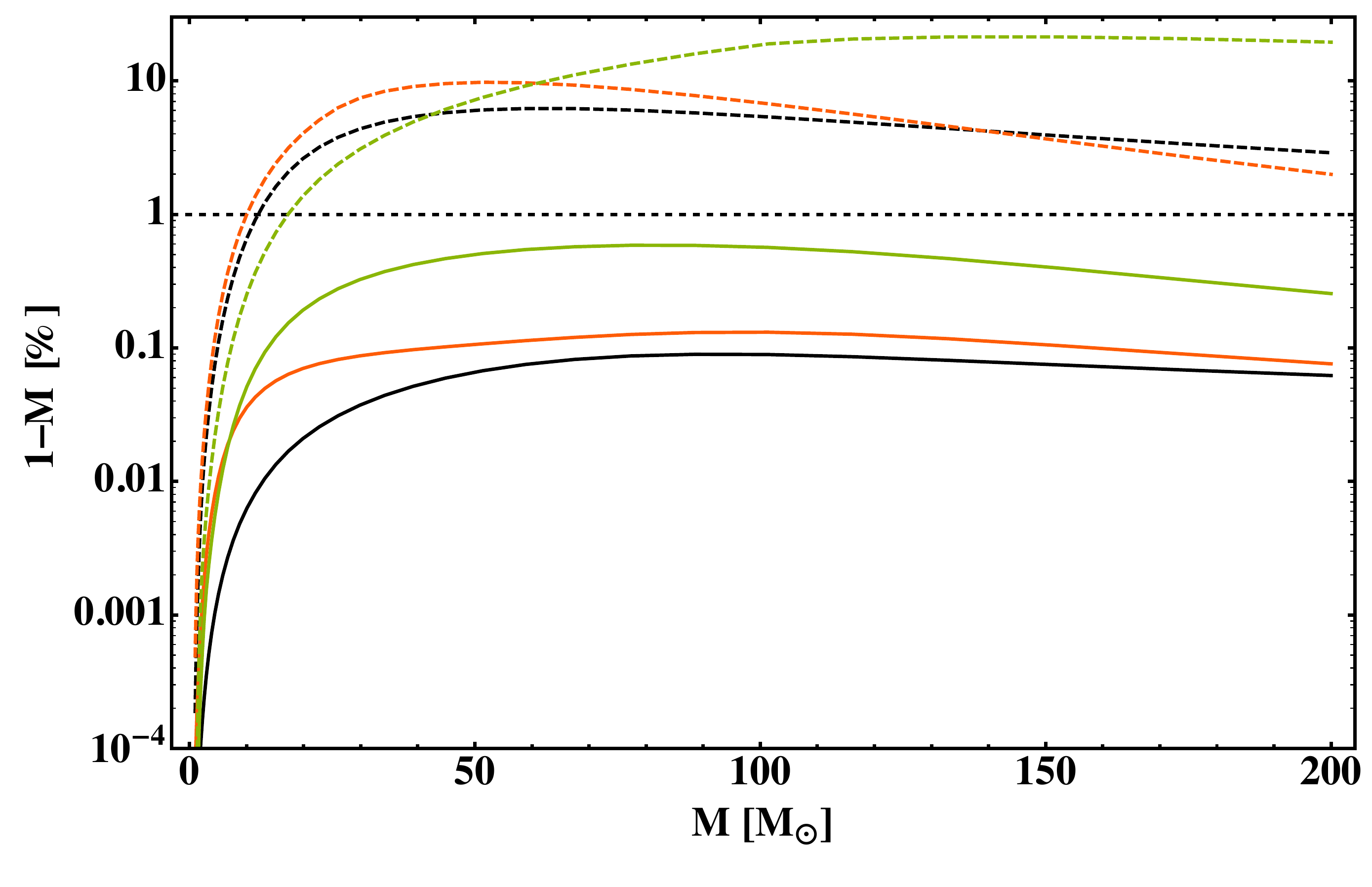}
    \caption{
    Mismatch of PhenomD (solid) or SEOBNRv2\_ROM (dashed)
    against cases A10 (orange), A15 (green) and B17 (black).
    }
\label{fig:highspin}
\end{figure}

To illustrate further the disagreement between PhenomD and SEOBNRv2 at unequal
masses and high spins, we consider in more detail the three \NR configurations
that
we have available. Fig.~\ref{fig:highspin} 
shows mismatches between pure \NR waveforms (\emph{not} the hybrids) for each of
these
cases, and against the PhenomD and SEOBNRv2 models, using the techniques discussed
in Sec.~\ref{sec:match}. The mismatch against 
SEOBNRv2 is above 1\% for all masses, and can be as high as 10\%. We have
reproduced these
plots using SEOBNRv2 waveforms generated from the LAL code, and the results are
indistinguishable;
the poor mismatches cannot be attributed to any errors in the ROM construction.

We therefore conclude that the merger and ringdown are not accurately
represented in the 
SEOBNRv2 model for high spins. This does not detract from the power of the
\EOB\NR approach,
but simply illustrates that we should not expect any merger-ringdown model to be
accurate
outside its region of \NR calibration. The same applies to our PhenomD model; we
cannot make
any statements on its accuracy for spins with $\hat\chi \gtrsim 0.85$, other
than for equal-mass
systems.

\section{Summary and discussion}

We have presented a new phenomenological model of the \GW signal from the inspiral,
merger and ringdown of aligned-spin \BH binaries, PhenomD. The new model 
is calibrated to hybrid EOB+\NR waveforms that cover the largest region of
parameter space of any 
aligned-spin model to date --- mass ratios up to 1:18 and spins up to $a/m \sim 0.85$. 
The inspiral and merger-ringdown are described by three separate models, allowing high 
accuracy over the full frequency range detectable by \aLIGO and \AdV, and also making
the model modular: the inspiral and merger-ringdown parts can easily be modified
or replaced if improved or extended models (e.g., to a yet larger region of parameter space)
become available. 

The inspiral part of our hybrids consists of \emph{uncalibrated} SEOBv2 waveforms. 
We have shown in Paper 1 that
the SEOBv2 waveforms are the most consistent with our \NR simulations over the full parameter
space that we consider, and we choose to use uncalibrated SEOBv2 to produce a model that
is fully independent of the \NR calibration done to produce SEOBNRv2. 

The merger-ringdown part of the hybrids (i.e., the \NR waveforms) have a common lowest
frequency of $Mf \sim 0.018$, and so this is the frequency at which we switch from the
inspiral to the merger-ringdown model. 

The final model has mismatches against both the 19 calibration hybrids and an additional 
29 verification hybrids, of typically better than 1\% for all masses. The mismatches are shown in 
Fig.~\ref{fig:PhenDvsHybIMR}, and demonstrate that we have faithfully modelled this
region of the aligned-spin parameter space. 

The model is parameterized by the binary's symmetric mass ratio, $\eta$, and a 
normalized reduced effective spin parameter, $\hat\chi$, defined in Eq.~(\ref{eqn:chihat}). 
A parameterization in terms of a weighted sum of the two \BH spins has been used in 
previous Phenom models~\cite{Ajith2011,Santamaria2010}, and is motivated by the leading-order 
spin effect on the
inspiral phasing~\cite{Cutler:1994ys,Poisson:1995ef,Ajith:2011ec}, and demonstrations of its efficacy for 
merger-ringdown~\cite{Puerrer2013}. In
this paper we show that the reduced-spin approximation becomes inaccurate only for 
high-spin unequal-mass systems, but in these configurations the parameter errors due to 
our approximation appear to be smaller than statistical errors in the spin and mass-ratio 
measurements with \aLIGO and \AdV. This implies that it will be difficult to measure both
\BH spins in \GW measurements; this will be considered in more detail in a forthcoming 
paper~\cite{Puerrer2015}. 

We have compared the new PhenomD model with the state-of-the-art SEOBNRv2 model,
and found that the two models agree well over their common region of calibration, which is
mass ratios up to 1:8, and spins up to $a/m \sim 0.5$ (and near-extremal spins for equal-mass
systems). At low masses the agreement is not good, but we show that this is due to 
differences between the calibrated and uncalibrated SEOBv2 inspiral descriptions. 

Outside the common calibration region, the two models show significant disagreement, in
terms of their mismatch. This is particularly true for high aligned spins. Given that PhenomD
was calibrated to several high-spin unequal-mass simulations (spins of 0.75 or 0.85), while
SEOBNRv2 was calibrated to spins of no higher than 0.5 for unequal-mass configurations,
we conclude that SEOBNRv2 does not accurately capture the merger and ringdown for these
systems. We expect, however, that its performance will become comparable to PhenomD when 
calibrated to additional \NR waveforms.

The broader conclusion we draw from these results is that high-aligned-spin systems deserve 
greater attention in future modelling efforts. The PhenomD model was calibrated to only two
high-aligned-spin binaries, but it is clear that a larger number of \NR simulations in this region
of parameter space will benefit \GW astronomy. 

The PhenomD model involves 17 coefficients that are mapped across the parameter 
space with polynomials up to second order in $\eta$ and up to third order in $\hat\chi$. 
Although the total number of coefficients is similar to the previous PhenomC model, 
the development of a refined ansatz for each frequency region allows us to more accurately
model a wider range of features of the waveforms. This is described in more detail in Paper 1.
We have also carefully tuned each ansatz, and our parameter-space fits, to ensure that the
model produces physically reasonable results outside the calibration region, and that the 
waveforms show no pathological features when converted to the time-domain 
(Appendix~\ref{app:td}). These 
modifications significantly improve the model beyond previous Phemom models, in addition
to increasing the range of calibration and lowering the mismatch error. 

In previous work we have shown that models for generic (precessing) binaries can be produced
by ``twisting up'' an aligned-spin model. The PhenomP model exploits that idea, but to date
has been based on the PhenomC model, which limits its applicability to mass ratios 
$q \lesssim 4$. With the advent of PhenomD, we will be able to make PhenomP valid to much 
higher mass ratios and higher values of the parallel component of the spin. This simple
replacement of PhenomC with PhenomD in the LIGO-Virgo LAL code has already been 
tested, and will be made available in the near future.

\section*{Acknowledgements}

We thank P. Ajith for discussions.
SH, XJ, and in part AB were supported the Spanish Ministry of Economy and Competitiveness grants
FPA2010-16495, CSD2009-00064, FPA2013-41042-P, European Union FEDER funds, Conselleria d'Economia i Competitivitat del Govern de les Illes Balears and Fons Social Europeu.
MH was supported by Science and Technology Facilities Council grants ST/H008438/1
and ST/I001085/1, and FO and MP by ST/I001085/1.
SK was supported by Science and Technology Facilities Council.
{\tt BAM} simulations were carried out at Advanced Research Computing 
(ARCCA) at Cardiff, as part of the European PRACE petascale computing
initiative on the clusters Hermit, Curie and SuperMUC, and on the UK DiRAC Datacentric cluster.

\appendix 

\section{Time-domain conversion}
\label{app:td}

Our PhenomD model is formulated entirely in the frequency domain, which is a
great advantage for performing fast \GW searches and parameter estimation
studies. However, our construction process started with data in the time
domain, and physical signals are smooth functions in \emph{both} the
frequency and time domain. Therefore, it is desirable to check how our model
transforms from the frequency domain back into the time domain via a
straightforward inverse Fourier transformation. 

This serves also as an
independent, powerful sanity check. The previous PhenomC model
\cite{Santamaria2010}, for instance, quickly develops a pathological behavior
in the time domain once the parameters leave the calibration region, which is a
result of steep transitions caused by extrapolating fitting
coefficients. We do not find these features for our new PhenomD model.

Before applying the inverse Fourier transformation, we multiply our model with
a variant of the Planck taper function \cite{McKechan2010},
\begin{equation*}
 \mathcal T(f) = \begin{cases}
  0, & f \leq f_1 \\
                            \left[\exp\left(\frac{f_2 - f_1}{f-f_1} + \frac{f_2
- f_1}{f - f_2}\right) + 1 \right]^{-1}, & f_1 < f < f_2\\
 1, & f > f_2
                           \end{cases},
\end{equation*}
where $f_2$ is the smallest frequency that we want to represent in the
time-domain data (which become infinitively long for $f_2 \to 0$). In order to
avoid a sharp transition, which would introduce unphysical oscillations,
$\mathcal T$ uses an extra cushion, $f \in (f_1, f_2)$, in which the
frequency domain amplitude smoothly increases from zero to their correct value.
We typically set $f_1 = 0.8f_2$.

We perform the Fourier transformation numerically, which requires us to define a
suitable sampling rate in the time and frequency domain. From our model, we find
that the amplitude has dropped several orders of magnitude for
frequencies $M f > 0.25$, so we can choose any sampling with $\Delta t / M < 2$
which in turn is solely determined by the largest frequency we include in our
frequency-domain data.

The frequency-domain sampling, on the other hand, is determined by the total
length of the signal in the time domain, which is information we do not have
a priori access to. However, in the spirit of the stationary-phase
approximation that typically relates the time-domain phase derivative to the
frequency ($d \phi(t) / dt \approx 2 \pi f$), we approximate
\begin{align}
 \frac{d \phi(f)}{df} &= \phi'(f) \approx 2\pi t, \\
\Rightarrow ~~ \Delta f &< \frac{1}{t_{\rm max} - t_1} \approx \frac{\pi}{\vert 
\phi'(f_{\rm max}) - \phi'(f_1)\vert} . \label{eq:iFFT_deltaf}
\end{align}
In \eqref{eq:iFFT_deltaf}, we have introduced an extra factor of $1/2$ to
account for the negative-frequency content of real-valued signals (just like in
the usual sampling theorem), and when choosing $\Delta f$ we usually
apply another factor of $1/2$ as safety margin.

\begin{figure*}[tb]
 \begin{overpic}[height=3.5cm]{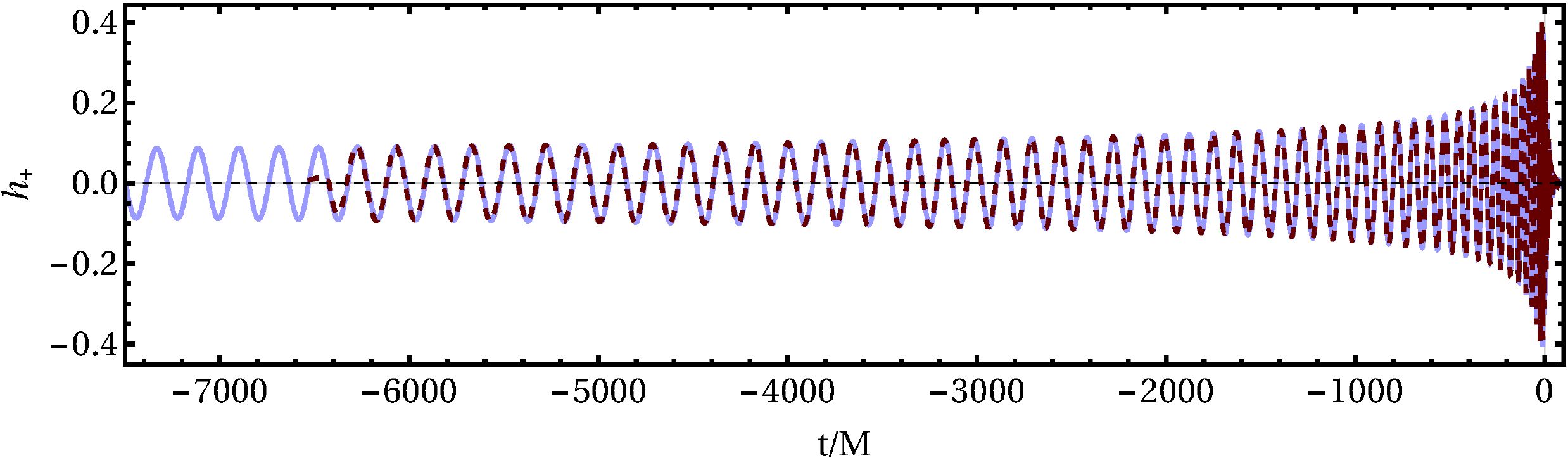}
  \put(11,24){$q=1,~\chi_1=\chi_2=0.98$}
 \end{overpic}%
 \includegraphics[height=3.5cm]{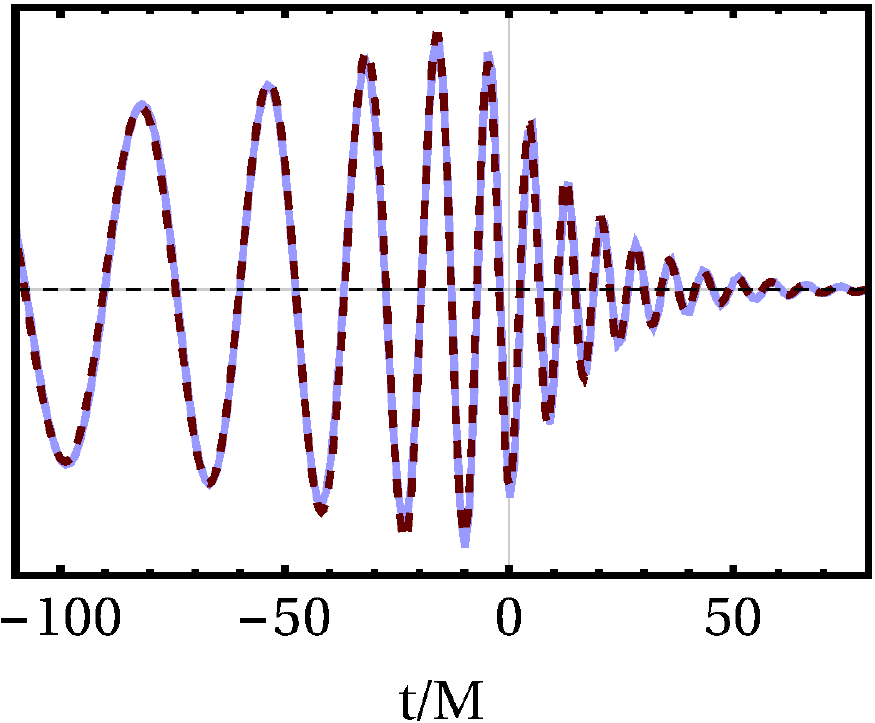} \\[8pt]
  \begin{overpic}[height=3.5cm]{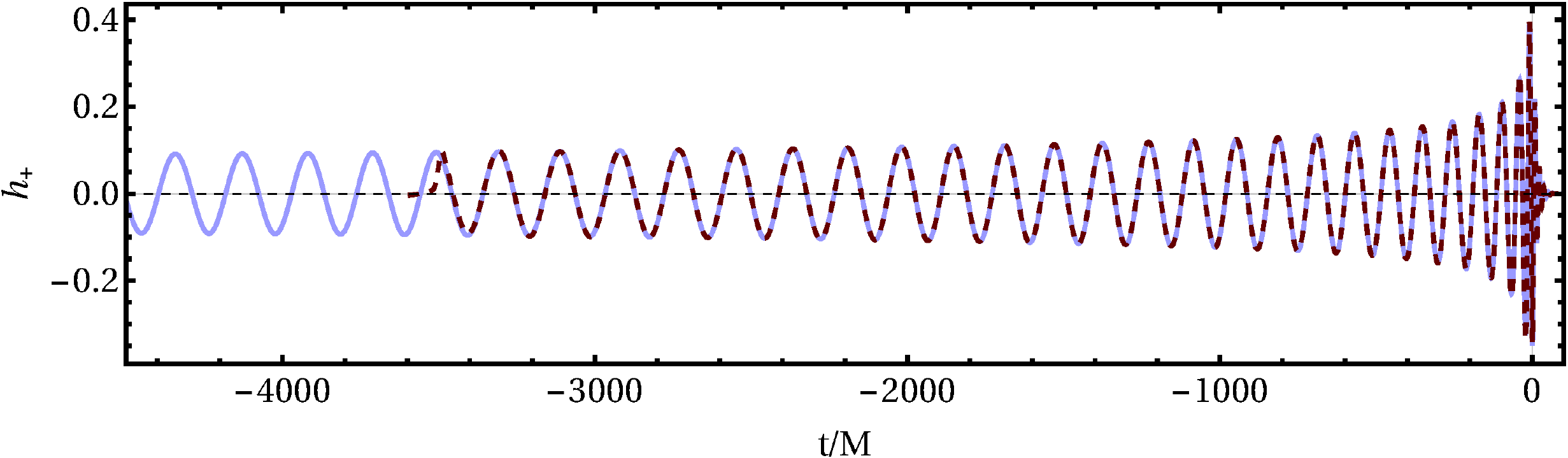}
   \put(11,24){$q=1,~\chi_1=\chi_2=-0.95$}
  \end{overpic}%
 \includegraphics[height=3.5cm]{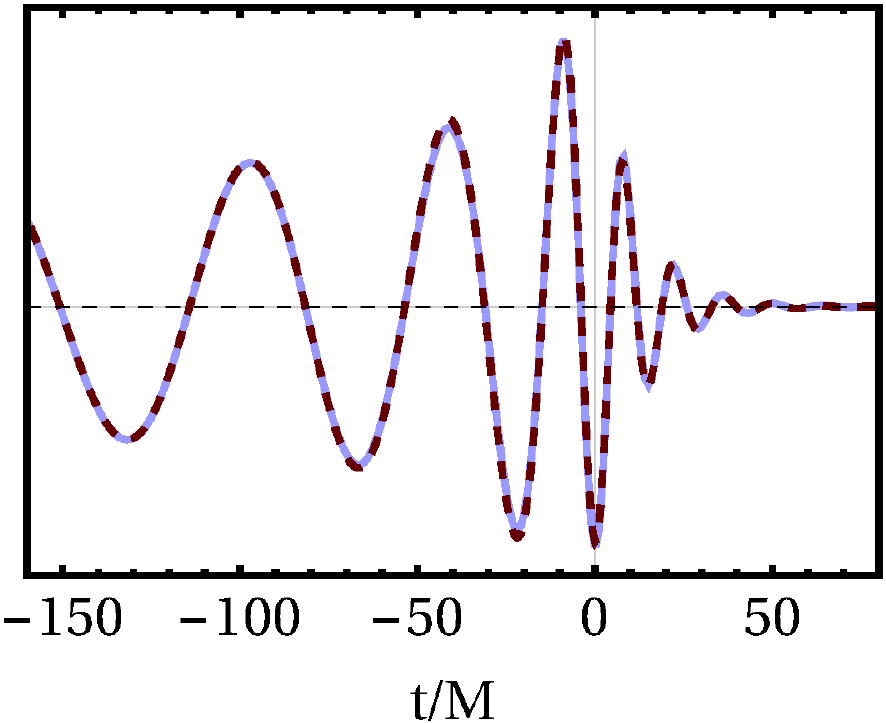} \\[8pt]
 \begin{overpic}[height=3.5cm]{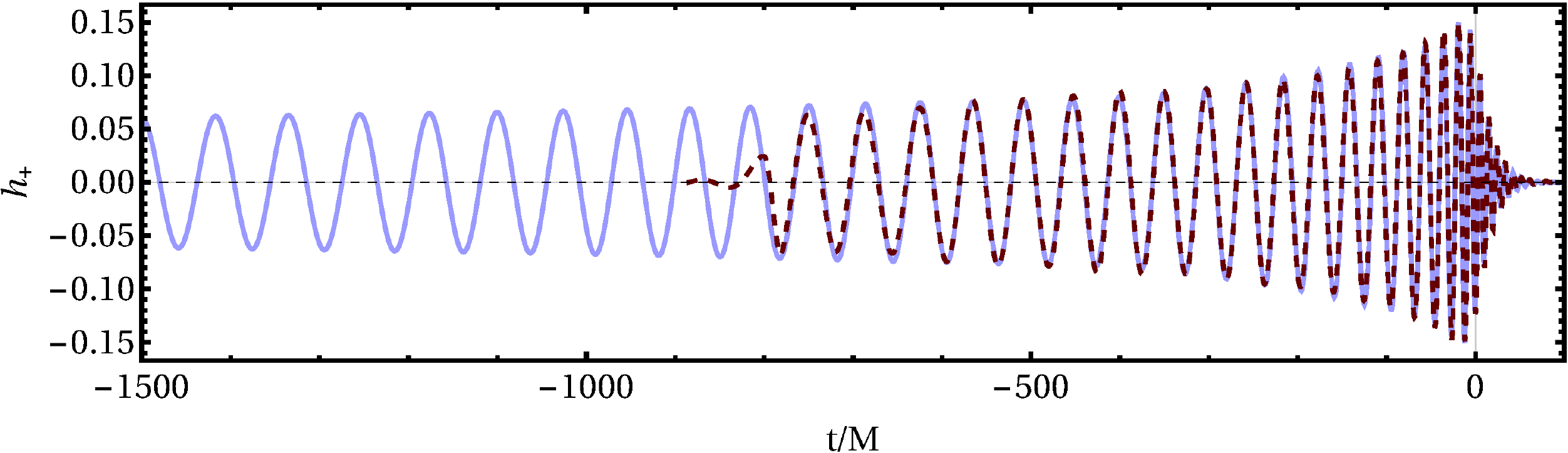}
  \put(11,24){$q=8,~\chi_1=\chi_2=0.85$}
 \end{overpic}%
 \includegraphics[height=3.5cm]{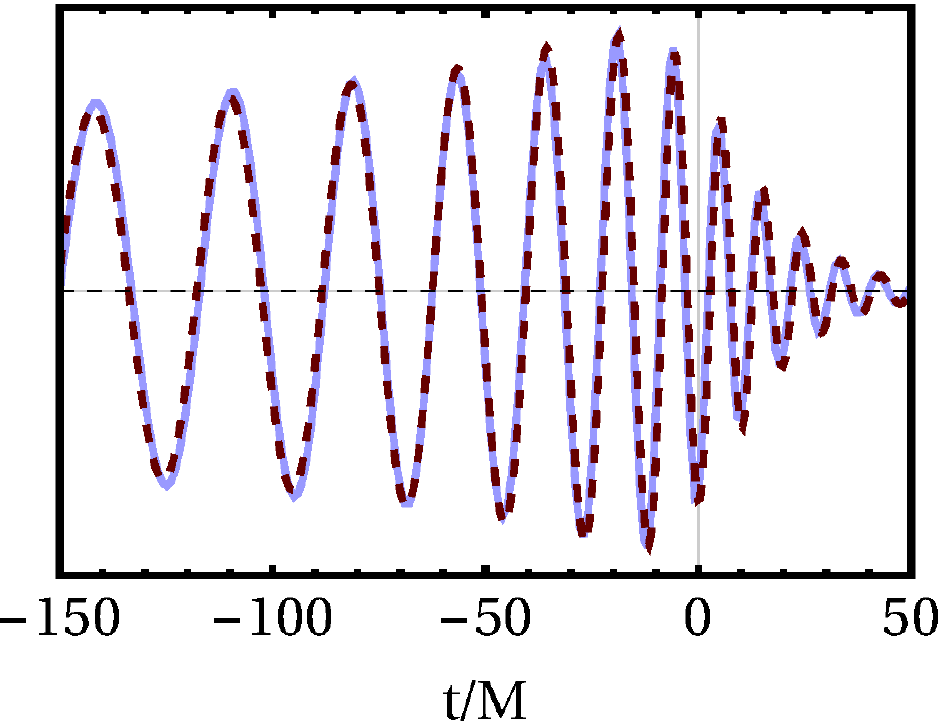}  \\[8pt]
  \begin{overpic}[height=3.5cm]{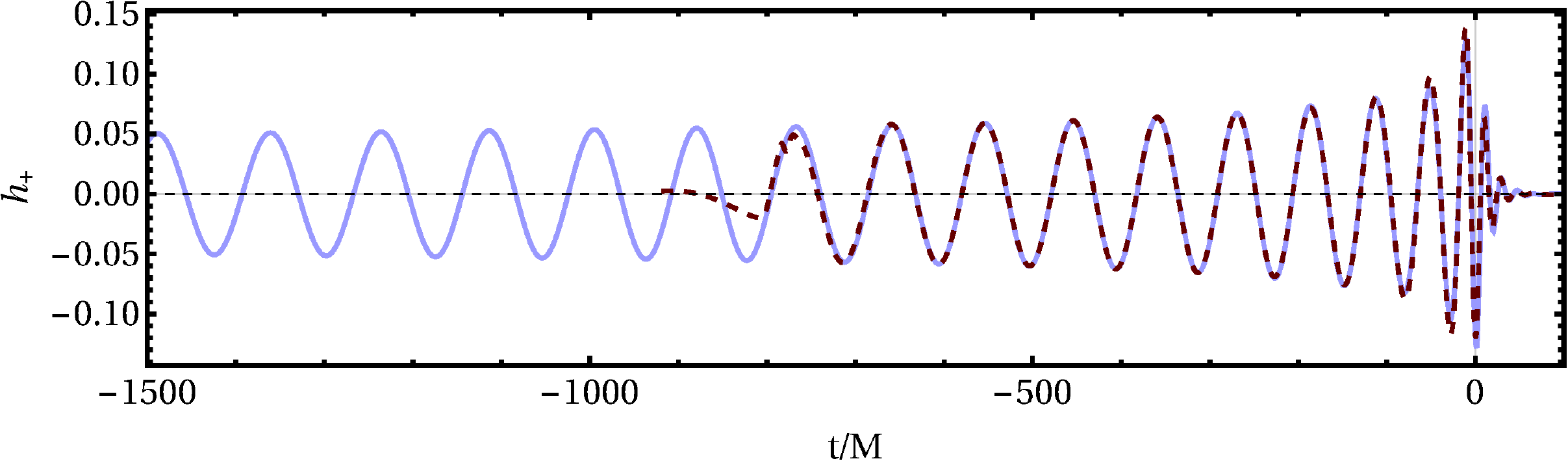}
   \put(11,24){$q=8,~\chi_1=\chi_2=-0.85$}
  \end{overpic}%
 \includegraphics[height=3.5cm]{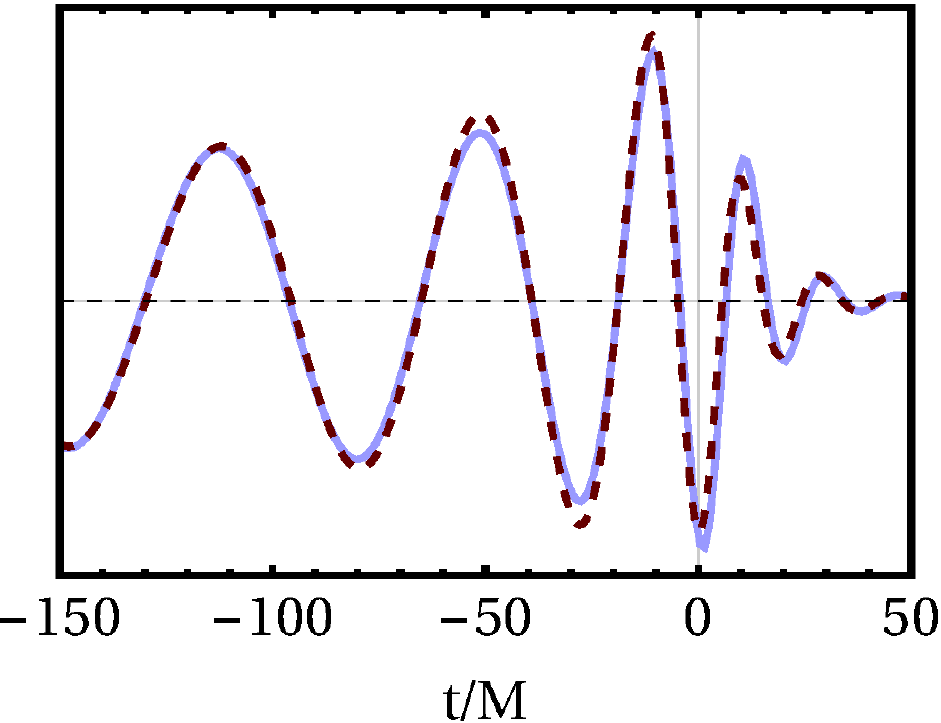} \\[8pt]
  \begin{overpic}[height=3.5cm]{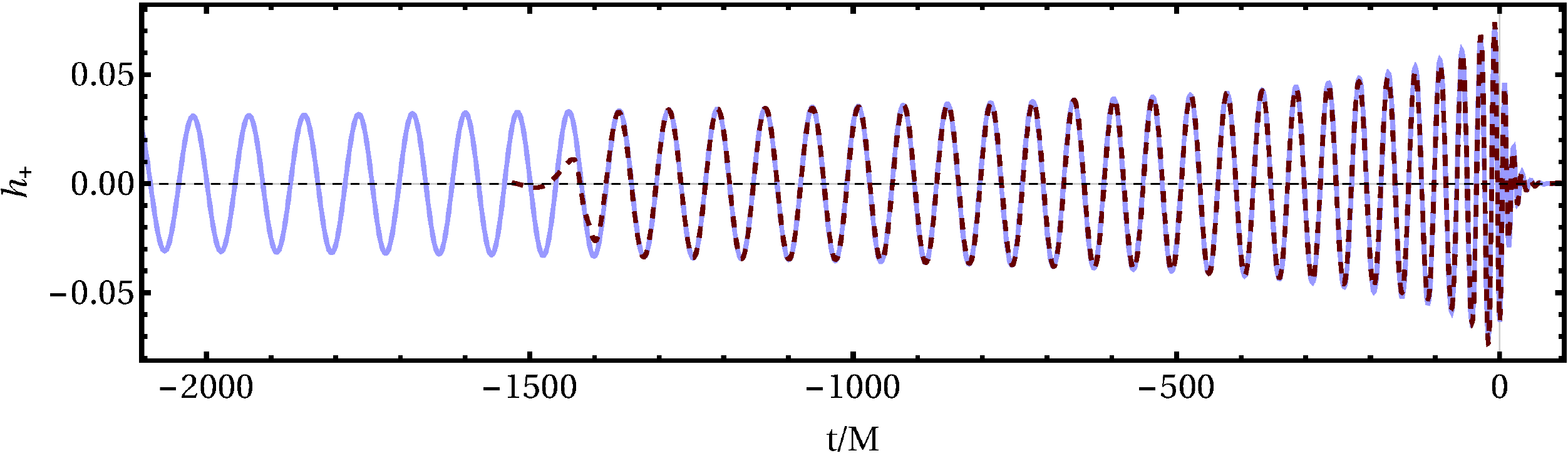}
   \put(11,24){$q=18,~\chi_1=0.4,~\chi_2=0.$}
  \end{overpic}%
 \includegraphics[height=3.5cm]{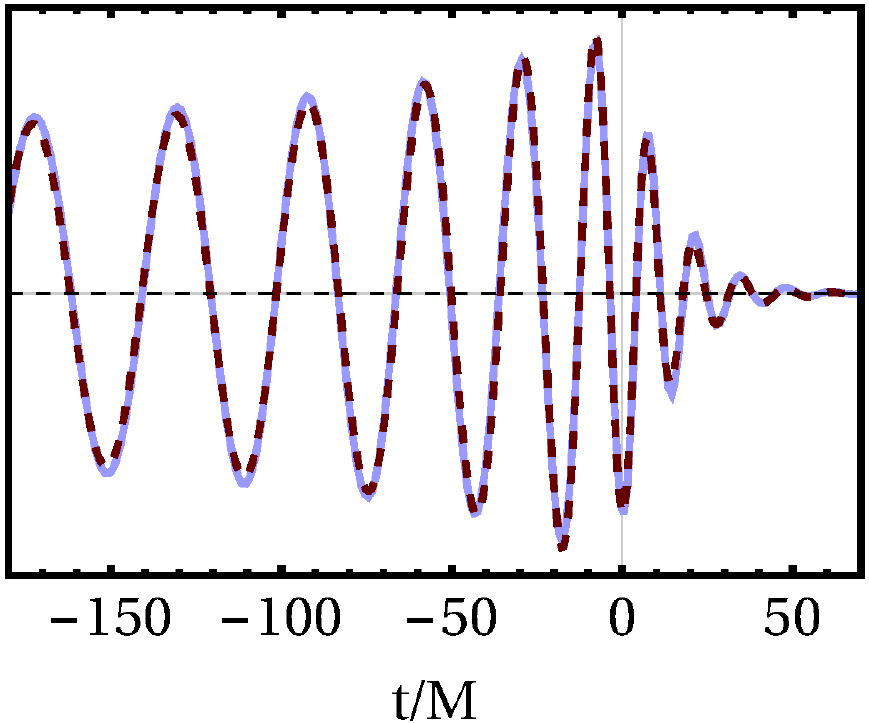} \\[8pt]
  \begin{overpic}[height=3.5cm]{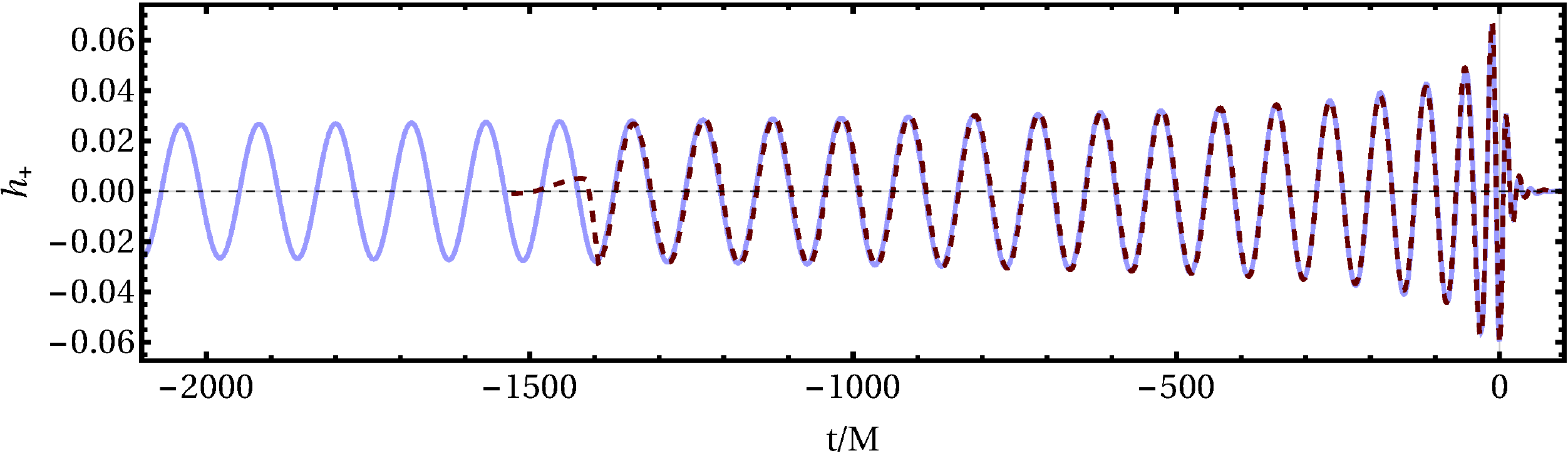}
   \put(11,24){$q=18,~\chi_1=-0.8,~\chi_2=0.$}
  \end{overpic}%
 \includegraphics[height=3.5cm]{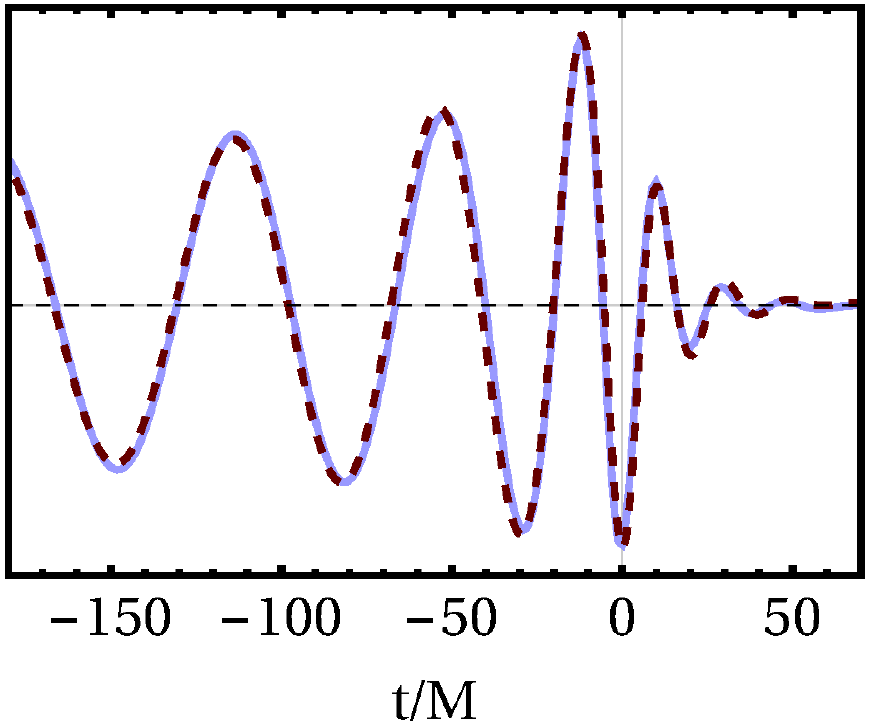}  
 \caption{Time-domain PhenomD waveforms (solid, light blue online) and \NR
waveforms (dashed, red online) for corners of the parameter space used for
calibration. We plot the plus polarization $h_+$ normalized by the extraction
radius, and the binary's parameters are indicated by the mass ratio $q =
m_1/m_2$ and the two spin parameters $\chi_1, \chi_2$.}
 \label{fig:TD_PhenD+NR}
\end{figure*}

The time-domain waveforms we obtain this way can be compared to the original
\NR data, and for corners of the parameter space used for calibration we show
the results in Fig.~\ref{fig:TD_PhenD+NR}. Note that a small overall time and
phase shift was applied to the model, as these parameters are not meant to
faithfully capture the arbitrary choices made in the original \NR simulations.
No other optimization has been applied. The agreement visible in
Fig.~\ref{fig:TD_PhenD+NR} throughout the late inspiral, merger and ringdown is
remarkable and a strong indication (in additional to the matches presented in
Sec.~\ref{sec:mismatches}) that our hybridization, fitting and interpolation
procedures accurately represent the original data.

\begin{figure*}[tb]
 \centering
 \includegraphics[width=0.9\textwidth]{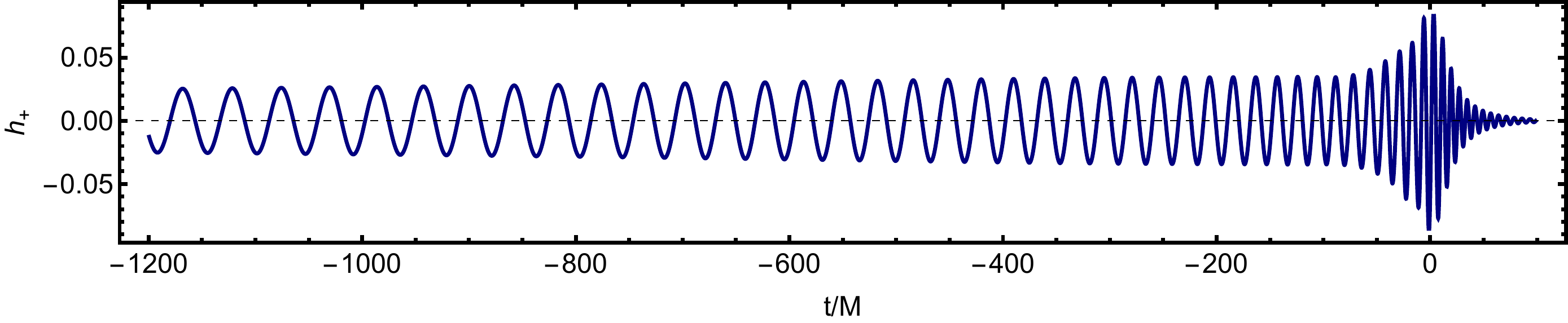}
 \caption{Time-domain representation of the PhenomD model outside its
calibration region, here for mass ratio 50 and spin parameters of $\chi_1 =
\chi_2 = 0.99$.}
\label{fig:TD_PhenDq50}
\end{figure*}

In addition to complementing the model validation, we may also use the
time-domain representations as a visual sanity check, even outside the model's
calibration region. As mentioned above, this proved to be a powerful test of
the previous PhenomC model that failed to produce reasonable time-domain
waveforms in many parts of the parameter space outside its calibration range.
PhenomD, however, does not show any pathological behavior outside its
calibration region, neither in the time nor frequency domain. We illustrate this
fact in Fig.~\ref{fig:TD_PhenDq50} by showing a case where the model parameters
have been extrapolated to mass ratio 50 and near-extremal spins $\chi_1 =
\chi_2 = 0.99$. While such a plot is by no means a guarantee that the waveforms
are accurate in this regions of the parameter space, it is reassuring that our
new model is much more robust in its extrapolation, which will allow \GW search
algorithms to use our model slightly outside its calibration region, even if we
cannot vouch for the level of accuracy there.

\section{PN coefficients} \label{sec:app_pncoeffs}

For the convenience of the reader, we list below the \PN coefficients implemented in
our model. We incorporated
spin-independent corrections up to 3.5PN order ($i=7$)
\cite{Buonanno2009,Blanchet2014}, linear spin-orbit corrections up to 3.5PN
order \cite{Bohe:2013cla} and quadratic spin corrections up to 2PN order
\cite{Poisson:1997ha,Arun:2008kb,Mikoczi:2005dn}. Our re-expansion strategy
follows the choices made in the current state of the
LIGO software library~\cite{lalsuite} as discussed in
Sec.~\ref{sec:inspiral_phase}.

\allowdisplaybreaks
Following (\ref{equ:phiTF2}), we express the frequency-domain phase as
\begin{equation*}
\begin{split}
\phi_{\text{TF2}} &= \, 2 \pi f t_c - \varphi_c - \pi/4\\
                        & + \frac{3}{128 \, \eta}(\pi f M)^{-5/3} \sum_{i=0}^{7}
\varphi_i(\Xi) (\pi f M)^{i/3}.
\end{split}
\end{equation*}
The individual masses and spin parameters, $m_i$ and $\chi_i$ ($i=1,2$), are
encoded in the following parameter combinations,
\begin{align}
  M &= m_1 + m_2, \\
  \eta &= m_1\, m_2 /M^2, \\
  \delta &= (m_1 - m_2) /M, \\
  \chi_s &= (\chi_1 + \chi_2)/2, \\
  \chi_a &= (\chi_1 - \chi_2)/2.
\end{align}
The expansion coefficients are then given by
\begin{align}
  \varphi_0 &= 1, \\
  \varphi_1 &= 0, \\
  \varphi_2 &= \frac{3715}{756}+\frac{55 \eta }{9}, \\
  \varphi_3 &= -16 \pi +\frac{113 \delta  \chi
_a}{3}+\left(\frac{113}{3}-\frac{76 \eta }{3}\right) \chi _s,
\end{align}
\begin{widetext}
 \begin{align}
   \varphi_4 &= \frac{15293365}{508032}+\frac{27145 \eta }{504}+\frac{3085 \eta
^2}{72}+\left(-\frac{405}{8}+200 \eta \right) \chi _a^2-\frac{405}{4} \delta 
\chi _a \chi _s+\left(-\frac{405}{8}+\frac{5 \eta }{2}\right) \chi _s^2, \\
\varphi_5 &= \left[ 1 + \log \left( \pi M f \right) \right] \left[ \frac{38645
\pi }{756}-\frac{65 \pi  \eta }{9}+\delta  \left(-\frac{732985}{2268}-\frac{140
\eta }{9}\right) \chi _a +\left(-\frac{732985}{2268}+\frac{24260  \eta
}{81}+\frac{340 \eta ^2}{9}\right) \chi _s \right], \\
\begin{split}
\varphi_6 & = \frac{11583231236531}{4694215680}-\frac{6848\gamma_E}
{21}-\frac{640 \pi ^2}{3}+\left(-\frac{15737765635}{3048192}+\frac{2255 \pi
^2}{12}\right) \eta +\frac{76055 \eta ^2}{1728}-\frac{127825 \eta
^3}{1296} \\ & \phantom{=} -\frac{6848}{63} \log (64 \pi M f)+\frac{2270}{3}
\pi \delta \chi _a+\left(\frac{2270 \pi }{3}-520 \pi \eta \right) \chi _s,
\end{split}
\\
\begin{split}
\varphi_7 &= \frac{77096675 \pi }{254016}+\frac{378515 \pi  \eta
}{1512}-\frac{74045 \pi  \eta ^2}{756}+\delta 
\left(-\frac{25150083775}{3048192}+\frac{26804935 \eta }{6048}-\frac{1985 \eta
^2}{48}\right) \chi _a \\
& \phantom{=} +\left(-\frac{25150083775}{3048192}+\frac{10566655595 \eta
}{762048}-\frac{1042165 \eta ^2}{3024}+\frac{5345 \eta ^3}{36}\right) \chi _s.
\end{split}
 \end{align}
\end{widetext}

As discussed in Sec.~\ref{sec:InsAmp} and Sec.~IV in Paper 1,
our inspiral amplitude model is based
on a re-expanded PN amplitude.
The expansion coefficients of Eq.~(\ref{equ:AmpPN}) are given by

\begin{align}
  \mathcal{A}_0 &= 1, \\
  \mathcal{A}_1 &= 0, \\
  \mathcal{A}_2 &= -\frac{323}{224}+\frac{451 \eta }{168}, \\
  \nonumber \\
  \mathcal{A}_3 &= \frac{27 \delta  \chi _a}{8}+
  \left(\frac{27}{8}-\frac{11 \eta }{6}\right) \chi _s,
\end{align}
\begin{widetext}
 \begin{align}
  \mathcal{A}_4 &= -\frac{27312085}{8128512}-\frac{1975055 \eta }{338688}
 +\frac{105271 \eta ^2}{24192} +\left(-\frac{81}{32}+8 \eta \right) \chi _a^2
 -\frac{81}{16} \delta  \chi _a \chi _s
 +\left(-\frac{81}{32}+\frac{17 \eta }{8}\right) \chi _s^2, \\
 \begin{split}
 \mathcal{A}_5 &=-\frac{85 \pi }{64}+\frac{85 \pi  \eta }{16}
 +\delta  \left(\frac{285197}{16128}
 -\frac{1579 \eta }{4032}\right) \chi _a
 +\left(\frac{285197}{16128}-\frac{15317 \eta }{672}
 -\frac{2227 \eta ^2}{1008}\right) \chi _s ,
 \end{split}
  \\
  \begin{split}
  \mathcal{A}_6 &= -\frac{177520268561}{8583708672}+\left(\frac{545384828789}{5007163392}
  -\frac{205 \pi ^2}{48} \right) \eta -\frac{3248849057 \eta ^2}{178827264}
  +\frac{34473079 \eta ^3}{6386688} \\
  & +\left(\frac{1614569}{64512}-\frac{1873643 \eta }{16128} +\frac{2167 \eta ^2}{42} \right)
  \chi _a^2 
  +\left(\frac{31 \pi }{12}-\frac{7 \pi  \eta }{3} \right) \chi _s
  +\left(\frac{1614569}{64512}-\frac{61391 \eta }{1344}
  +\frac{57451 \eta ^2}{4032}\right) \chi _s^2 \\
  & +\delta  \chi _a \left(\frac{31 \pi }{12}
  +\left(\frac{1614569}{32256}-\frac{165961 \eta }{2688}\right) \chi _s\right)
 \end{split}
 \end{align}
\end{widetext}

\section{Phenomenological Coefficients}
\label{app:phencoeff}

The values of the coefficients for the mapping functions given 
in Eq.~(\ref{eqn:mapping}) are shown in Tab.~\ref{tab:coefftable}.
These values are calculated under the parametrization
($\eta$, $\chi_{\rm{PN}}$).

\newpage

\begin{sidewaystable}[tb]
\begin{tabular}{llllllllllll}
\hline
\hline
\multicolumn{1}{c}{$\Lambda^i$} & \multicolumn{1}{c}{$\lambda_{00}$} & \multicolumn{1}{c}{$\lambda_{10}$} & \multicolumn{1}{c}{$\lambda_{01}$} & \multicolumn{1}{c}{$\lambda_{11}$} & \multicolumn{1}{c}{$\lambda_{21}$} & \multicolumn{1}{c}{$\lambda_{02}$} & \multicolumn{1}{c}{$\lambda_{12}$} & \multicolumn{1}{c}{$\lambda_{22}$} & \multicolumn{1}{c}{$\lambda_{03}$} & \multicolumn{1}{c}{$\lambda_{13}$} & \multicolumn{1}{c}{$\lambda_{23}$} \\ \hline
 $\rho_1$ & 3931.9 & -17395.8 & 3132.38 & 343966. & -1.21626$\times 10^6$ & -70698. & 1.38391$\times 10^6$ & -3.96628$\times 10^6$ & -60017.5 & 803515. & -2.09171$\times 10^6$ \\
 $\rho_2$ & -40105.5 & 112253. & 23561.7 & -3.47618$\times 10^6$ & 1.13759$\times 10^7$ & 754313. & -1.30848$\times 10^7$ & 3.64446$\times 10^7$ & 596227. & -7.42779$\times 10^6$ & 1.8929$\times 10^7$ \\
 $\rho_3$ & 83208.4 & -191238. & -210916. & 8.71798$\times 10^6$ & -2.69149$\times 10^7$ & -1.98898$\times 10^6$ & 3.0888$\times 10^7$ & -8.39087$\times 10^7$ & -1.4535$\times 10^6$ & 1.70635$\times 10^7$ & -4.27487$\times 10^7$ \\
 $v_2$ & 0.814984 & 2.57476 & 1.16102 & -2.36278 & 6.77104 & 0.757078 & -2.72569 & 7.11404 & 0.176693 & -0.797869 & 2.11624 \\
 $\gamma_1$ & 0.0069274 & 0.0302047 & 0.00630802 & -0.120741 & 0.262716 & 0.00341518 & -0.107793 & 0.27099 & 0.000737419 & -0.0274962 & 0.0733151 \\
 $\gamma_2$ & 1.01034 & 0.000899312 & 0.283949 & -4.04975 & 13.2078 & 0.103963 & -7.02506 & 24.7849 & 0.030932 & -2.6924 & 9.60937 \\
 $\gamma_3$ & 1.30816 & -0.00553773 & -0.0678292 & -0.668983 & 3.40315 & -0.0529658 & -0.992379 & 4.82068 & -0.00613414 & -0.384293 & 1.75618 \\
 $\sigma_1$ & 2096.55 & 1463.75 & 1312.55 & 18307.3 & -43534.1 & -833.289 & 32047.3 & -108609. & 452.251 & 8353.44 & -44531.3 \\
 $\sigma_2$ & -10114.1 & -44631. & -6541.31 & -266959. & 686328. & 3405.64 & -437508. & 1.63182$\times 10^6$ & -7462.65 & -114585. & 674402. \\
 $\sigma_3$ & 22933.7 & 230960. & 14961.1 & 1.19402$\times 10^6$ & -3.10422$\times 10^6$ & -3038.17 & 1.87203$\times 10^6$ & -7.30915$\times 10^6$ & 42738.2 & 467502. & -3.06485$\times 10^6$ \\
 $\sigma_4$ & -14621.7 & -377813. & -9608.68 & -1.71089$\times 10^6$ & 4.33292$\times 10^6$ & -22366.7 & -2.50197$\times 10^6$ & 1.02745$\times 10^7$ & -85360.3 & -570025. & 4.39684$\times 10^6$ \\
 $\beta_1 $& 97.8975 & -42.6597 & 153.484 & -1417.06 & 2752.86 & 138.741 & -1433.66 & 2857.74 & 41.0251 & -423.681 & 850.359 \\
 $\beta_2 $& -3.2827 & -9.05138 & -12.4154 & 55.4716 & -106.051 & -11.953 & 76.807 & -155.332 & -3.41293 & 25.5724 & -54.408 \\
 $\beta_3 $& -2.51564$\times 10^{-5}$ & 1.97503$\times 10^{-5}$ & -1.83707$\times 10^{-5}$ & 2.18863$\times 10^{-5}$ & 8.25024$\times 10^{-5}$ & 7.15737$\times 10^{-6}$ & -5.578$\times 10^{-5}$ & 1.91421$\times 10^{-4}$ & 5.44717$\times 10^{-6}$& -3.22061$\times 10^{-5}$ & 7.97402$\times 10^{-5}$ \\
 $\alpha _1$ & 43.3151 & 638.633 & -32.8577 & 2415.89 & -5766.88 & -61.8546 & 2953.97 & -8986.29 & -21.5714 & 981.216 & -3239.57 \\
 $\alpha _2$ & -0.0702021 & -0.162698 & -0.187251 & 1.13831 & -2.83342 & -0.17138 & 1.71975 & -4.53972 & -0.0499834 & 0.606207 & -1.68277 \\
 $\alpha _3$ & 9.59881 & -397.054 & 16.2021 & -1574.83 & 3600.34 & 27.0924 & -1786.48 & 5152.92 & 11.1757 & -577.8 & 1808.73 \\
 $\alpha _4$ & -0.0298949 & 1.40221 & -0.0735605 & 0.833701 & 0.224001 & -0.0552029 & 0.566719 & 0.718693 & -0.0155074 & 0.157503 & 0.210768 \\
 $\alpha _5$ & 0.997441 & -0.00788445 & -0.0590469 & 1.39587 & -4.51663 & -0.0558534 & 1.75166 & -5.99021 & -0.0179453 & 0.59651 & -2.06089 \\ \hline \hline
\end{tabular}
\caption{
    \label{tab:coefftable} 
    Coefficient values for the mapping functions given in Eq.~(\ref{eqn:mapping}).
    These values are calculated under the parametrisation ($\eta$, $\chi_{\rm{PN}}$)
    }
\end{sidewaystable}

\bibliography{PhenomPaper2}

\end{document}